\documentclass[pre,aps,twocolumn,amsmath,amssymb,amsfonts,floatfix,superscriptaddress]{revtex4}

\usepackage{graphicx}% include figure files
\usepackage{dcolumn}% align table columns on decimal point
\usepackage{bm}% bold math
\usepackage[hidelinks]{hyperref}
\usepackage{latexsym}
\usepackage{float}
\usepackage{supertabular}
\usepackage{longtable}
\usepackage{makecell} % this allows line breaks in a cell
\usepackage{amsmath}
\usepackage{amssymb}
\usepackage{textcase}
\usepackage[caption=false]{subfig}
\usepackage[english]{babel}

\usepackage{color}
\usepackage{verbatim}
\usepackage{comment,xcolor}

\newcommand{\etal}{\hbox{\textit{et al.}}}

\newcommand{\PFCalpha}{\hbox{PFC-$\alpha$}}
\newcommand{\PFCbeta}{\hbox{PFC-$\beta$}}
\newcommand{\PFCgamma}{\hbox{PFC-$\gamma$}}
\newcommand{\PFCdelta}{\hbox{PFC-$\delta$}}
\newcommand{\PFCepsilon}{\hbox{PFC-$\epsilon$}}

\newcommand{\bx}{{\bm x}}
\newcommand{\bk}{{\bm k}}
\newcommand{\rd}{{\rm d}}
\newcommand{\rdbx}{\rd\bx}
\newcommand{\rdt}{\rd{t}}
\newcommand{\cF}{{\cal F}}
\newcommand{\cFex}{{\cal F_{\text{ex}}}}

\newcommand{\cUext}{U_{\text{ext}}}
\newcommand{\cFa}{{{\cal F}_{1}}}
\newcommand{\cFb}{{{\cal F}_{2}}}
\newcommand{\cFc}{{{\cal F}_{3}}}

\newcommand{\cFe}{{{\cal F}_{5}}}
\newcommand{\cFPFCalpha}{{\cal F_{\text{$\alpha$}}}}
\newcommand{\cFPFCbeta}{{\cal F_{\text{$\beta$}}}}
\newcommand{\cFPFCgamma}{{\cal F_{\text{$\gamma$}}}}

\newcommand{\cFPFCepsilon}{{\cal F_{\text{$\epsilon$}}}}
\newcommand{\cG}{{\cal G}}
\newcommand{\cH}{{\cal H}}
\newcommand{\deltaX}{{\delta{\mkern-1mu}X}}

\newcommand{\cO}{{\cal O}}
\newcommand{\cL}{{\cal L}}
\newcommand{\cLgrad}{{{\cal L}_{\text{grad}}}}
\newcommand{\cLgradJ}{{{\cal L}_{\text{grad-8}}}}
\newcommand{\cLgradinv}{{{\cal L}_{\text{grad}}^{-1}}}

\newcommand{\deltacF}{{\delta{\mkern-1mu}\cF}}
\newcommand{\deltacG}{{\delta{\mkern-1mu}\cG}}
\newcommand{\deltaOmega}{{\delta{\mkern-1mu}\Omega}}
\newcommand{\deltacFex}{{\delta{\mkern-1mu}\cFex}}
\newcommand{\deltacFa}{{\delta{\mkern-1mu}\cFa}}
\newcommand{\deltacFb}{{\delta{\mkern-1mu}\cFb}}
\newcommand{\deltacFc}{{\delta{\mkern-1mu}\cFc}}

\newcommand{\deltacFe}{{\delta{\mkern-1mu}\cFe}}
\newcommand{\deltacFPFCalpha}{{\delta{\mkern-1mu}\cFPFCalpha}}
\newcommand{\deltacFPFCbeta}{{\delta{\mkern-1mu}\cFPFCbeta}}
\newcommand{\deltacFPFCgamma}{{\delta{\mkern-1mu}\cFPFCgamma}}

\newcommand{\deltarho}{{\delta{\mkern-2mu}\rho}}
\newcommand{\deltan}{{\delta{\mkern-1mu}n}}
\newcommand{\Deltarho}{{\Delta{\mkern-1.0mu}\rho}}
\newcommand{\Deltamu}{{\Delta{\mkern-0.5mu}\mu}}
\newcommand{\Dbx}{{\Delta{\mkern-1.0mu}\bx}}
\newcommand{\sigmamax}{\sigma_{\text{max}}}
\newcommand{\nliq}{n_{\text{liq}}}
\newcommand{\Omegaliq}{\Omega_{\text{liq}}}

\begin{document}

\title{Deriving phase field crystal theory from dynamical density functional theory: consequences of the approximations}

\author{Andrew J. Archer}
\email{A.J.Archer@lboro.ac.uk}
\affiliation{Department of Mathematical Sciences, Loughborough University, Loughborough LE11 3TU, U.K.}

\author{Daniel J. Ratliff}
\email{D.J.Ratliff@lboro.ac.uk}
\affiliation{Department of Mathematical Sciences, Loughborough University, Loughborough LE11 3TU, U.K.}

\author{Alastair M. Rucklidge}
\email{A.M.Rucklidge@leeds.ac.uk}
\affiliation{School of Mathematics, University of Leeds, Leeds LS2 9JT, U.K.}

\author{Priya Subramanian}
\email{P.Subramanian@leeds.ac.uk}
\affiliation{School of Mathematics, University of Leeds, Leeds LS2 9JT, U.K.}

\begin{abstract}
Phase field crystal (PFC) theory is extensively used for modelling the phase
behaviour, structure, thermodynamics and other related properties of solids.
PFC theory can be derived from dynamical density functional theory (DDFT) via a
sequence of approximations. Here, we carefully identify all of these
approximations and explain the consequences of each. One approximation that is
made in standard derivations is to neglect a term of form
$\nabla\cdot[n\nabla\cL n]$, where $n$ is the scaled density profile and $\cL$
is a linear operator. We show that this term makes a significant contribution
to the stability of the
crystal, and that dropping this term from the theory forces another
approximation, that of replacing the logarithmic term from the ideal gas
contribution to the free energy with its truncated Taylor expansion, to yield a
polynomial in~$n$. However, the consequences of doing this are: (i)~the
presence of an additional spinodal in the phase diagram, so the liquid is
predicted first to freeze and then to melt again as the density is increased; and
(ii)~other periodic structures, such as stripes, are erroneously predicted to
be thermodynamic equilibrium structures. In general, $\cL$~consists of a
non-local convolution involving the pair direct correlation function. A second
approximation sometimes made in deriving PFC theory is to replace $\cL$ by a
gradient expansion involving derivatives. We show that this leads to the
possibility of the density going to zero, with its logarithm going to~$-\infty$
whilst being balanced by the fourth derivative of the density going
to~$+\infty$. This subtle singularity leads to solutions failing to exist above
a certain value of the average density. We illustrate all of these conclusions
with results for a particularly simple model two-dimensional fluid, the
generalised exponential model of index~4 (GEM-4), chosen because a DDFT is
known to be accurate for this model. The consequences of the subsequent PFC
approximations can then be examined. These include the phase diagram being both
qualitatively incorrect, in that it has a stripe phase, and quantitatively
incorrect (by orders of magnitude) regarding the properties of the crystal
phase. Thus, although PFC models are very successful as phenomenological models
of crystallisation, we find it impossible to derive the PFC as a theory for the
(scaled) density distribution when starting from an accurate DDFT, without
introducing spurious artefacts. However, we find that making a simple one-mode
approximation for the \emph{logarithm} of the density distribution~$\log\rho(\bx)$ 
(rather than for~$\rho(\bx)$), is surprisingly
accurate. This approach gives a tantalising hint that accurate PFC-type theories
may instead be derived as theories for the field $\log\rho(\bx)$, 
rather than for the density profile itself.

 \end{abstract}

%\pacs{*** For PRX. Need the `Physics Subject Headings', 100-word concise and compelling
%justification, Referee suggestions (Ken Elder, Zhi-Feng Huang, Hartmut Lowen, Nikolas
%Provotas, Heike Emmerich), 250 word popular summary, PRX subject areas (Soft Matter,
%Nonlinear Dynamics), cover letter, which editor to submit to (Bulbul Chakraborty)}

\maketitle

\section{Introduction}

The phase field crystal (PFC) theory for matter is widely used and
has been successfully applied to describe a broad range of phenomena,
including, for example,
grain boundary dynamics \cite{Elder2002, Elder2004}, crystal nucleation
\cite{Backofen2010,Toth2010}, crystal growth \cite{Archer2012}, glass
formation~\cite{Berry2008a}, crack propagation~\cite{Elder2004} and many other properties of
condensed matter. For more background and examples of situations to which
the PFC theory has been applied, see the excellent review~\cite{Emmerich2012}.
The PFC theory was originally proposed, in the spirit of `regular' phase
field theory (PFT), as a diffuse-interface theory for the time evolution
of an order parameter field~\cite{Elder2002}. The equations of PFT are obtained via
symmetry, thermodynamic
and other arguments and the result is a theory that is widely used
in materials science and other disciplines to model the structure of materials.
For more background on PFT see for example 
Ref.~\cite{Boettinger2002} and references therein.

The central and original idea in extending PFT to arrive
at PFC theory is to incorporate aspects of the microscopic structure of the
material
into the model~\cite{Elder2002}. The result is a theory that operates on
atomic length scales and diffusive time scales~\cite{Emmerich2012}.
By this we mean that PFC theory is a theory for a field that exhibits numerous
maxima, each of which is identified as the average location of the atoms
(or more generally `particles') in the system. This idea is powerful
because, by building into the theory more information about
the underlying material structure, it enables the inclusion of much more
of the physics coming
from particle correlations to be incorporated. Over the years
several variants of PFC theory have been developed that are able
to describe a range different crystalline (and even quasicrystalline) structures
\cite{Jaatinen2009, Pisutha-Arnond2013b, Wu2010, Barkan2011, Achim2014, Subramanian2016, Jiang2017, Savitz2018}.

Thus, the original PFC \cite{Elder2002} may be viewed as the simplest
partial differential equation model one can conceive of
for a conserved order parameter exhibiting peaks arranged
with crystalline ordering. It is obtained from a (scaled)
free energy~$\cFPFCalpha$ that is a functional of the dimensionless order
parameter~$n$:
 \begin{equation}
 \cFPFCalpha[n] = \int \left(\frac{1}{2}n\left((k_s^2+\nabla^2)^2 - r\right)n
                    + \frac{1}{4}n^4\right)\rdbx,
 \label{eq:PFCFalpha}
 \end{equation}
where $n(\bx,t)$ is a field that depends on position in
space $\bx$ and on time $t$, and $k_s$~is an
inverse length scale that determines the lattice spacing of the crystal.
The parameter~$r$ defines how near the system is to freezing.
The time evolution of the conserved field~$n$ is given by the dynamics
 \begin{equation}
 \frac{\partial n}{\partial t} = \nabla^2 \left(\frac{\deltacFPFCalpha}{\deltan}\right)
                               = - \nabla^2 \left(rn - (k_s^2+\nabla^2)^2n - n^3\right),
 \label{eq:PFCalphadynamics}
 \end{equation}
where $\frac{\deltacFPFCalpha}{\deltan}$ is the functional derivative of
$\cFPFCalpha$ with respect to~$n(\bx)$.

Given the ingredients in the model, it is therefore no surprise that PFC theory
is closely related to the Swift--Hohenberg equation~\cite{Swift1977}:
 \begin{equation}
 \frac{\partial n}{\partial t} = - \frac{\deltacFPFCalpha}{\deltan}
                               = rn - (k_s^2+\nabla^2)^2n - n^3,
 \label{eq:SwiftHohenberg}
 \end{equation}
which is one of the archetypal equations in pattern formation theory.
As one can see above,
both the Swift--Hohenberg equation and PFC theory
can be expressed as a different type of dynamics
based on the same free energy functional~\cite{Emmerich2012}.
The Swift--Hohenberg equation~(\ref{eq:SwiftHohenberg}) is based on an underlying dynamics
that seeks to minimise the free energy over time, whilst the PFC
dynamics (\ref{eq:PFCalphadynamics}), which also decreases the free energy over time,
in addition enforces a conservation of the average value of the order parameter
in the system. Thus, the PFC equation~(\ref{eq:PFCalphadynamics})
is sometimes referred to as the conserved Swift--Hohenberg
equation~\cite{Thiele2013, Sagui1994, Knobloch2015, Matthews2000, Emmerich2012}.

In the years after PFC theory was originally proposed it was realised
that it could be derived from
classical dynamical density functional theory (DDFT) \cite{Elder2007,
Teeffelen2009, Huang2010a, Emmerich2012, Archer2012}, via
a sequence of several different approximations.
Below, we say much more on what these approximations are. DDFT
is a theory for the time evolution of the ensemble average one-body
(number) density profile $\rho(\bx,t)$, for a non-equilibrium system
of interacting classical particles. DDFT is based
on equilibrium density functional theory (DFT)
\cite{Evans1979a, Evans1992, Hansen2013} and for an equilibrium
system, DDFT
is equivalent to DFT. DDFT was originally developed as a
theory for Brownian particles with over-damped stochastic equations
of motion~\cite{Marconi1999, Marconi2000, Archer2004,Archer2004a},
but it has also been extended to describe the dynamics of under-damped
systems and atomic or molecular systems where the particle
dynamics is governed by Newton's equations of
motion~\cite{Archer2006, Archer2009, Goddard2012, Goddard2013, DuranOlivencia2017, Schmidt2018}.
This body of work shows that when such systems are not too far from equilibrium,
then the dynamics predicted by the original DDFT is still often correct
in the long-time limit where the particle dynamics is dominated by diffusive
processes. This is because
DDFT corresponds to a dynamics given by the continuity
equation
 \begin{equation}
 \frac{\partial \rho}{\partial t} = -\nabla \cdot \mathbf{j},
 \label{eq:continuity}
 \end{equation}
where the current $\mathbf{j}\propto-\nabla \mu(\bx,t)$,
with $\mu(\bx,t)$ a local (non-equilibrium) chemical
potential~\cite{Marconi1999, Marconi2000, Archer2004,Archer2004a}.
Eq.~(\ref{eq:continuity}) is of course expected since the total number of particles in the
system $N=\int \rho(\bx,t)\rdbx$ is a conserved quantity.

Refs.~\cite{Elder2007, Teeffelen2009, Huang2010a,
Emmerich2012, Archer2012} give various different derivations of the PFC model,
starting from DFT and/or DDFT. Here, starting from DDFT, we systematically
show how all
the various different theories are related and we identify and highlight
the significance of each of the approximations that are made in the derivation
of PFC theory.
We show that there is a particular term of the form $\nabla\cdot[n\nabla\cL n]$,
where $\cL$ is a linear operator, that is almost universally
neglected because it is `of higher order'~\cite{Huang2010a}, but this term is actually
important for stabilizing crystalline structures: its contribution is of the same order
as some of the terms that are retained. As we explain in detail,
neglecting this term 
essentially forces one to make the
Taylor expansion of the ideal gas
logarithmic term in the free energy in order to recover something physically
reasonable. We show that neglecting this
term, as is done in PFC theory, {and the subsequent replacement of the 
logarithm by its Taylor series,} leads to the spurious appearance
in the phase diagram of 
an extra spinodal and also alters the relative stabilities of the
crystal state compared to a stripe phase and also other phases,
leading in two dimensions (2D) to the stripe phase becoming
the global free energy minimum state for certain parameter values.
Essentially, all this behaviour originates because the function
$\log(1+n)$ has one root, but when
it is replaced by a truncated Taylor expansion, the resulting polynomial
generally has two roots. Our arguments
also directly apply in three dimensions to explain why lamellar
phases occur as equilibrium phases in PFC theory.
Recall that most PFC theories predict that as one moves in the phase diagram away
from the region where there is coexistence between the liquid and the crystal,
moving deeper into the crystalline portion of the phase diagram,
such stripe/lamellar phases appear as equilibrium structures and are global
minima of the
free energy~\cite{Emmerich2012}. Of course, particles with isotropic pair
interactions generally never
`freeze' to form striped phases, unless they have an unusual and special
form for the pair potential between the particles~\cite{Imperio2004, Imperio2006,
Archer2007}. DDFT, taken together with a reliable approximation for the Helmholtz
free energy functional of course does not predict such stripe phases
for crystallisation from simple liquids.

The linear operator $\cL$ has the form of a
non-local convolution involving the pair direct correlation function
plus another simpler term (see Eq.~\eqref{eq:DDFTlinear} below).
Another approximation that is often made in deriving PFC theories
is to approximate $\cL$ by a gradient expansion involving derivatives.
We show below that if one makes this approximation whilst simultaneously
retaining the logarithmic term from the ideal gas free energy, this
results in a theory that still predicts reasonably accurately the freezing
transition, but as one increases the average density, moving
deeper into the region of the phase diagram
where the crystal phase occurs, there is a point where
$\rho(\bx)\to0$ at the points in space $\bx$ between the density peaks,
where the density is a minimum. On increasing the average density beyond
this point in the phase diagram, there is no solution to the theory. We
analyse in detail this singular behaviour. As $\rho(\bx)\to0$
we have $\log\rho(\bx)\to-\infty$, of course. In the equation for the
equilibrium density profile this divergence is initially balanced by the term involving
the fourth  derivative, $\partial^4\rho/\partial x^4\to+\infty$. However, when the
average density in the system is increased beyond the value at which this
divergence occurs, we find there is no solution. 

We illustrate these conclusions by finding the predicted structures and
phase diagram for the 2D version of the GEM-4
(Generalised Exponential Model of index 4)~\cite{Mladek2006, Prestipino2014},
chosen because DDFT based on a simple approximation (the so-called
random phase approximation (RPA) \cite{Likos2001}) for the
Helmholtz free energy functional can be very accurate for predicting
the equilibrium structures formed in this model and also the
thermodynamics~\cite{Mladek2007, Prestipino2014, Archer2014}.
At higher temperatures, the 2D GEM-4 system exhibits
just a single fluid phase and at higher densities a single crystal phase.
At lower temperatures, where the RPA DDFT is no longer
accurate, there is a hexatic phase and multiple crystalline phases
as the density is increased \cite{Prestipino2014}. Here we do not consider
this regime, restricting ourselves to the regime where there is just one
fluid and one crystal phase, which are predicted accurately by
the RPA DDFT. This 
enables us to investigate the effect of making subsequent
approximations to the DDFT, including those made to derive PFC theory.
We find that the PFC type theories
spuriously predict three additional phases that are in reality not
present in the phase diagram (i.e., are not thermodynamically stable). 
These are (i)~a stripe phase,
(ii)~what we refer to as
`down hexagons' (in contrast to the true crystal structure,
which we refer to as `up hexagons') and then at even higher
densities a melting to form
(iii)~another uniform liquid phase. We show
how the approximations made in deriving the PFC result in these
structures being predicted.

The final contribution of this paper is to show that there is a very simple
and accurate ansatz one can make for the form of the equilibrium crystal
density profile in DDFT (and so also for DFT, of course). The ansatz is
$\rho(\bx)=\rho_0e^{\phi(\bx)}$, where $\rho_0$ is a constant and the
field $\phi(\bx)$ is approximated by a sinusoid of the form
$\phi(\bx)\approx\phi_0+\phi_1e^{i\bk\cdot\bx}+$complex conjugate
(in one dimension), plus other similar terms (in higher dimensions),
where $\phi_0$ and $\phi_1$ are constants. The results presented here are
for the GEM-4 model and show why this approximation is unexpectedly accurate:
the approximation is able to replicate almost exactly the numerical solution
to the DDFT problem, from small to arbitrarily large amplitude density variations.
We expect this ansatz also to be reliable
for other systems. This form of one-mode theory gives a hint for future
directions to develop accurate PFC-type theories, since using a one-mode
approximation in PFC is often fairly accurate.

%*** Do we want to state our main conclusions?

This paper is structured as follows: In Sec.~\ref{sec:2} we present our
systematic step-by-step derivation of PFC, starting from DDFT.
After each approximation, we carefully state the model, i.e., we
give the corresponding free energy functional and also the
expression for the chemical potential, which is a quantity that is a constant
at all points in space for equilibrium states. In order to keep track
of the different orders in which the approximations can be made, we give each
model a name, starting with \PFCalpha\ for the original PFC model
in Eq.~(\ref{eq:PFCalphadynamics}) above, and
with DDFT-0 for the original formulation of DDFT below. The different
DDFT approximations result in five different versions, DDFT-1 to
DDFT-5. Similarly, we explain the various different approximations
that can be made to each of these, leading to a corresponding
PFC theory, which we name \PFCalpha\ to \PFCepsilon.
Note that the criterion we use here for distinguishing between
whether we refer to a theory as a DDFT or a PFC is based on
whether the free energy
which is minimised by the dynamical equations (i.e., the Lyapunov
functional) has the logarithmic ideal gas term or not: if it does not
have the logarithm, we refer to it as a~\hbox{PFC}.
{Table~\ref{tab:DDFTvsPFC} below is there to help the reader navigate the
various models and the approximations made in each one.}
Sec.~\ref{sec:2} concludes with a
summarising discussion. In Sec.~\ref{sec:3} we present results for the GEM-4
system comparing predictions for the density profiles and thermodynamics of
equilibria for two of the different DDFT theories and also two of the PFC
theories. In this section we also present the phase diagrams for the GEM-4
system predicted by these different DDFT and PFC theories. By comparing all of
these we are able to assess the accuracy of the different theories and the
validity of the various approximations. In Sec.~\ref{sec:4} we discuss the
implications of the main two approximations and analyse the singular behaviour
displayed by some models. In Sec.~\ref{sec:5} we introduce the ansatz
$\rho(\bx)=\rho_0e^{\phi(\bx)}$ and derive the new one-mode approximation
for~\hbox{DDFT}. We draw our conclusions in Sec.~\ref{sec:6}. The paper
includes two appendices in which we describe the numerical (continuation)
methods we use to calculate the density profiles.

\section{Derivation of the Phase Field Crystal model from DDFT}\label{sec:2}

In this section we progress from the original formulation of DDFT (which we
call DDFT-0) through a series of approximations (DDFT-1, \dots, DDFT-5), as
listed in Table~\ref{tab:DDFTvsPFC}. Our main starting point is DDFT-2. From
this point, there are three main approximations that can be made (or not made):
(i)~the Ramakrishan--Yussouff (RY) or the random phase approximation (RPA)
for the free energy;
(ii)~the gradient expansion of the convolution term; and (iii)~the
Taylor expansion of the logarithmic term. Making (or not making) the first two of these
approximations results in DDFT-3, DDFT-4 and DDFT-5. Then, making the third
approximation from DDFT-2 results in \PFCbeta, from DDFT-3 results in
\PFCgamma, and so on up to~\PFCepsilon. The \PFCepsilon\ model can be rescaled
to recover the original version of~\hbox{PFC}, \PFCalpha, see 
Eqs.~\eqref{eq:PFCFalpha} and~\eqref{eq:PFCalphadynamics}. The various models are
summarised in Table~\ref{tab:DDFTvsPFC}. Amongst the models we present below,
DDFT-5 is equivalent to the model derived by Huang~\etal~\cite{Huang2010a} and
advocated by van Teeffelen \etal~\cite{Teeffelen2009} (named PFC1 in that
paper), and DDFT-3 and \PFCepsilon\ are equivalent to the models named
DDFT and PFC2 by van Teeffelen \etal~\cite{Teeffelen2009}.

%\begingroup
%\squeezetable
\begin{table*}
\begin{center}
\begin{ruledtabular}
\begin{tabular}{l|ccccc|cccc}
Name  % & Comments 
      & \makecell{Truncate\\at $\cO(c^{(4)})$}
      & \makecell{LDA (\ref{eq:definecs}) for\\$c^{(3)}$ and $c^{(4)}$}
      & \makecell{RY/RPA:\\ $c^{(3)}=c^{(4)}=0$}
      & \makecell{Gradient\\expansion\\of~$\cL$ (\ref{eq:cLgraddefn})}
      & \makecell{Constant mobility,\\
                  expand $\log(1+n)$}
      & Dynamics & \makecell{Free\\energy} & \makecell{Chemical\\potential} & $Q$, $C$, $R$ \\ [2pt]
\noalign{\vspace{4pt}}
\hline
\noalign{\vspace{4pt}}
\PFCalpha % & \makecell[l]{Phenomenological model:\\conserved Swift--Hohenberg\\equation}
      & Yes
      & N/A
      & Yes
      & Yes
      & Yes
      & (\ref{eq:PFCalphadynamics})
      & (\ref{eq:PFCFalpha})
      & ---
      & \makecell{$Q=0$\\[0.5ex] $C=-1$}
      \\ [4pt]
\noalign{\vspace{4pt}}
\hline
\noalign{\vspace{4pt}}
DDFT-0 % & Original DDFT
      &
      &
      &
      &
      &
      & (\ref{eq:DDFT0dynamics})
      & (\ref{eq:separatedF})
      & (\ref{eq:chempotdefn})
      & ---
      \\ [4pt]
DDFT-1 % & --- %Truncated at $\cO(c^{(4)})$, otherwise as DDFT-0
      & Yes
      &
      &
      &
      &
      & (\ref{eq:DDFT1dynamics}), (\ref{eq:DDFT1deltaF})
      & (\ref{eq:DDFTF1})
      & (\ref{eq:ChemPot}), (\ref{eq:betadFdn1})
      & ---
      \\ [4pt]
\noalign{\vspace{4pt}}
\hline
\noalign{\vspace{4pt}}
DDFT-2 % & --- %$c^{(3)}$ and $c^{(4)}$ simplified as in (\ref{eq:definecs}): zero wavenumber only
      & Yes
      & Yes
      &
      &
      &
      & (\ref{eq:DDFT2dynamics})
      & (\ref{eq:DDFTF2})
      & (\ref{eq:betadFdn2})
      & (\ref{eq:qandc})
      \\ [4pt]
DDFT-3 % & --- % as DDFT-2 with RY/RPA: $c^{(3)}=c^{(4)}=0$
      & Yes
      & N/A
      & Yes
      &
      &
      & (\ref{eq:DDFT3dynamics})
      & (\ref{eq:DDFTF3})
      & (\ref{eq:betadFdn3})
      & \makecell{$Q=\tfrac{1}{2}$\\[0.5ex]$C=0$\\[0.5ex] $R=0$}
      \\ [4pt]
DDFT-4 % & --- % as DDFT-2 with gradient expansion of~$\cL$ (\ref{eq:cLgraddefn})
      & Yes
      & N/A
      &
      & Yes
      &
      & (\ref{eq:DDFT4dynamics})
      & as (\ref{eq:DDFTF2})
      & as (\ref{eq:betadFdn2})
      & (\ref{eq:qandc})
      \\ [4pt]
DDFT-5 % & --- % as DDFT-2 with RY/RPA and gradient expansion of~$\cL$
      & Yes
      & N/A
      & Yes
      & Yes
      &
      & (\ref{eq:DDFT5dynamics})
      & (\ref{eq:DDFTF5})
      & (\ref{eq:betadFdn5})
      & \makecell{$Q=\tfrac{1}{2}$\\[0.5ex]$C=0$\\[0.5ex] $R=0$}
      \\ [4pt]
\noalign{\vspace{4pt}}
\hline
\noalign{\vspace{4pt}}
\PFCbeta  % & --- % \makecell[l]{Constant mobility,
                %          $\nabla\cdot\left[n\nabla\cL{n}\right]$ term absent,\\
                %          expanded $\log(1+n)$ leading to additional nonlinear terms}
      & Yes
      & Yes
      &
      &
      & Yes
          & (\ref{eq:PFCdynamics}), (\ref{eq:PFCbetadynamics})
          & (\ref{eq:DDFTFPFCbeta})
          & (\ref{eq:betadFdnbeta})
          & (\ref{eq:PFCqandc})
          \\ [4pt] 
\PFCgamma % & --- % as \PFCbeta\ with RY/RPA: $c^{(3)}=c^{(4)}=0$
      & Yes
      & N/A
      & Yes
      &
      & Yes
          & (\ref{eq:PFCgammadynamics})
          & (\ref{eq:DDFTFPFCgamma})
          & (\ref{eq:betadFdngamma})
          & \makecell{$Q=\tfrac{1}{2}$\\[0.5ex]$C=-\frac{1}{3}$}
          \\ [4pt] 
\PFCdelta % & --- % as \PFCbeta\ with gradient expansion of~$\cL$ (\ref{eq:cLgraddefn})
      & Yes
      & N/A
      &
      & Yes
      & Yes
          & as (\ref{eq:PFCbetadynamics})
          & as (\ref{eq:DDFTFPFCbeta})
          & as (\ref{eq:betadFdnbeta})
          & (\ref{eq:PFCqandc})
          \\ [4pt] 
\PFCepsilon % & equivalent to \PFCalpha % \makecell[l]{as \PFCbeta\ with RY/RPA and gradient expansion of~$\cL$}
      & Yes
      & N/A
      & Yes
      & Yes
      & Yes
            & (\ref{eq:PFCepsilondynamics})
            & as (\ref{eq:DDFTFPFCgamma})
            & as (\ref{eq:betadFdngamma})
            & \makecell{$Q=\tfrac{1}{2}$\\[0.5ex]$C=-\frac{1}{3}$}
            \\ [4pt] 
\end{tabular}
\end{ruledtabular}
\end{center}
\caption{Various versions of DDFT and PFC, in order of appearance, along with 
references to the equations defining the dynamics, the free energy and the 
chemical potential. We also give the quadratic ($Q$), cubic ($C$) and
quartic ($R$) coefficients. \PFCalpha\ is
the phenomenological model, also known as the conserved Swift--Hohenberg equation;
\PFCepsilon\ is equivalent to~\PFCalpha.
}
 \label{tab:DDFTvsPFC}
\end{table*}
%\endgroup

\subsection{Dynamic Density Functional Theory: DDFT-0}

The starting point for all of our derivations is the key DDFT
equation~\cite{Marconi1999, Marconi2000, Archer2004,Archer2004a}:
 \begin{equation}
 \frac{\partial \rho}{\partial t} = \nabla \cdot \left[ \beta M(\rho) \nabla \frac{\deltacF}{\deltarho} \right],
 \label{eq:DDFT0dynamics}
 \end{equation}
where $\beta=(k_BT)^{-1}$ (with $k_B$ being Boltzmann's constant and $T$ being
temperature), $M(\rho)$~is the positive $\rho$-dependent mobility. The
Helmholtz free energy~$\cF[\rho]$ depends on the density profile~$\rho(\bx,t)$
integrated over space; hence $\cF[\rho]$ depends on time but not on
position~\cite{Archer2004}. The expression ${\deltacF}/{\deltarho}$ is the
functional derivative of~$\cF$ with respect to~$\rho(\bx,t)$, which therefore
depends on both time and on position. DDFT usually takes $M(\rho)=D \rho$,
i.e., the mobility is proportional to density~\cite{Marconi1999,
Marconi2000, Archer2004,Archer2004a}, where $D$~is the diffusion coefficient.
We henceforth scale time so that $D=1$. With boundary conditions that do not
allow material to enter or leave the system, $N=\int\!\rho(\bx)\rdbx$ (or
equivalently, the mean density) is a constant of the motion and is the total
number of particles in the system.

With suitable boundary conditions, one can readily show that the Helmholtz free
energy decreases monotonically with time:
 \begin{equation}
 \frac{\rd\cF}{\rdt} = - \int \! \beta M(\rho)
         \left| \nabla \frac{\deltacF}{\deltarho} \right|^2 \rdbx
 \leq 0,
 \label{eq:Fdecreases}
 \end{equation}
so (assuming that $\cF[\rho]$ is bounded below) the system typically
evolves to a (local)
minimum of~$\cF$, which is a
dynamically stable equilibrium of~(\ref{eq:DDFT0dynamics}). Here, `dynamically
stable' means that small perturbations
away from the equilibrium decay,
and `equilibrium' means that ${\partial \rho}/{\partial t} =0$
and ${\rd\cF}/{\rdt}=0$. Owing to the dynamics being governed by a
continuity equation \eqref{eq:continuity}, such perturbations cannot
change the mean density.
Local minima of~$\cF$ that are not the global minimum are thermodynamically
metastable. The system can also have dynamically unstable equilibria, for
which~$\cF$ is a saddle or maximum. From~(\ref{eq:Fdecreases}), we see that
all equilibria of~(\ref{eq:DDFT0dynamics}) satisfy
$\nabla(\deltacF/\deltarho)=0$, so
 \begin{equation}
 \frac{\deltacF}{\deltarho} = \text{constant} = \mu,
 \label{eq:chempotdefn}
 \end{equation}
where $\mu$ is the chemical potential of the equilibrium. This is of course
the Euler--Lagrange equation for the problem of finding stationary points of
the functional~$\cF[\rho]$, subject to the constraint of fixed mean density.
Note however that when evolving~(\ref{eq:DDFT0dynamics}) forward in time
from an arbitrary
initial condition, $\mu$~is not necessarily known \emph{a priori}. 

The theory can also be cast in terms of the grand potential (also called the 
Landau free energy) functional~\cite{Evans1979a, Evans1992, Hansen2013}:
 \begin{equation}                                                              
 \Omega[\rho] = \cF[\rho] - \mu N = \cF[\rho] - \mu \int \! \rho(\bx)\rdbx.
 \label{eq:omegadefn}
 \end{equation}                                   
From this it follows that the functional derivative of $\Omega$ is
 \begin{equation}
 \frac{\deltaOmega}{\deltarho} = \frac{\deltacF}{\deltarho} - \mu,
 \label{eq:dOmega}
 \end{equation}
and that this is zero at equilibrium: equilibria are extreme values
of~$\Omega$. Like the Helmholtz free energy, the grand potential decreases
monotonically with time, since Eq.~\eqref{eq:Fdecreases} is also
true if one replaces $\cF$ by~$\Omega$.
Therefore, for two phases to coexist, they must have the same specific grand potential
(i.e., the same pressure) and the same chemical
potential. Thus, the global minimum of $\Omega$ for a given $\mu$ and $T$
is the thermodynamic equilibrium state of the
system~\cite{Evans1979a, Evans1992, Hansen2013}.

Following the usual approach in DFT,
we separate the Helmholtz free energy into three parts: the
`ideal gas' contribution, which is proportional to the temperature but takes
no account of particle interactions,
an excess ($\cFex$) over the ideal gas contribution arising from the particle
interactions, and the contribution due to an
external potential~$\cUext(\bx)$, as follows~\cite{Evans1979a, Evans1992, Hansen2013}:
 \begin{equation}
 \cF[\rho] = k_BT \int \rho\left(\log(\Lambda^d\rho) - 1\right)\rdbx + \cFex[\rho] + \int\rho\cUext\rdbx,
 \label{eq:separatedF}
 \end{equation}
where the integral is taken over the volume~$V$ in three dimensions ($d=3$)
(or the area in 2D, $d=2$) and where $\Lambda$ is the thermal
de~Broglie wavelength. Since for our purposes here
the value of $\Lambda$ is irrelevant (changing $\Lambda$ will shift the values
of $\cF$ and~$\mu$ by constants), we henceforth
set $\Lambda=1$. We also consider bulk systems and so we assume that $\cUext=0$.
With the separation in Eq.~(\ref{eq:separatedF}), we have
 \begin{equation}
 \beta\frac{\deltacF}{\deltarho} = \log\rho +
                                   \beta\frac{\deltacFex}{\deltarho},
 \label{eq:separatedFdrho}
 \end{equation}
which gives
 \begin{equation}
 \beta\nabla\frac{\deltacF}{\deltarho}=\frac{1}{\rho}\nabla\rho+\dots,
 \label{eq:rhoinverse}
 \end{equation}
 where on the right hand side we only explicitly write the contribution
 from the ideal gas part of the free energy. Inserting this into
 Eq.~(\ref{eq:DDFT0dynamics}) with $M=D\rho$ we obtain
 \begin{equation}
 \frac{\partial \rho}{\partial t} = \nabla^2 \rho+\dots,
 \label{eq:rhodiffusion}
 \end{equation}
in which the coefficient in front of the term $\nabla^2 \rho$ is~$D$,
but our choice of time scaling has~$D=1$.
Note that this term is \emph{linear} in~$\rho$, in spite of
it originating from a \emph{nonlinear} logarithmic contribution to the free energy.

We refer to the model up to this point as DDFT-0.

\subsection{Expansion of~$\cFex$: DDFT-1}

To proceed, we must have an expression for the excess Helmholtz free energy
functional $\cFex[\rho]$. We use a functional Taylor expansion,
which is also that used in all derivations of
PFC theory. This gives
the free energy functional of the system of interest in terms of properties of a
reference system, which are assumed to be known.
The reference system that is chosen is a uniform liquid, with constant density~$\rho_0$.
The density profile of the system of interest may be varying in space
and with an average density that may be different from~$\rho_0$.
The functional Taylor series expansion of the excess free energy can be written
in terms of the density difference $\Deltarho(\bx,t)=\rho(\bx,t)-\rho_0$
as follows \cite{Evans1992, Hansen2013}:
 \begin{equation}
 \begin{split}
 \cFex[\rho] &= \cFex[\rho_0]
   - k_BT \!\! \int \! c^{(1)}(\bx_1)\Deltarho(\bx_1)\rdbx_1 \\
 & \quad{}-\frac{k_BT}{2!} \!\! \int \!
                          c^{(2)}(\bx_1,\bx_2)
                          \Deltarho(\bx_1)
                          \Deltarho(\bx_2)
                          \rdbx_1\rdbx_2 \\
 & \quad{}-\frac{k_BT}{3!} \!\! \int \!
                          c^{(3)}(\bx_1,\bx_2,\bx_3) \times{}\\
 &\qquad\qquad\qquad      \Deltarho(\bx_1)
                          \Deltarho(\bx_2)
                          \Deltarho(\bx_3)
                          \rdbx_1\rdbx_2\rdbx_3\\
 & \quad{} + \text{similar fourth order term} + \dots.
 \end{split}
 \label{eq:expandFex}
 \end{equation}
The expressions $c^{(n)}$ in the equation above are
proportional to the first and higher
functional derivatives of $\cFex$ with respect to density,
all evaluated at $\rho=\rho_0$:
 \begin{equation}
 c^{(n)}(\bx_1,\dots,\bx_n) = -\beta
    \frac{\delta^n\!\cFex}{\deltarho(\bx_1)\dots\deltarho(\bx_n)}[\rho_0].
 \label{eq:defncn}
 \end{equation}
These functions~$c^{(n)}$ are known as {\emph{direct correlation
functions}}~\cite{Evans1979a, Evans1992, Hansen2013}, and
are related to $n$-point density correlation functions. In the two-point case,
$c^{(2)}$ is the pair direct correlation function and is related to the
pair correlation function (i.e., the radial distribution function) through
the Ornstein--Zernike equation~\cite{Evans1979a, Evans1992, Hansen2013}.
These direct correlation
functions depend on our choice of~$\rho_0$ and depend directly
on temperature through the linear factor of~$\beta$ in the definition
\eqref{eq:defncn} and also
indirectly via the fact that the correlations in a liquid change
with temperature. Note also that~$c^{(1)}[\rho_0]$ is a constant
when $\rho_0$ is a constant.

For a homogeneous liquid with distant (or periodic) boundaries,
these functions depend on displacements but
not on absolute position, so
(through a slight abuse of notation) we also write
 \begin{equation}
 c^{(n)}(\bx_1,\dots,\bx_n) = c^{(n)}(\Dbx_2,\Dbx_3\dots,\Dbx_n),
 \label{eq:shiftorigin}
 \end{equation}
where $\Dbx_j=\bx_j-\bx_1$~\cite{Hansen2013}.
We also take the liquid to be isotropic.

We are considering density perturbations away from the liquid state, so it
is convenient to write
 \begin{equation}\label{eq:rho_sub}
 \rho(\bx,t)=\rho_0 (1+n(\bx,t)).
 \end{equation}
We do not assume that $n$~is small, but it is often the case that
the average of $n(\bx,t)$ over the whole system is small.
Note also that $\rho(\bx,t)=\rho_0$ is a stationary solution
of~(\ref{eq:DDFT0dynamics}). Substituting Eq.~(\ref{eq:expandFex}) into
Eq.~(\ref{eq:DDFT0dynamics}) and writing only the terms up to~$c^{(1)}$, we
get:
 \begin{equation}
 \frac{\partial n}{\partial t} = \nabla^2 n
   - \nabla^2 c^{(1)} - \nabla \cdot \left[ n \nabla c^{(1)} \right]
+ \dots \\
 \label{eq:DDFT1dynamicsc1}
 \end{equation}
That the uniform liquid state is an equilibrium of~(\ref{eq:DDFT0dynamics}) implies that
$n=0$ is a solution of equation~(\ref{eq:DDFT1dynamicsc1}): all terms
not written down involve~$\Deltarho$ and so are zero for the uniform liquid
with density $\rho_0$.
Recall that~$c^{(1)}[\rho_0]$ is a constant, which means
terms involving gradients of this can be
dropped. {Whilst this constant term does not influence the structure (density profile)
both in and out of equilibrium, it does affect the thermodynamics (i.e., free
energy value) and so also mechanical properties~\cite{Wang2018b}.}
With this, we can write the equation for the time evolution
of~$n(\bx,t)$ (up to $\cO(c^{(4)})$) as:
 \begin{equation}
 \begin{split}
 \frac{\partial n}{\partial t} =& \nabla^2 n
   - \rho_0 \nabla^2  \!\!\int\! c^{(2)}(\bx,\bx_2)n(\bx_2)\rdbx_2 \\
   &{}- \rho_0 \nabla \cdot \left[ n \nabla
                               \!\!\int\! c^{(2)}(\bx,\bx_2)n(\bx_2)\rdbx_2
                         \right]\\
   &{}- \frac{\rho_0^2}{2} \nabla^2 \!\!\int\! c^{(3)}(\bx,\bx_2,\bx_3)n(\bx_2)n(\bx_3)\rdbx_2\rdbx_3\\
   &{}- \frac{\rho_0^2}{2} \nabla \cdot \left[ n \nabla
                               \!\!\int\! c^{(3)}(\bx,\bx_2,\bx_3)n(\bx_2)n(\bx_3)\rdbx_2\rdbx_3
                         \right]\\
   &{}- \frac{\rho_0^3}{6} \nabla^2 \!\!\int\! c^{(4)}(\bx,\bx_2,\bx_3,\bx_4) \times{}\\
   & \qquad\qquad\qquad n(\bx_2)n(\bx_3)n(\bx_4)\rdbx_2\rdbx_3\rdbx_4 \\
   &{}- \frac{\rho_0^3}{6} \nabla \cdot \bigg[ n \nabla
                               \!\!\int\! c^{(4)}(\bx,\bx_2,\bx_3,\bx_4) \times{}\\
   & \qquad\qquad\qquad n(\bx_2)n(\bx_3)n(\bx_4)\rdbx_2\rdbx_3\rdbx_4
                         \bigg] + \dots\\
 \end{split}
 \label{eq:DDFT1dynamicspretruncate}
 \end{equation}
where we have suppressed writing the time dependence of~$n$ throughout and
the $\bx$ dependence of $n$ when it is not inside an integral.
We have written this equation so that the first line is linear in~$n$,
the next two lines are quadratic in~$n$, the fourth and fifth lines
are cubic in~$n$, and the last line is quartic in~$n$.

Since the first line is linear in $n$, and both terms involve a Laplacian, we
can write the linearised version of (\ref{eq:DDFT1dynamicspretruncate}) in
terms of the negative Laplacian of a linear operator~$\cL$:
 \begin{equation}
 \frac{\partial n}{\partial t} = - \nabla^2 \cL n,
 \label{eq:DDFTlinear}
 \end{equation}
where
 \begin{equation}
 \cL n(\bx) = - n(\bx) + \rho_0 \!\!\int\! c^{(2)}(\bx,\bx_2)n(\bx_2)\rdbx_2.
 \label{eq:defnL}
 \end{equation}
The non-local operator~$\cL$ is most conveniently considered in terms of
its Fourier transform, or equivalently, in terms of how it acts
on modes of the form~$\exp(i\bk\cdot\bx)$. If
 \begin{equation}
 \cL e^{i\bk\cdot\bx} = \sigma(\bk) e^{i\bk\cdot\bx},
 \label{eq:defnsigma}
 \end{equation}
then $\sigma(\bk)$ is the eigenvalue
of~$\cL$ with eigenfunction $\exp(i\bk\cdot\bx)$.
With this, the linear
equation~(\ref{eq:DDFTlinear}) can readily be solved in terms of linear
combinations of functions like
$\exp\left({k^2\sigma(\bk) t + i\bk\cdot\bx}\right)$,
where $k^2\sigma(\bk)$ is the growth rate for a mode with wavevector~$\bk$,
and $k=|\bk|$.
If $\sigma(\bk)$ is negative for all~$\bk$,
all small amplitude density modulations decay to zero, and the liquid
state is dynamically stable.

\begin{figure}
\begin{center}
\includegraphics{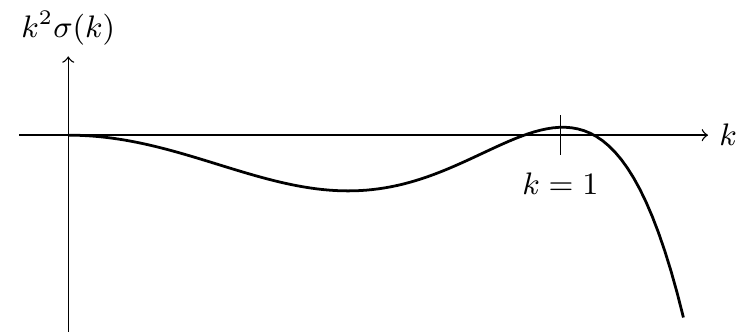}
\end{center}

\caption{Illustrative example of the growth rate $k^2\sigma(k)$ as a
function of wavenumber~$k$. Small amplitude
modes with $k^2\sigma(k)<0$ decay exponentially in time, while those
with $k^2\sigma(k)>0$ grow exponentially. Throughout we scale lengths
so that the maximum growth rate occurs at $k=1$.}
 \label{fig:growthrate}
 \end{figure}

Recall that for a bulk liquid, $c^{(2)}(\bx,\bx_2)=c^{(2)}(\Dbx_2)$, with
$\Dbx_2=\bx_2-\bx$, and for spherically symmetric (isotropic) particles,
$c^{(2)}(\bx,\bx_2)=c^{(2)}(|\Dbx_2|)$. Therefore,
in this case $\sigma(\bk)=\sigma(k)$,
i.e., $\sigma$ depends only on the wavenumber~$k=|\bk|$.
The eigenvalue~$\sigma(k)$ can be expressed as:
 \begin{equation}
 \begin{split}
 \sigma(k) &= - 1 + \rho_0 \!\!\int\! c^{(2)}(|\bx_2-\bx|) e^{i\bk\cdot(\bx_2-\bx)}\rdbx_2\\
           &= -1 + \rho_0 {\hat c}^{(2)}(k),
 \label{eq:sigmactwo}
 \end{split}
 \end{equation}
where ${\hat c}^{(2)}$ is the Fourier transform of $c^{(2)}$.
Recall from~(\ref{eq:defncn}) that $c^{(2)}$ is
proportional to~$\beta$, so if ${\hat c}^{(2)}$ has any positive Fourier components,
decreasing the temperature (increasing~$\beta$) can be expected to
lead to a range of
wavenumbers with positive growth rates, and the liquid being dynamically
unstable to modes with wavenumbers centered around the maximum
of~$k^2\sigma(k)$, see Fig.~\ref{fig:growthrate}. We have scaled
lengths so that the maximum growth rate occurs at wavenumber $k=1$.
This argument, of course, assumes that the product $\beta^{-1}c^{(2)}$
is independent of temperature. This is not true in general, but for some
systems it is a good approximation (at least over a limited range
of temperatures) -- see Ref.~\cite{Somerville2018}
for a recent discussion on this for a particular colloidal system.
Recall too that for an equilibrium liquid the static structure factor
$S(k)=[1-\rho_0{\hat c}^{(2)}(k)]^{-1}$. $S(k)$
is proportional to the Fourier transform of the radial distribution
function \cite{Hansen2013}. So, for
the stable uniform liquid, we have $\sigma(k)=-1/S(k)$.

We refer to the state point at which the uniform liquid becomes linearly
unstable to density modulations with wavenumber $k\neq0$ as the
\emph{spinodal point}, in keeping with the terminology of~\cite{Trudu2006}.
The more common usage of the term `spinodal' relates to the onset of the
zero-wavenumber phase separation instability of liquid--liquid or gas--liquid
phase separation~\cite{Hansen2013, Archer2004}.
At the spinodal point, the density and temperature are such that the
liquid is dynamically marginally stable, that is, the maximum of~$k^2\sigma(k)$
is zero. Therefore, at higher temperatures, small amplitude density
modulations decay, and at
lower temperatures, small amplitude density modulations grow. For a given fixed
value of~$\rho_0$, the spinodal temperature is~$T_s$, with a
corresponding $\beta_s=(k_BT_s)^{-1}$. Similarly, either increasing the
density~$\rho_0$ of the liquid or increasing the chemical potential~$\mu$ 
can also lead to crossing the spinodal.

With~(\ref{eq:defnL}), we can eliminate $c^{(2)}$ in favour of~$\cL$
in~(\ref{eq:DDFT1dynamicspretruncate}), and obtain (truncating
at~${\cO}(c^{(4)})$):
 \begin{equation}
 \begin{split}
 \frac{\partial n}{\partial t} =& - \nabla^2 \left(\cL n + \tfrac{1}{2}n^2\right)
                                  - \nabla \cdot \left[ n \nabla \cL n \right]\\
   &{}- \frac{\rho_0^2}{2} \nabla^2 \!\!\int\! c^{(3)}(\bx,\bx_2,\bx_3)n(\bx_2)n(\bx_3)\rdbx_2\rdbx_3\\
   &{}- \frac{\rho_0^2}{2} \nabla \cdot \left[ n \nabla
                               \!\!\int\! c^{(3)}(\bx,\bx_2,\bx_3)n(\bx_2)n(\bx_3)\rdbx_2\rdbx_3
                         \right]\\
   &{}- \frac{\rho_0^3}{6} \nabla^2 \!\!\int\! c^{(4)}(\bx,\bx_2,\bx_3,\bx_4)\times{}\\
   &\qquad\qquad\qquad n(\bx_2)n(\bx_3)n(\bx_4)\rdbx_2\rdbx_3\rdbx_4\\
   &{}- \frac{\rho_0^3}{6} \nabla \cdot \bigg[ n \nabla
                               \!\!\int\! c^{(4)}(\bx,\bx_2,\bx_3,\bx_4) \times{}\\
   & \qquad\qquad\qquad n(\bx_2)n(\bx_3)n(\bx_4)\rdbx_2\rdbx_3\rdbx_4
                         \bigg]\\\
 \end{split}
 \label{eq:DDFT1dynamics}
 \end{equation}
where we have used the result $\nabla \cdot \left[n \nabla n\right] = \frac{1}{2}\nabla^2n^2$.
{For an ideal gas, with $\cL n=-n$ and $c^{(2)}=c^{(3)}=c^{(4)}=0$,
the first line of the equation above reduces to the diffusion equation,
$\frac{\partial n}{\partial t}=\nabla^2n$, similar to~(\ref{eq:rhodiffusion}).} 

At this point, we have made no approximations beyond expanding the free energy
in Eq.~(\ref{eq:expandFex}) and truncating at~${\cO}(c^{(4)})$.
We refer to the model at this stage, truncated in this way, as~\hbox{DDFT-1}.
In the new variables, and incorporating~$c^{(2)}$ into~$\cL$,
the Helmholtz free energy $\cF$ can be expressed (up to fourth
order) in terms of a scaled free energy $\cFa=\cF/\rho_0$, where
 \begin{equation}
 \begin{split}
 \beta\cFa[n] &=
     \int \! \big([1+n(\bx_1)] \log [1+n(\bx_1)] - n(\bx_1)\big)\rdbx_1\\
 & \quad{}- \frac{1}{2} \! \int \!
       \left(n^2(\bx_1) + n(\bx_1)\cL n(\bx_1) \right)\rdbx_1 \\
 & \quad{}-\frac{\rho_0^2}{6} \!\! \int \!
                          c^{(3)}(\bx_1,\bx_2,\bx_3) \times{}\\
 &\qquad\qquad\quad       n (\bx_1)
                           n (\bx_2)
                           n (\bx_3)
                          \rdbx_1\rdbx_2\rdbx_3\\
 & \quad{}-\frac{\rho_0^3}{24} \! \int \!
                          c^{(4)}(\bx_1,\bx_2,\bx_3,\bx_4) \times{}\\
 &\qquad\qquad\quad       n (\bx_1)
                           n (\bx_2)
                           n (\bx_3)
                           n (\bx_4)
                          \rdbx_1\rdbx_2\rdbx_3\rdbx_4,
 \end{split}
 \label{eq:DDFTF1}
 \end{equation}
and where we have also dropped terms that do not contribute to~(\ref{eq:DDFT1dynamics}).
In these variables, the DDFT that leads to the dynamics~(\ref{eq:DDFT1dynamics}) is
 \begin{equation}
 \frac{\partial n}{\partial t} = \nabla \cdot \left[ \beta (1+n) \nabla \frac{\deltacFa}{\deltan} \right].
 \label{eq:DDFT1deltaF}
 \end{equation}
Note that, because of the $\log(1+n)$ term in~(\ref{eq:DDFTF1}), $n$~is constrained
so that $1+n$ is always non-negative. Also, because of Eq.~\eqref{eq:rho_sub},
we have 
 \begin{equation}
\frac{\deltacF}{\delta\rho}=\frac{\deltacFa}{\deltan}.
 \label{eq:derivs_eq}
 \end{equation}
Moreover, states that satisfy
 \begin{equation}
 \frac{\deltacFa}{\deltan} = \Deltamu,
 \label{eq:ChemPot}
 \end{equation}
where $\Deltamu=\mu-\mu_0$ and where [see~(\ref{eq:chempotdefn}), 
(\ref{eq:separatedF}) and~(\ref{eq:expandFex})]
 \begin{equation}
 \mu_0=k_BT\log\Lambda^d\rho_0-k_BTc^{(1)}[\rho_0],
  \label{eq:mu_rho_0}
 \end{equation}
are equilibrium solutions of~(\ref{eq:DDFT1dynamics}), or equivalently, extrema
of~$\cFa$. Henceforth, we redefine $\mu$ to be $\beta\Deltamu/\rho_0$, which is a shifted
and rescaled chemical potential. 
For the free energy in~(\ref{eq:DDFTF1}), we have
 \begin{equation}
 \begin{split}
 \beta\frac{\deltacFa}{\deltan} &=
    \log\left(1+n(\bx)\right)
    - n(\bx) - \cL n(\bx) \\
 & \quad{}-\frac{\rho_0^2}{2} \!\! \int \!
                          c^{(3)}(\bx,\bx_2,\bx_3)
                           n (\bx_2)
                           n (\bx_3)
                          \rdbx_2\rdbx_3\\
 & \quad{}-\frac{\rho_0^3}{6} \! \int \!
                          c^{(4)}(\bx,\bx_2,\bx_3,\bx_4) \times{}\\
 &\qquad\qquad\quad
                           n (\bx_2)
                           n (\bx_3)
                           n (\bx_4)
                          \rdbx_2\rdbx_3\rdbx_4.
 \end{split}
 \label{eq:betadFdn1}
 \end{equation}
At equilibrium, this expression (the rescaled chemical potential~$\mu$) does
not vary in space. The reference liquid $n=0$ has $\cFa=0$ and $\mu=0$. 
The zero value for $\cFa$ arises (in part) from dropping $\cFex[\rho_0]$ 
from~(\ref{eq:expandFex}), while the zero value for the rescaled
chemical potential is a consequence of \eqref{eq:mu_rho_0},
which is equivalent to dropping $c^{(1)}$ from~(\ref{eq:expandFex}) and
setting $\Lambda=1$ in \eqref{eq:separatedF}.

\subsection{Simplification of~$c^{(3)}$ and~$c^{(4)}$: DDFT-2}

As the next step,
Huang~\etal~\cite{Huang2010a}
kept only the zero-wavenumber components of~$c^{(3)}$ and~$c^{(4)}$, or
equivalently, they took
 \begin{equation}
 \begin{split}
 c^{(3)}(\bx,\bx_2,\bx_3)&=c^{(3)}_0 \delta(\bx-\bx_2)\delta(\bx-\bx_3),\\
 c^{(4)}(\bx,\bx_2,\bx_3,\bx_4)&=c^{(4)}_0 \delta(\bx-\bx_2)\delta(\bx-\bx_3)\delta(\bx-\bx_4),
 \label{eq:definecs}
 \end{split}
 \end{equation}
where $c^{(3)}_0$ and $c^{(4)}_0$ are constants (our sign convention is
opposite to that of~\cite{Huang2010a}). This is equivalent to making a local density
approximation (LDA)~\cite{Evans1992} for these terms in the free energy.
We could in principle include terms
involving~$c^{(5)}$ and higher as well, treated in the same way: these would
contribute a more general function of~$n$ in the free energy, treated with
the~\hbox{LDA}. However, since we are investigating the effect of
approximations that have not yet been discussed, we keep as simple a free
energy as possible at this point, consistent with truncating at~$\cO(c^{(4)})$. With this, the
free energy in~(\ref{eq:DDFTF1}) becomes
 \begin{equation}
 \begin{split}
 \beta\cFb[n] &=
     \int \! \big([1+n(\bx)] \log [1+n(\bx)] - n(\bx)\big)\rdbx\\
 & \quad{}+ \! \int \! \bigg(
           - \frac{1}{2} \left(n^2(\bx) + n\cL n(\bx) \right) \\
 & \qquad\qquad{}-\frac{\rho_0^2}{6}c^{(3)}_0 n^3 (\bx)
                 -\frac{\rho_0^3}{24}c^{(4)}_0 n^4 (\bx)
          \bigg)\rdbx,\\
 \end{split}
 \label{eq:DDFTF2}
 \end{equation}
and the four terms involving~$c^{(3)}$ and~$c^{(4)}$
in~(\ref{eq:DDFT1dynamics}) become
 \begin{equation}
 \begin{split}
   &- \frac{\rho_0^2}{2}c^{(3)}_0 \nabla^2 n^2,\quad
   - \frac{\rho_0^2}{2}c^{(3)}_0 \nabla \cdot \left[ n \nabla n^2
                         \right],\\
   &- \frac{\rho_0^3}{6}c^{(4)}_0 \nabla^2 n^3
   \quad\text{and}\quad
    - \frac{\rho_0^3}{6}c^{(4)}_0 \nabla \cdot \left[ n \nabla n^3
                         \right].
 \end{split}
 \label{eq:cthreedelta}
 \end{equation}
Using $\nabla \cdot \left[ n \nabla n^2\right]=\frac{2}{3}\nabla^2n^3$
and $\nabla \cdot \left[ n \nabla n^3\right]=\frac{3}{4}\nabla^2n^4$,
Huang
\etal~\cite{Huang2010a} combined~(\ref{eq:cthreedelta}) and~(\ref{eq:DDFT1dynamics})
to get
 \begin{equation}
 \frac{\partial n}{\partial t} = - \nabla^2 \left(\cL n + Qn^2 + Cn^3 + Rn^4\right)
                                  - \nabla \cdot \left[ n \nabla \cL n \right]
 \label{eq:DDFT2dynamics}
 \end{equation}
where
 \begin{equation}
 Q = \frac{1}{2} + \frac{\rho_0^2}{2}c^{(3)}_0 ,\quad
 C =  \frac{\rho_0^2}{3}c^{(3)}_0 + \frac{\rho_0^3}{6}c^{(4)}_0
 \quad\text{and}\quad
 R =  \frac{\rho_0^3}{8}c^{(4)}_0.
 \label{eq:qandc}
 \end{equation}
We also have a chemical potential
 \begin{equation}
 \begin{split}
 \mu=\beta\frac{\deltacFb}{\deltan} &=
    \log\left(1+n(\bx)\right)
    - n(\bx) - \cL n(\bx)\\
 & \quad\quad{}- \frac{\rho_0^2}{2} c^{(3)}_0 n^2 (\bx) - \frac{\rho_0^3}{6} c^{(4)}_0 n^3 (\bx),
 \end{split}
 \label{eq:betadFdn2}
 \end{equation}
which does not vary in space at equilibrium.
Up to this point, we refer to the model as~\hbox{DDFT-2}.

Here, we retain the $n^4$ term (as did Huang \etal~\cite{Huang2010a}), 
because otherwise the dynamics
in~(\ref{eq:DDFT2dynamics}) would not be consistent with the free
energy~(\ref{eq:DDFTF2}) and the DDFT dynamics~(\ref{eq:DDFT1deltaF})
(with $\cF_2$ instead of~$\cF_1$).

The next three models involve making (or not making) two approximations:
(i)~assuming the Ramakrishan--Yussouff or random phase approximation,
which leads to a quadratic excess Helmholtz free energy functional,
and (ii)~making
a gradient expansion of the linear operator~$\cL$.

\subsection{Quadratic excess free energy: DDFT-3}

Often, the free energy functional in~(\ref{eq:expandFex}) is truncated
at~$\cO(\Deltarho^2)$. This is known as the Ramakrishan--Yussouff (RY)
approximation~\cite{Ramakrishnan1979, Teeffelen2009, Emmerich2012}, which
effectively sets $c^{(3)}=c^{(4)}=0$. A mathematically equivalent approximation
arises in the treatment of soft purely repulsive particles modelling soft matter,
namely the RPA~\cite{Likos2001}.
Here, two soft isotropic particles at $\bx_1$ and~$\bx_2$
separated by a distance~$x_{12}=|\bx_1-\bx_2|$ interact
through a potential energy~$u(x_{12})$,
which depends only on the magnitude of the distance
and is finite for all values of $x_{12}$.
The excess free energy [c.f.\ Eq.~(\ref{eq:expandFex})] is then
 \begin{equation}
 \cFex[\rho] = \frac{1}{2} \! \int \!
               u(|\bx_1-\bx_2|) \rho(\bx_1) \rho(\bx_2) \rdbx_1\rdbx_2.
 \label{eq:RPAfreeenergy}
 \end{equation}
This amounts to setting $c^{(3)}=c^{(4)}=0$ and
 \begin{equation}
 c^{(2)}(\bx_1,\bx_2) = - \beta  u(|\bx_1-\bx_2|)
 \label{eq:RPAc2u}
 \end{equation}
in~\hbox{DDFT-2}. The eigenvalues~$\sigma(k)$ can thus
be related to the Fourier transform of~$u$
through~(\ref{eq:sigmactwo}) \cite{Likos2001, Archer2012}:
 \begin{equation}
 \begin{split}
 \sigma(k) &= - 1 - \rho_0 \beta \!\!\int\! u(|\bx-\bx_2|) e^{i\bk\cdot(\bx_2-\bx)}\rdbx_2\\
           &= - 1 - \rho_0 \beta {\hat u}(k).
 \label{eq:sigmauhat}
 \end{split}
 \end{equation}
Setting $c^{(3)}=c^{(4)}=0$ implies from~(\ref{eq:qandc})
that $Q=\frac{1}{2}$, $C=0$ and $R=0$, and
results in a free energy
 \begin{equation}
 \begin{split}
 \beta\cFc[n] &=
     \int \! \big((1+n(\bx_1)) \log (1+n(\bx_1)) - n(\bx_1)\big)\rdbx_1\\
 & \quad{}- \frac{1}{2} \! \int \!
       \left(n^2(\bx_1) + n(\bx_1)\cL n(\bx_1) \right)\rdbx_1.
 \end{split}
 \label{eq:DDFTF3}
 \end{equation}
With this choice of free energy, the dynamics in~(\ref{eq:DDFT2dynamics}) becomes:
 \begin{equation}
 \frac{\partial n}{\partial t} = - \nabla^2 \left(\cL n + \tfrac{1}{2}n^2\right)
                                 - \nabla \cdot \left[ n \nabla \cL n \right],
 \label{eq:DDFT3dynamics}
 \end{equation}
along with an analogous version of (\ref{eq:betadFdn2}), for the
chemical potential:
 \begin{equation}
 \mu=\beta\frac{\deltacFc}{\deltan} =
    \log\left(1+n(\bx)\right)
    - n(\bx) - \cL n(\bx)
 \label{eq:betadFdn3}
 \end{equation}
We refer to this model as~\hbox{DDFT-3}; it is equivalent to
DDFT-1 with the RY approximation, and to DDFT-0 with $\cFex$ given by the
RPA approximation.

Before moving on to make further approximations, it is worth noting a useful
property that DDFT-3 and the subsequent theories derived from it possess. If the
pair potential $u(x_{12})$ in Eq.~\eqref{eq:RPAc2u} can be written as
$u(x_{12})=\epsilon\psi(x_{12})$, where $\epsilon$ is a parameter that controls
the overall strength of the potential, then from Eqs.~\eqref{eq:defnL},
\eqref{eq:RPAc2u} and~\eqref{eq:betadFdn3} we obtain:
 \begin{equation}
 \mu=
    \log\left(1+n(\bx)\right)
    +\rho_0\beta\epsilon \!\!\int\! \psi(|\bx-\bx_2|) n(\bx_2)\rdbx_2.
 \label{eq:convol_form}
 \end{equation}
The consequence of this is that for a given $\psi$, the behaviour of the
model depends only on the combination of parameters $\rho_0\beta\epsilon$
and the value of~$\mu$.
If one changes the value of the reference density $\rho_0$ to some other value,
then this is entirely equivalent to solving the system with the original
reference density $\rho_0$ at a different value of~$\beta\epsilon$. 
We should emphasize that this is only true
if $\psi$ does not change with density, which in general is not true,
but is approximately the case for some systems.

\subsection{Gradient expansion of the linear term: DDFT-4}

Returning to DDFT-2, Huang~\etal~\cite{Huang2010a}
(following~\cite{Elder2004}) 
expanded $\cL$ in powers of the gradient operator~$\nabla$, 
replacing $\cL$ by the simplest linear operator
that allows a positive growth rate for modes with a wavenumber~$k_s$.
Scaling lengths so that $k_s=1$ results in:
 \begin{equation}
 \cLgrad n = r n - \gamma(1+\nabla^2)^2 n,
 \label{eq:cLgraddefn}
 \end{equation}
so $\sigma(k)=r - \gamma(1-k^2)^2$ from~(\ref{eq:defnsigma}).
This approximation is equivalent (within
scaling) to a local gradient expansion of~(\ref{eq:defnL}), expanding the
Fourier transform of $c^{(2)}$ about its maximum:
 \begin{equation}
 \rho_0 {\hat c}^{(2)}(k) = 1 + r - \gamma(1-k^2)^2,
 \label{eq:defnGradientExpansion}
 \end{equation}
where the function
$\rho_0{\hat c}^{(2)}(k)$ and its second derivative evaluated at~$k=1$ are
$1+r$ and $-8\gamma$, respectively.
Here, $r$~is a parameter,
notionally increasing with~$\beta$ (and with~$\rho_0$) and equal to zero 
at the spinodal point, when
$\beta=\beta_s$. This parameter controls the growth rate of waves with
wavenumber~$1$: effectively, $r$~is the height of the maximum at $k=1$
in the growth rate curve in Fig.~\ref{fig:growthrate}. The second
parameter~$\gamma$ can be used to fit the curvature of
${\hat c}^{(2)}(k)$ at $k=1$.

With this gradient expansion, the dynamics is
 \begin{equation}
 \begin{split}
 \frac{\partial n}{\partial t} &= - \nabla^2 \left(\cLgrad n + Qn^2 + Cn^3 + Rn^4\right) \\
                               &\qquad {} - \nabla \cdot \left[ n \nabla \cLgrad n \right].
 \end{split}
 \label{eq:DDFT4dynamics}
 \end{equation}
We refer to this model as~\hbox{DDFT-4}: $\cLgrad$ is now a (local) partial
differential operator and (\ref{eq:DDFT4dynamics}) is a partial differential
equation.
The free energy and chemical potential can be found from
(\ref{eq:DDFTF2}) and (\ref{eq:betadFdn2}), setting $\cL=\cLgrad$.
The lower bound $n\geq-1$ is still respected. This model is equivalent to that
written down by~\cite{Huang2010a}.

Higher powers (or other functions) of the Laplacian can be retained in~$\cLgrad$, to
improve the accuracy of the match between the eigenvalues of~$\cL$ and~$\cLgrad$, as
done for example by~\cite{Jaatinen2009,Pisutha-Arnond2013b}, or to introduce additional
unstable length scales, as done for example by~\cite{Wu2010, Barkan2011, Achim2014, Subramanian2016} 
and others. See also Eq.~\eqref{eq:cLgradJaatinen} below and the associated
discussion.

\subsection{RY and gradient expansion: DDFT-5}

Finally, we can make both the RY/RPA approximation ($c^{(3)}_0=c^{(4)}=0$)
and replace the linear operator~$\cL$ by~$\cLgrad$ to get
the model advocated in Ref.~\cite{Teeffelen2009}.
The free energy and evolution equation are
 \begin{equation}
 \begin{split}
 \beta\cFe[n] &=
     \int \! \big((1+n(\bx_1)) \log (1+n(\bx_1)) - n(\bx_1)\big)\rdbx_1\\
 & \quad{}- \frac{1}{2} \! \int \!
       \left(n^2(\bx_1) + n(\bx_1)\cLgrad n(\bx_1) \right)\rdbx_1,
 \end{split}
 \label{eq:DDFTF5}
 \end{equation}
and
 \begin{equation}
 \frac{\partial n}{\partial t} = - \nabla^2 \left(\cLgrad n + \tfrac{1}{2}n^2\right)
                                 - \nabla \cdot \left[ n \nabla \cLgrad n \right],
 \label{eq:DDFT5dynamics}
 \end{equation}
along with an analogous version of Eq.~(\ref{eq:betadFdn3}) for the
chemical potential:
 \begin{equation}
 \mu=\beta\frac{\deltacFe}{\deltan} =
    \log\left(1+n(\bx)\right)
    - n(\bx) - \cLgrad n(\bx)
 \label{eq:betadFdn5}
 \end{equation}
This model is named PFC1 in~\cite{Teeffelen2009}, but here we
call it DDFT-5 for consistency.

\subsection{{PFC models}}

The final simplification that can be made (or not made) is to discard the
$\nabla\cdot\left[n\nabla\cL{n}\right]$ (or
$\nabla\cdot\left[n\nabla\cLgrad{n}\right]$) term from the dynamical
equations for the four DDFT models
DDFT-2, \dots, DDFT-5, resulting in four PFC models \PFCbeta, \dots, \PFCepsilon.
Huang \etal~\cite{Huang2010a} justify making this simplification on the
grounds that this term is not truly quadratic in~$n$: the presence of $\cL{n}$
in the expression
means that it is effectively of higher order. However, we show below that this
term does in fact make an important contribution to the free energy: at least
as important as the $c^{(3)}$~term.

In addition, dropping this term implies significant changes to the DDFT
dynamics, the mobility and the nonlinear terms in the free energy. In fact, the $(1+n)$
factor in the mobility in~(\ref{eq:DDFT1deltaF}), the logarithm in the ideal
gas free energy in~(\ref{eq:separatedF}) and the
$\nabla\cdot\left[n\nabla\cL{n}\right]$ term in~(\ref{eq:DDFT1dynamics}) are
inextricably linked. This can be seen in the progression from
(\ref{eq:separatedF}) to~(\ref{eq:rhodiffusion}): the functional derivative of
the ideal gas term in~(\ref{eq:separatedF}) (the first term on the right hand side)
leads to the $\log\rho$ term
in~(\ref{eq:separatedFdrho}), the gradient of this leads to
$\rho^{-1}\nabla\rho$ in~(\ref{eq:rhoinverse}), and the mobility being
$M=D\rho$ cancels the $\rho^{-1}$, leading to a diffusion equation
in~(\ref{eq:rhodiffusion}). 
{If the $\nabla\cdot\left[n\nabla\cL{n}\right]$ term is dropped from~(\ref{eq:DDFT2dynamics}),
the equation for~$n$ becomes of the form $\frac{\partial n}{\partial t} = 
\nabla^2\frac{\deltacG}{\deltan}$ for some functional~$\cG[n]$.
We can see the implications of this by returning to~(\ref{eq:DDFT0dynamics}) and 
taking the steps needed to get to this modified version of~(\ref{eq:DDFT2dynamics}).
Clearly the mobility in~(\ref{eq:DDFT0dynamics}) has been taken to be constant. 
If we now think of the ideal gas part of the free energy in~(\ref{eq:separatedF})
and~(\ref{eq:separatedFdrho}), but with a constant mobility in the dynamical equation,
we end up with the ideal gas term contribution to the equation 
for~$\rho$ being the form
 \begin{equation}
 \frac{\partial\rho}{\partial t} = \frac{1}{\rho}\nabla^2\rho - 
                     \frac{1}{\rho^2}\left|\nabla\rho\right|^2
 \end{equation}
instead of the diffusion equation~(\ref{eq:rhodiffusion}).
This unlikely equation can be avoided,
and the diffusion equation recovered at leading order,
by expanding the logarithm in~(\ref{eq:DDFTF2}) in a Taylor
series. Thus,
dropping the $\nabla\cdot\left[n\nabla\cL{n}\right]$ term is equivalent to
taking constant mobility and expanding the logarithm.}

It is because of these substantial changes that we opt to use the term `DDFT'
for all models based on free energies that have the logarithmic ideal gas
term, the non-constant mobility and 
the $\nabla\cdot\left[n\nabla\cL{n}\right]$ term retained. In contrast,
we use the term `PFC' for models
based on expanding the logarithm, having a constant mobility and the
$\nabla\cdot\left[n\nabla\cL{n}\right]$ term dropped. One consequence of expanding the
logarithm 
{up to $\cO(n^4)$, as is done in most PFC derivations~\cite{Emmerich2012},
is that the ideal gas part of the free energy contributes 
cubic and quartic (as well as quadratic) terms to the free energy,}
so going from DDFT-2 to \PFCbeta\ turns out not to be just a matter of
dropping the $\nabla\cdot\left[n\nabla\cL{n}\right]$ term.

So, a consistent free energy--dynamics
derivation~\cite{Elder2004,Teeffelen2009} involves going back to DDFT-2 and
replacing the logarithm in~(\ref{eq:DDFTF2}) by:
 \begin{equation}
 (1+n)\log(1+n)={n}+\tfrac{1}{2}n^2-\tfrac{1}{6}n^3+\tfrac{1}{12}n^4,
 \label{eq:expandlogarithm}
 \end{equation}
resulting in a free energy
 \begin{equation}
 \begin{split}
 \beta\cFPFCbeta[n] &=
     \int \! \Big( - \tfrac{1}{2}n\cL n
                   - \tfrac{1}{6}n^3
                   + \tfrac{1}{12}n^4\\
     & \qquad\quad  {} - \frac{\rho_0^2}{6}  c^{(3)}_0 n^3
                       - \frac{\rho_0^3}{24} c^{(4)}_0 n^4\Big)\rdbx_1,
 \label{eq:DDFTFPFCbeta}
 \end{split}
 \end{equation}
where we have suppressed writing
the $\bx_1$ dependency of~$n(\bx_1)$.
Taking the mobility $M(\rho)$
in~(\ref{eq:DDFT0dynamics}) to be a constant ($M=D\rho_0$) implies
(after scaling)
 \begin{equation}
 \frac{\partial n}{\partial t} = \nabla^2 \left[\beta\frac{\deltacFPFCbeta}{\deltan} \right],
 \label{eq:PFCdynamics}
 \end{equation}
similar to~(\ref{eq:PFCalphadynamics}).
This leads to the PFC dynamical equation:
 \begin{equation}
 \frac{\partial n}{\partial t} =
     - \nabla^2 \left[
                \cL{n} + Qn^2 + Cn^3
                \right]
 \label{eq:PFCbetadynamics}
 \end{equation}
and to a chemical potential
 \begin{equation}
 \mu=\beta\frac{\deltacFPFCbeta}{\deltan} =
                - \cL{n} - Qn^2 - Cn^3,
 \label{eq:betadFdnbeta}
 \end{equation}
where $Q$ is as in (\ref{eq:qandc}) but $C$~is different:
 \begin{equation}
 Q = \frac{1}{2} + \frac{\rho_0^2}{2}c^{(3)}_0
 \quad\text{and}\quad
 C = - \frac{1}{3} + \frac{\rho_0^3}{6}c^{(4)}_0.
 \label{eq:PFCqandc}
 \end{equation}
We refer to this model as \PFCbeta, and recall that the factor of~$\beta$
in front of~$\cFPFCbeta$ is the inverse temperature.

The end result here is that \PFCbeta~(\ref{eq:PFCbetadynamics}) is not the same
as DDFT-2~(\ref{eq:DDFT2dynamics}) with the
$\nabla\cdot\left[n\nabla\cL{n}\right]$ term removed: the cubic coefficient~$C$
is different and the quartic contribution $Rn^4$ in (\ref{eq:DDFT2dynamics})
is absent. For the cubic coefficient, the contribution proportional to
$c^{(3)}_0$ in~(\ref{eq:qandc}) comes from the non-constant mobility, while the
$-\frac{1}{3}$ term in~(\ref{eq:PFCqandc}) comes from expanding the logarithm.
The contribution to~$C$ proportional to~$c^{(4)}_0$ is the same. Moreover, the
$\frac{1}{2}$ in $Q$ in (\ref{eq:qandc}) and~(\ref{eq:PFCqandc}), while having
the same numerical value, arises for two different reasons: 
non-constant mobility versus expanding the logarithm. An
additional difference between the DDFT and PFC models is that in the PFC
models, the constraint that $n\geq-1$ (i.e., $\rho\geq0$) is not enforced.

As in the DDFT derivations, we can now make (or not make) the RY/RPA approximation
and the gradient expansion. We consider first the RY/RPA approximation,
setting $c^{(3)}=c^{(4)}=0$ in \PFCbeta. The
free energy is
 \begin{equation}
 \beta\cFPFCgamma[n] =
     \int \! \left( - \tfrac{1}{2}n\cL n
                   - \tfrac{1}{6}n^3
                   + \tfrac{1}{12}n^4
             \right)\rdbx_1,
 \label{eq:DDFTFPFCgamma}
 \end{equation}
the dynamics is
 \begin{equation}
 \frac{\partial n}{\partial t} =
     - \nabla^2 \left[
                \cL{n} + \tfrac{1}{2}n^2 - \tfrac{1}{3}n^3
                \right]
 \label{eq:PFCgammadynamics}
 \end{equation}
and the chemical potential is
 \begin{equation}
 \mu=\beta\frac{\deltacFPFCgamma}{\deltan} =
                - \cL{n} - \tfrac{1}{2}n^2 + \tfrac{1}{3}n^3.
 \label{eq:betadFdngamma}
 \end{equation}
We refer to this model as \PFCgamma, and it is effectively the
same as \PFCbeta\ but with $Q=\frac{1}{2}$ and $C=-\frac{1}{3}$.

Finally, the gradient expansion can be made, replacing~$\cL$ by~$\cLgrad$ in
all expressions in this subsection, resulting in \PFCdelta\ (without~RY/RPA) and
\PFCepsilon\ (with~RY/RPA).

We refer to these models collectively as the PFC models, and
have chosen the
names \PFCalpha\ \hbox{etc.} to distinguish these from the PFC1 and
PFC2 models of Ref.~\cite{Teeffelen2009}.
The quadratic term in the dynamics ($Qn^2$) can be removed
(provided $C\neq0$) by adding a constant to~$n(\bx)$, but we choose not to do
this as it implies a change to what was meant by $\rho_0$ in the reference
liquid. In addition, a negative~$C$ can be scaled to~$-1$.
With these changes, \PFCepsilon\ is equivalent to
the original \PFCalpha\
model~(\ref{eq:PFCalphadynamics}) of~\cite{Elder2002,Elder2004}:
 \begin{equation}
 \frac{\partial n}{\partial t} = - \nabla^2 \left(r n - \gamma(1+\nabla^2)^2n + Qn^2 + Cn^3\right),
 \label{eq:PFCepsilondynamics}
 \end{equation}
where we have written out~$\cLgrad$ explicitly, and
$Q=\frac{1}{2}$ and $C=-\frac{1}{3}$ (or $Q=0$ and $C=-1$ after scaling
and adding a constant to~$n$ -- returning to the conserved Swift--Hohenberg equation).

The implication of dropping the $\nabla\cdot\left[n\nabla\cL{n}\right]$
term in the dynamics~(\ref{eq:DDFT3dynamics}) for DDFT-3 is now apparent: without this
term, Eq.~(\ref{eq:DDFT3dynamics}) reduces to~(\ref{eq:PFCgammadynamics}) but with the
cubic term removed. The absence of the cubic term here implies a free energy as
in~(\ref{eq:DDFTFPFCgamma}) that is not bounded below, i.e., a free energy that is
non-physical, and so the $\nabla\cdot\left[n\nabla\cL{n}\right]$  term can have the
effect of stabilizing patterns. In addition, dropping the
$\nabla\cdot\left[n\nabla\cL{n}\right]$ term is consistent only with a theory with a
constant mobility.

\subsection{Summary}

To summarise, we have carefully laid out the various approximations made in the
progression from the DDFT-0 starting point~(\ref{eq:DDFT0dynamics}) to the final
PFC~(\ref{eq:PFCalphadynamics},\ref{eq:PFCgammadynamics}) written 
down in~\cite{Elder2002,Elder2004}. We have largely followed earlier
derivations~\cite{Teeffelen2009,Huang2010a,Emmerich2012}, seeking to clarify
the approximations that are made. Along the way, we have identified four
intermediate versions of DDFT, listed for clarity in Table~\ref{tab:DDFTvsPFC}.
The change in name from DDFT to PFC could be made at any point in this
progression, but we prefer to make the name change at the point where the
$\nabla\cdot\left[n\nabla\cL{n}\right]$ term is dropped (along with all the 
other changes that are implied by this), since removing this
term marks a considerable alteration to the free energy expression and to
the dynamics.

The PFC model~(\ref{eq:PFCbetadynamics}) is appealing in its simplicity, and
it gives insight into a variety of crystallisation phenomena, but the
derivations of the model from DDFT presented here, as well as the derivation from 
Ref.~\cite{Huang2010a}, are both problematic. 
Just dropping the
$\nabla\cdot\left[n\nabla\cL{n}\right]$ term, as done by Huang
\etal~\cite{Huang2010a}, means that the dynamics is not equivalent to a DDFT
with mobility proportional to~$\rho$. On the other hand, the
alternative is to expand the logarithm up to
$\cO(n^4)$ in (\ref{eq:expandlogarithm}) in order to provide a \emph{nonlinear}
stabilizing term ($\frac{1}{12}n^4$) in the free
energy~(\ref{eq:DDFTFPFCgamma}). 
However, in the original formulation, the
logarithm comes from the ideal gas term in~(\ref{eq:separatedF}), and leads to
a \emph{linear} diffusive term in the dynamics. The stabilizing nonlinear terms
in~(\ref{eq:DDFT2dynamics}) are provided by $c^{(3)}_0$, $c^{(4)}_0$ (in DDFT-2) and 
by the $\nabla\cdot\left[n\nabla\cL{n}\right]$~term -- these are
all absent in~\PFCgamma.

Indeed, all these models only make physical sense if their free energies are bounded
below. The free energies for DDFT-0 and DDFT-1 are too general to make any
comment, but that for DDFT-2~(\ref{eq:DDFTF2}) \hbox{etc.} can be
discussed. The $(1+n) \log (1+n) - n$ term is bounded below by zero, and the
$-\tfrac{1}{2}\int{n}\cL{n}\rdbx$ term is bounded below because the eigenvalues
of~$\cL$ are bounded above:
 \begin{equation}
 - \int \! n(\bx)\cL n(\bx) \rdbx \geq - \sigmamax \int \! n^2(\bx) \rdbx,
 \end{equation}
where $\sigmamax$ is the maximum over~$k$ of~$\sigma(k)$ (we have in mind
a $\sigma(k)$ as in Fig.~\ref{fig:growthrate}).
In any case, this term, along with the other quadratic and cubic terms,
is dominated by the quartic in~$n$,
which is bounded below provided $c^{(4)}_0<0$. If $c^{(4)}_0=0$, then
$c^{(3)}_0<0$ will do, recalling that $n\geq-1$.
For DDFT-3, with the RY/RPA approximation $c^{(3)}_0=c^{(4)}=0$,
the boundedness of the free energy~(\ref{eq:DDFTF3}) depends on the
$n^2 + n\cL{n}$ combination. From~(\ref{eq:defnL}), the relevant term is
 \begin{equation}
 \begin{split}
  -\int \! &
       \left(n^2(\bx_1) + n(\bx_1)\cL n(\bx_1) \right)\rdbx_1 = \\
  & \qquad - \rho_0 \!\!\int\! n(\bx_1)c^{(2)}(\bx_1,\bx_2)n(\bx_2)\rdbx_1\rdbx_2.
 \end{split}
 \end{equation}
In general, this is not
bounded below, but it is in certain circumstances. For example,
it is if $\sigmamax<-1$,
and it is if $c^{(2)}(\bx_1,\bx_2)\leq0$ (or $u(|\bx_1-\bx_2|)\geq0$ for RPA)
for all $\bx_1$ and~$\bx_2$,
which is the case in the numerical examples below.
The PFC models are not constrained to have $n\geq-1$, but $\cFPFCbeta$
(\ref{eq:DDFTFPFCbeta}) is bounded by the $n^4$ term as long as its coefficient
is positive; $\cFPFCgamma$
(\ref{eq:DDFTFPFCgamma}) is always bounded below, because the expansion of
the logarithm in~(\ref{eq:expandlogarithm}) was
truncated after an even powered term.

Throughout we have made the simplest choices in the approximations, but other
authors have made many other choices. For example, the original PFC
paper~\cite{Elder2002}, as well as later
papers~\cite{Elder2007,Huang2010a,Robbins2012a,Alster2017a,Elder2017a},
included a two-component (binary) version of the PFC model. Recently, Wang
\etal~\cite{Wang2018c} took a much closer look at $c^{(3)}$ and $c^{(4)}$,
expressing these in terms of isotropic tensors and so allowing these functions
to introduce bond angle dependence into the free energy. Some choices of
$c^{(3)}$ and $c^{(4)}$ lead to nonlinear terms that include gradients, which
can affect the selection of the final stable crystal~\cite{Wu2010a}. The
gradient expansion approach has been generalised in two ways: (i) higher
order terms or rational functions were considered
by~\cite{Jaatinen2009,Pisutha-Arnond2013b} in order to improve the fit between
the functional form and the Fourier transform of~$c^{(2)}$, and (ii) PFC
models with two unstable length scales have been put forward by several
authors~\cite{Pisutha-Arnond2013b, Wu2010, Barkan2011, Achim2014,
Subramanian2016, Jiang2017}, since these allow more complex
crystals (face-centered cubic, icosahedral quasicrystals, \dots) to be
stabilized. We discuss the model of~\cite{Jaatinen2009} in more detail below. 
Alternative approaches involving weighted densities are also 
possible~\cite{Jaatinen2010a}.

\section{Comparison of DDFT and PFC}\label{sec:3}

We are interested in the effects of the approximations made in going from DDFT 
to~\hbox{PFC}. A full assessment of the validity of the RY/RPA approximation for
$\cFex$, which in itself constitutes a major simplification, is beyond the scope of
the present study. The general conclusion on the validity of the RY/RPA
approximation is that it depends on the nature of
the interactions between the particles; there are examples in the literature
where this approximation is reliable and others where it
works badly -- see for example the discussion in
Refs~\cite{Evans2009, Lowen2009} and references therein. 

Here, we consider one particular system where the
RPA is accurate and then we focus on the effects of 
approximating $\cL$ by $\cLgrad$ and of making the
suite of other approximations inherent 
in going from DDFT to PFC: expanding the logarithm, assuming constant mobility 
and dropping the $\nabla\cdot\left[n\nabla\cL{n}\right]$ term. To this end, we 
start with DDFT-3 and solve~(\ref{eq:betadFdn3}), rewritten here as:
 \begin{equation}
 \text{DDFT-3:}\quad
    \log\left(1+n(\bx)\right)
    - n(\bx) - \cL n(\bx) = \mu.
 \label{eq:betadFdn3_rewritten}
 \end{equation}
The system that we consider is particles interacting via the
GEM-4~\cite{Mladek2006, Prestipino2014} potential: this
is a model for soft-matter particles and in particular for dendrimers and other polymers
in suspension, treating the polymers via an effective pair potential
between their centers of mass. This potential
is soft, i.e., finite for all values of $x_{12}$
\cite{Mladek2006, Prestipino2014, Likos2001, Likos2006, Lenz2012}, and is
 \begin{equation}
 u(x_{12}) = \epsilon e^{-\left(x_{12}/R\right)^4},
 \label{eq:GEM4Potential}
 \end{equation}
where the parameter~$\epsilon$ controls the strength of the potential and $R$
controls its spatial range. We consider
here the system in 2D \cite{Prestipino2014, Archer2014}. 
As long as the temperature and density are high enough that the particle
cores regularly overlap (the regime in which the system freezes), the RPA
approximation \eqref{eq:RPAfreeenergy} is known to be rather accurate
for the GEM-4 system and gives a good account of the phase diagram and the
structure of the liquid and solid phases
\cite{Mladek2007, Prestipino2014, Archer2014}.

From Eqs.~(\ref{eq:defnL}), (\ref{eq:RPAc2u}) and \eqref{eq:GEM4Potential} 
we obtain the linear operator~$\cL$:
 \begin{equation}
 \cL n(\bx) = - n(\bx) - \rho_0\beta\epsilon \!\!\int\! 
                         e^{-|\bx-\bx_2|^4/R^4} n(\bx_2)\rdbx_2.
 \label{eq:defnL_GEM4}
 \end{equation}
Recall from~(\ref{eq:defnsigma}) that~$\cL$ has eigenvalues~$\sigma(k)$ with
eigenfunctions $e^{i\bk\cdot\bx}$. We can choose the combined parameter 
$\rho_0\beta\epsilon$ and soft-particle radius $R$ so
that the maximum in $\sigma(k)$ occurs
at $k=1$ when the system is at the linear stability threshold, i.e.,
this is a maximum with $\sigma(1)=0$, 
similar to Fig.~\ref{fig:growthrate}. In 2D, to satisfy this
condition we must have $\rho_0\beta\epsilon=0.2455$ and
$R=5.0962$ -- see Appendix~\ref{app:LinearGEM4} for details.

\begin{figure}
\begin{center}
\includegraphics[width=8.6truecm]{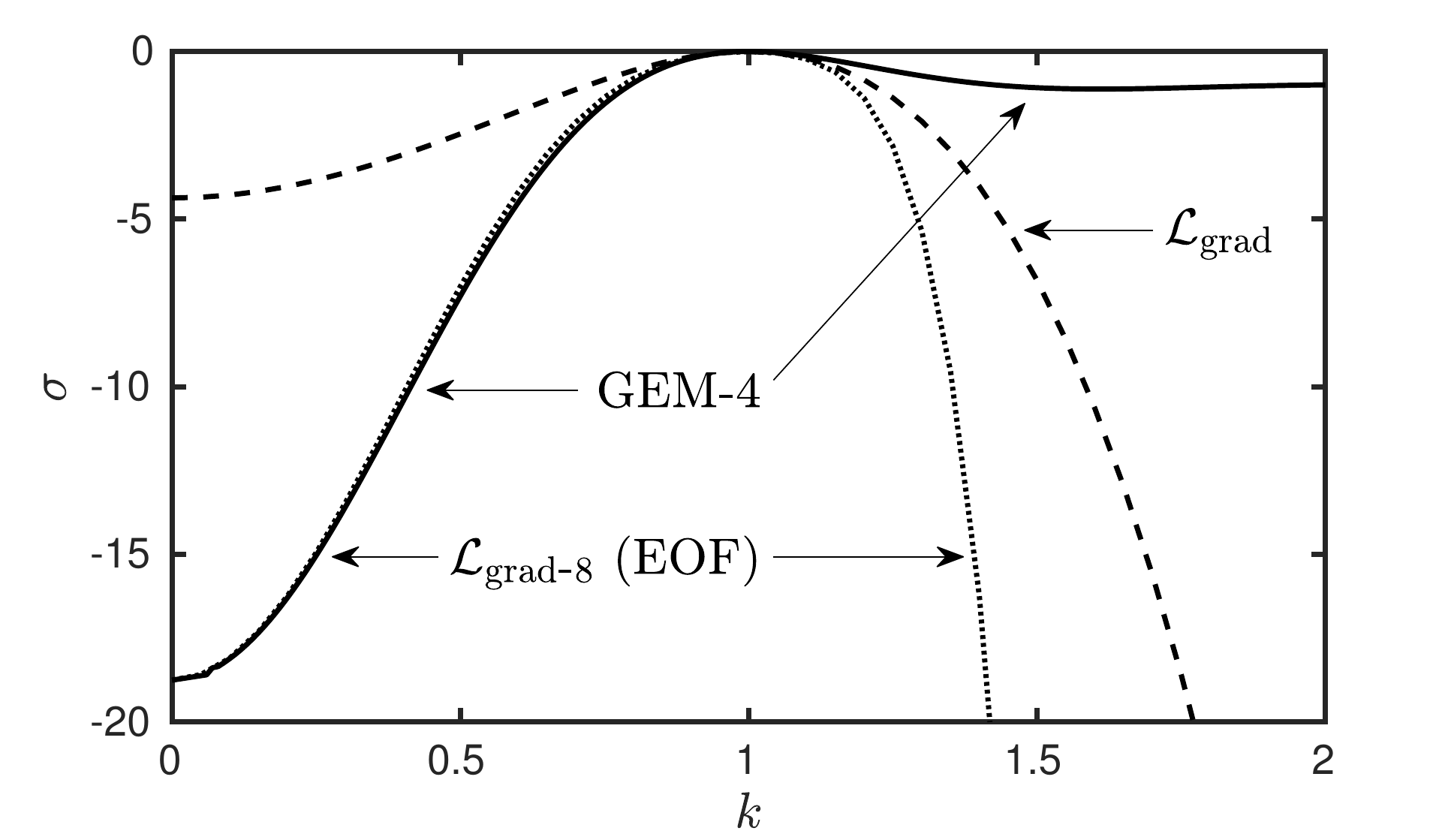}
\caption{The eigenvalue $\sigma(k)$ of~$\cL$
plotted as a function of wavenumber~$k$ for the GEM-4 potential (solid line)
for $R=5.0962$ and $\rho_0\beta\epsilon=0.2455$, which is at the threshold where
the system becomes linearly
unstable. This has $\sigma(0)=-18.75$. We also display $\sigma(k)$ from
the gradient  expansion of $\cL$ (dashed line), i.e., a Taylor expansion in Fourier
space around $k=1$, which is the PFC relation for $\cLgrad$,
$\sigma(k)=-\gamma(1-k^2)^2$, with
$\gamma=4.37$. Recall that the growth rate of Fourier modes
with wavenumber $k$ is is $k^2\sigma(k)$.
The dotted line labelled $\cLgradJ$ (EOF) is the curve for~(\ref{eq:cLgradJaatinen}),
the eighth-order fitting proposed in~\cite{Jaatinen2009}: it nearly coincides with
the GEM-4 curve for $k\leq1$.}
 \label{fig:dispersionGEM4}
\end{center}
 \end{figure}

With this choice of parameters, the eigenvalue $\sigma(k)$ is shown as 
a solid line in Fig.~\ref{fig:dispersionGEM4}. The figure also shows (dashed 
line) the eigenvalue for the gradient expansion of~$\cL$ around $k=1$:
 \begin{equation}
 \cLgrad n(\bx) = - \gamma(1+\nabla^2)^2 n(\bx),
 \label{eq:cLgradGEM4}
 \end{equation}
where $\gamma=4.37$ is chosen to match the second derivative
$\frac{\rd^2\sigma}{\rd{k}^2}$ at $k=1$, as done for
example in Refs.~\cite{Elder2004, Wu2007, Jaatinen2009}.
The dotted line in Fig.~\ref{fig:dispersionGEM4} is the eigenvalue
for~(\ref{eq:cLgradJaatinen}),
the eighth-order fitting model proposed in~\cite{Jaatinen2009} and discussed
in more detail below. 

\begin{figure}[t]
\begin{center}

\includegraphics[width=8.6truecm]{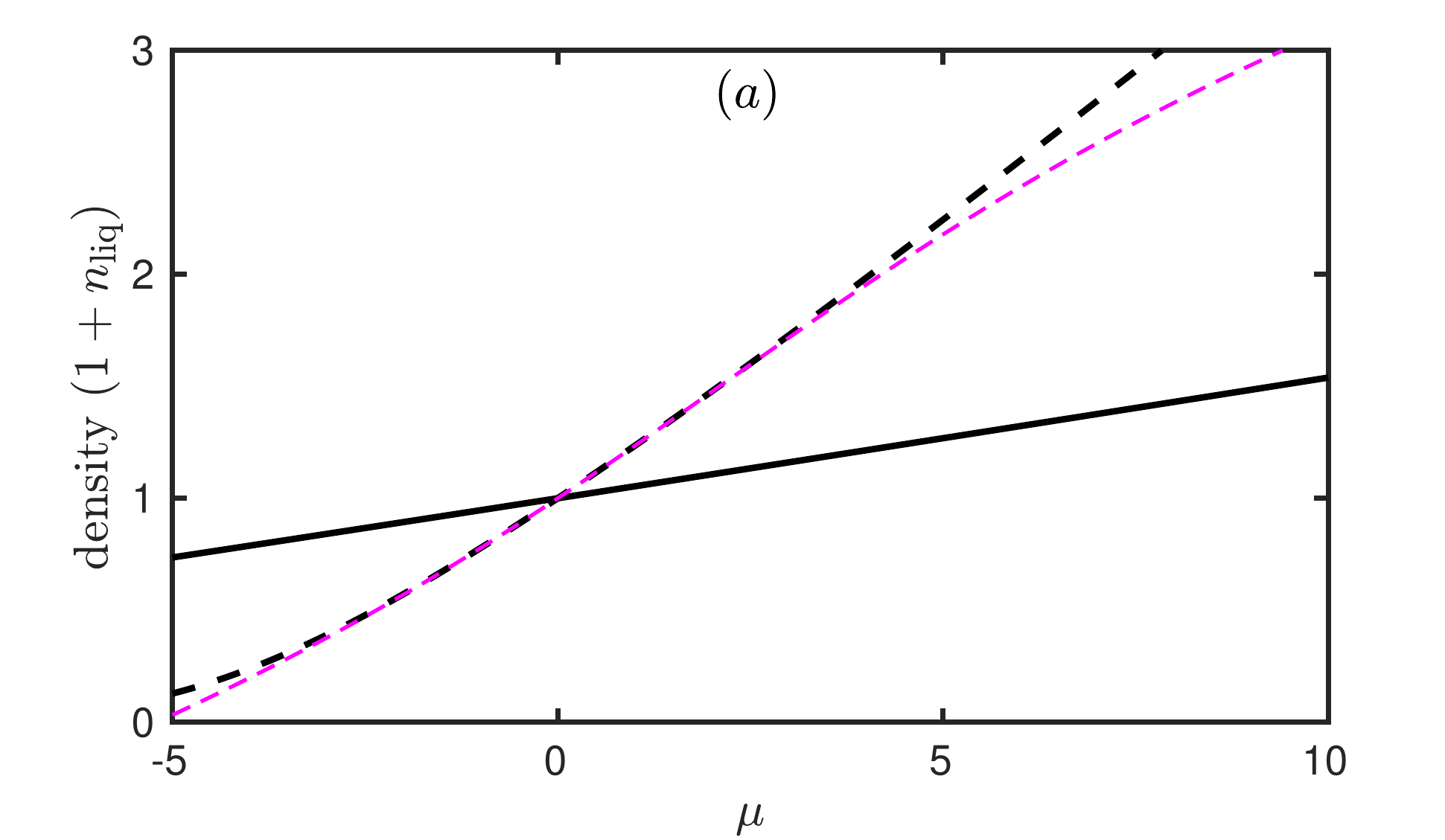}

\includegraphics[width=8.6truecm]{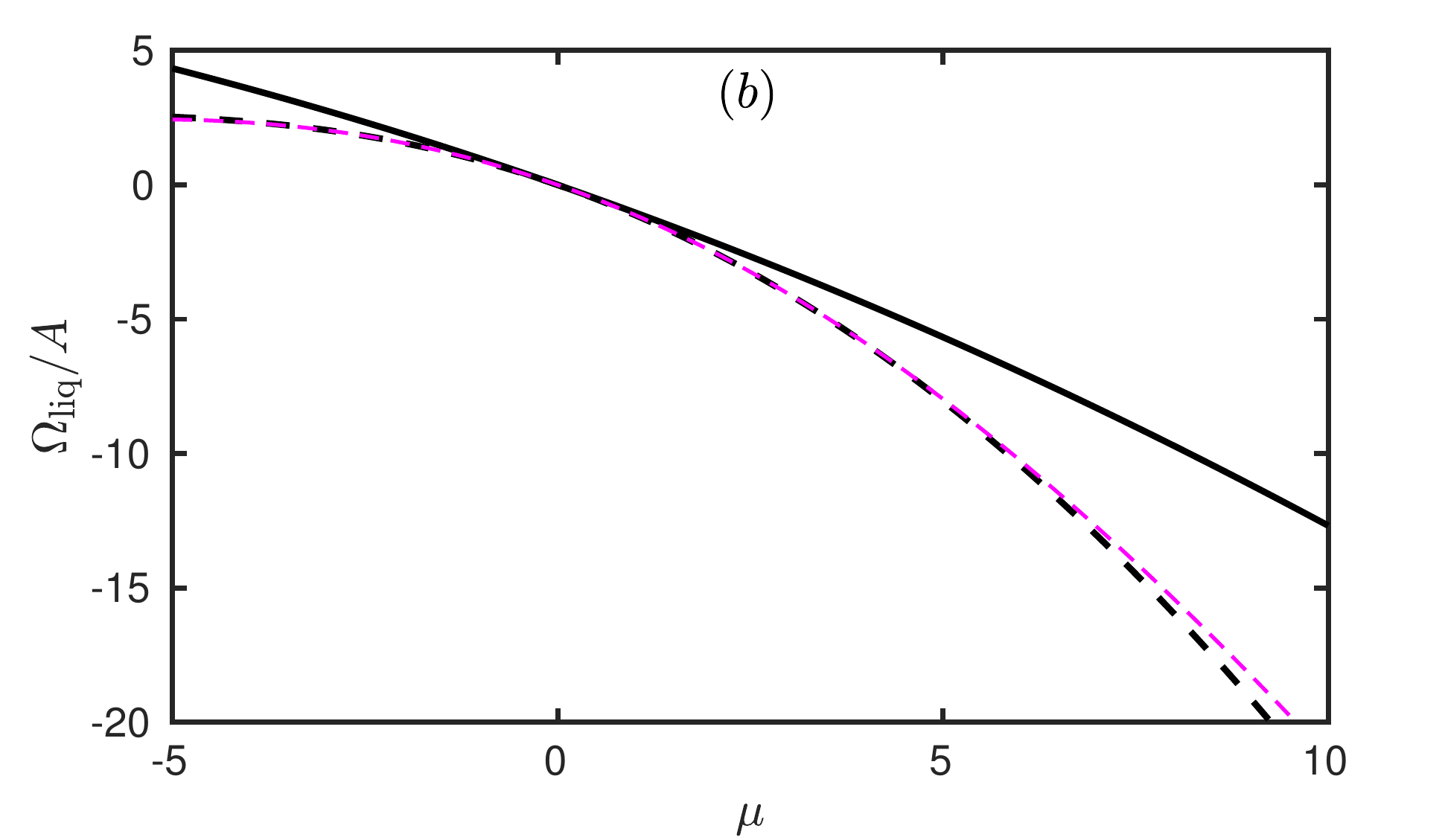}
\caption{(a) Liquid density ($1+\nliq$) and (b) specific grand potential $\Omegaliq/A$ as a
function of the scaled chemical potential $\mu$, for DDFT-3
(solid black line), DDFT-5 (dashed black line), \PFCgamma\ (indistinguishable from
DDFT-3) and \PFCepsilon\ (dashed magenta line). }
 \label{fig:liquiddensity}
\end{center}
 \end{figure}

In what follows we compare solutions of (\ref{eq:betadFdn3_rewritten}) for 
DDFT-3 with solutions of the analogous equations for DDFT-5, \PFCgamma\ and 
\PFCepsilon:
 \begin{align}
 &\text{DDFT-5:}&
    \log\left(1+n\right)
    - n - \cLgrad n &= \mu,
 \label{eq:betadFdn5_rewritten}\\
 &\text{\PFCgamma:}&
    -\tfrac{1}{2}n^2 + \tfrac{1}{3} n^3 - \cL n &= \mu,
 \label{eq:betadFdngamma_rewritten}\\
 &\text{\PFCepsilon:}&
    -\tfrac{1}{2}n^2 + \tfrac{1}{3} n^3 - \cLgrad n &= \mu.
 \label{eq:betadFdnepsilon_rewritten}
 \end{align}
See Appendix~\ref{app:continuation} for details of the pseudo-arclength
continuation numerical method we use for solving these equations.
Also, in the supplementary material we include a \textsc{Matlab} code for
solving DDFT-3.
Note that throughout what follows, we refer to the quantity
$1+n(\bx)$ as the `density'. 

Since the DDFT-5, \PFCgamma\ and 
\PFCepsilon\ represent different forms of Taylor expansion
around the reference state with density $\rho_0$,
there are a variety of ways comparison between solutions can be made.
Here, we opt to fix~$\cL$ and~$\cLgrad$ as in~(\ref{eq:defnL_GEM4})
and~(\ref{eq:cLgradGEM4}) with the specified values of
$\rho_0\beta\epsilon$, $R$ and~$\gamma$. This implies that at
$\mu=0$ the reference state with
$n=0$ is at the spinodal point and is marginally unstable to modes
with wavenumber~$k=1$. We then
vary~$\mu$ starting from $\mu=0$ and follow
the liquid, stripe and hexagonal
solutions of (\ref{eq:betadFdn3_rewritten}) and
(\ref{eq:betadFdn5_rewritten})--(\ref{eq:betadFdnepsilon_rewritten})
in appropriately sized two-dimensional domains. For a given value of
$\mu$ the different solutions have different values for the mean density
$1+\bar{n}=1+\frac{1}{A}\!\!\int\! n(\bx)\rdbx$, where $A$~is the area of
the domain. For each state we calculate
the specific grand potential:
 \begin{equation}
 \frac{\Omega[n]}{A} = \frac{\cF[n]}{A} - \mu (1+\bar{n}),
 \label{eq:specificgrandpotential}
 \end{equation} 
where $\cF$ is $\cFc$, $\cFe$,
$\cFPFCgamma$ or $\cFPFCepsilon$, as appropriate. We also
minimise~$\Omega/A$ with respect to 
the domain size~$A$ by applying the approach described
in Appendix~\ref{app:continuation}. For a given value of
the chemical potential $\mu$ and the combined parameter~$\rho_0\beta\epsilon$,
the thermodynamic equilibrium state is that with the minimum
value of~$\Omega/A$. Note that equilibria with the same~$\mu$ do not
necessarily have the same value of~$\bar n$, which is important when considering
which equilibria might result from initial conditions via the dynamics.

\begin{figure*}[t]
\begin{center}
\hbox to \hsize{\includegraphics[width=6.3truecm]{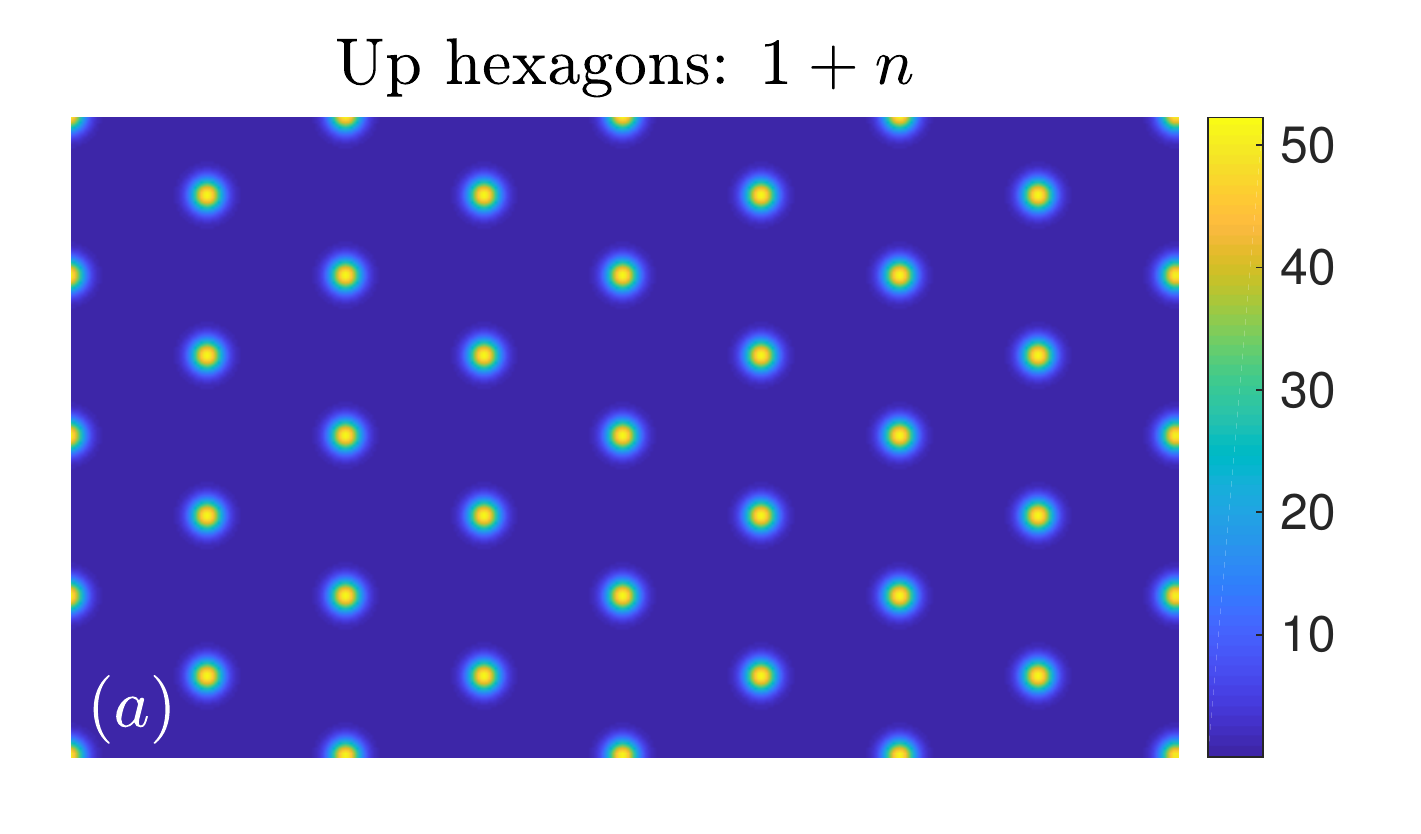}\hfill
                \includegraphics[width=6.3truecm]{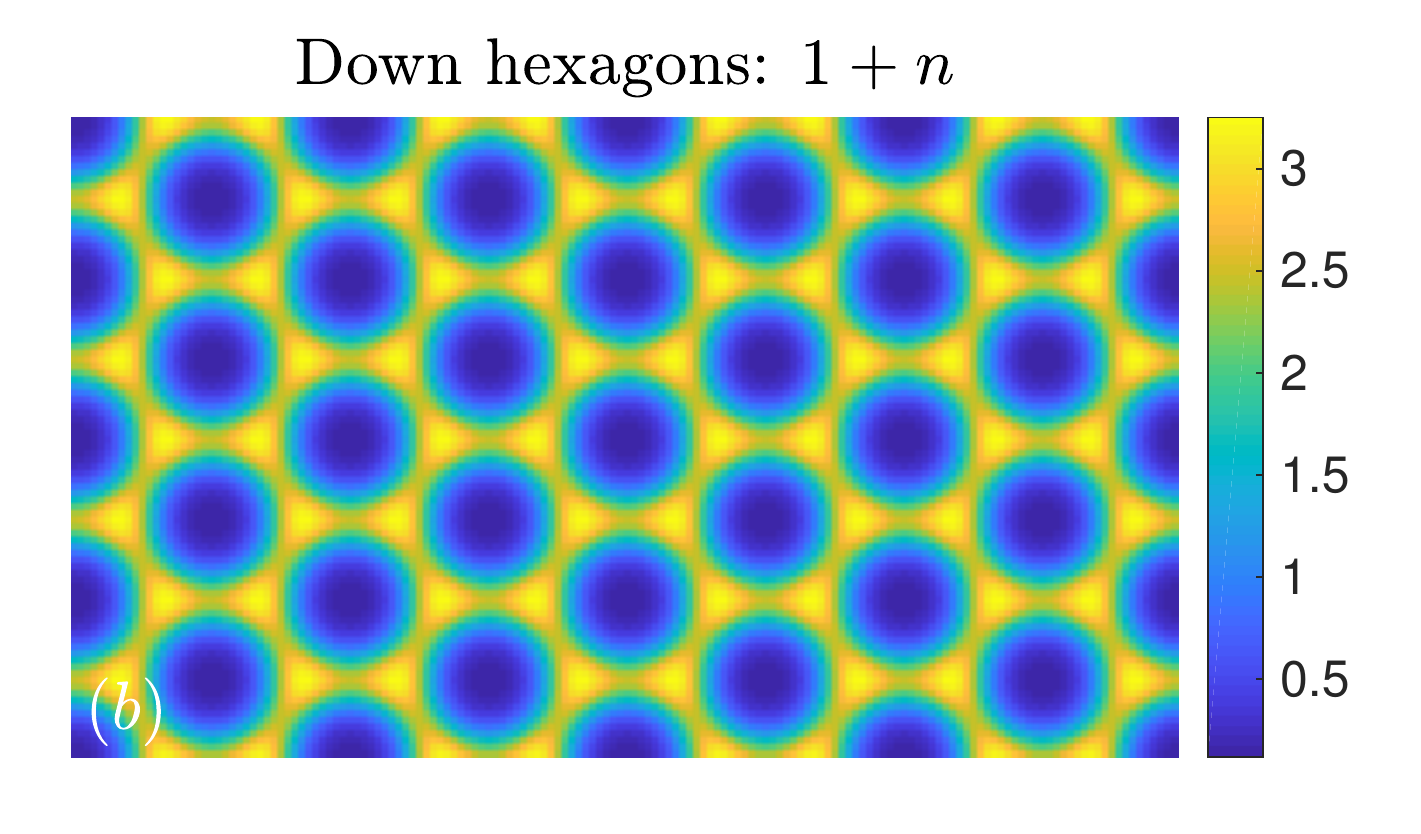}\hfill
                \raisebox{1.5ex}{\includegraphics[width=3.7truecm]{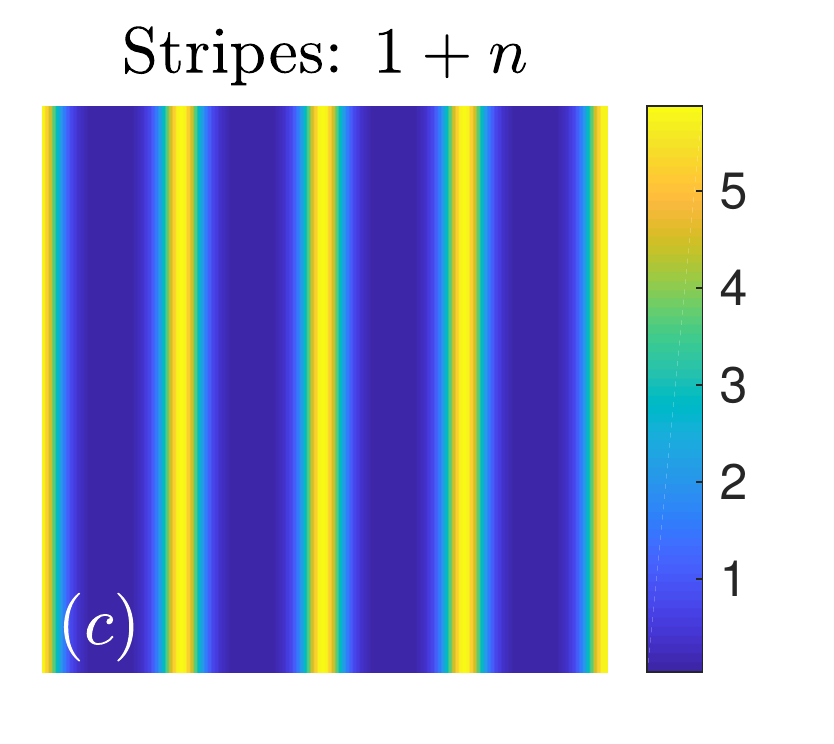}}}
\hbox to \hsize{\includegraphics[width=6.3truecm]{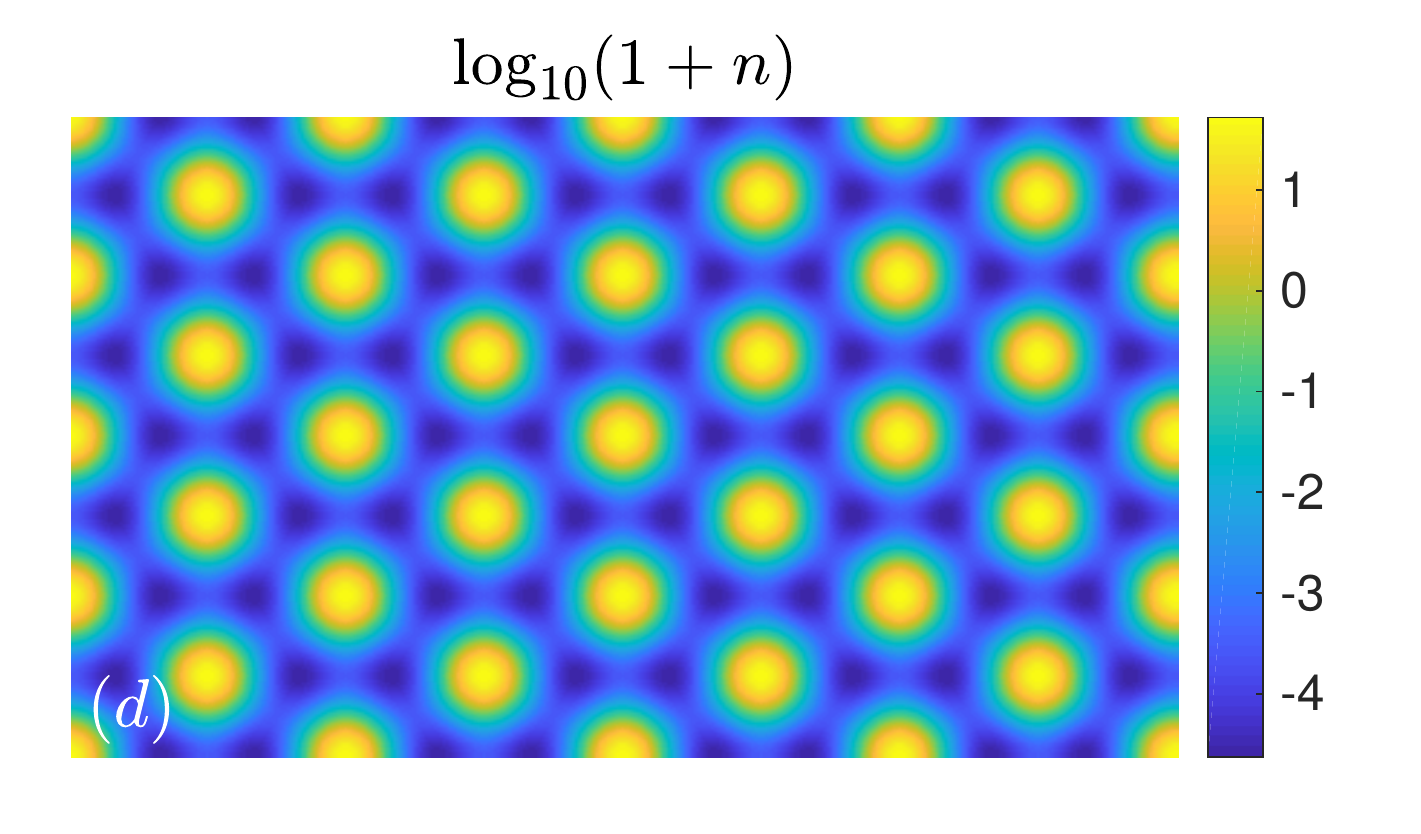}\hfill
                \includegraphics[width=6.3truecm]{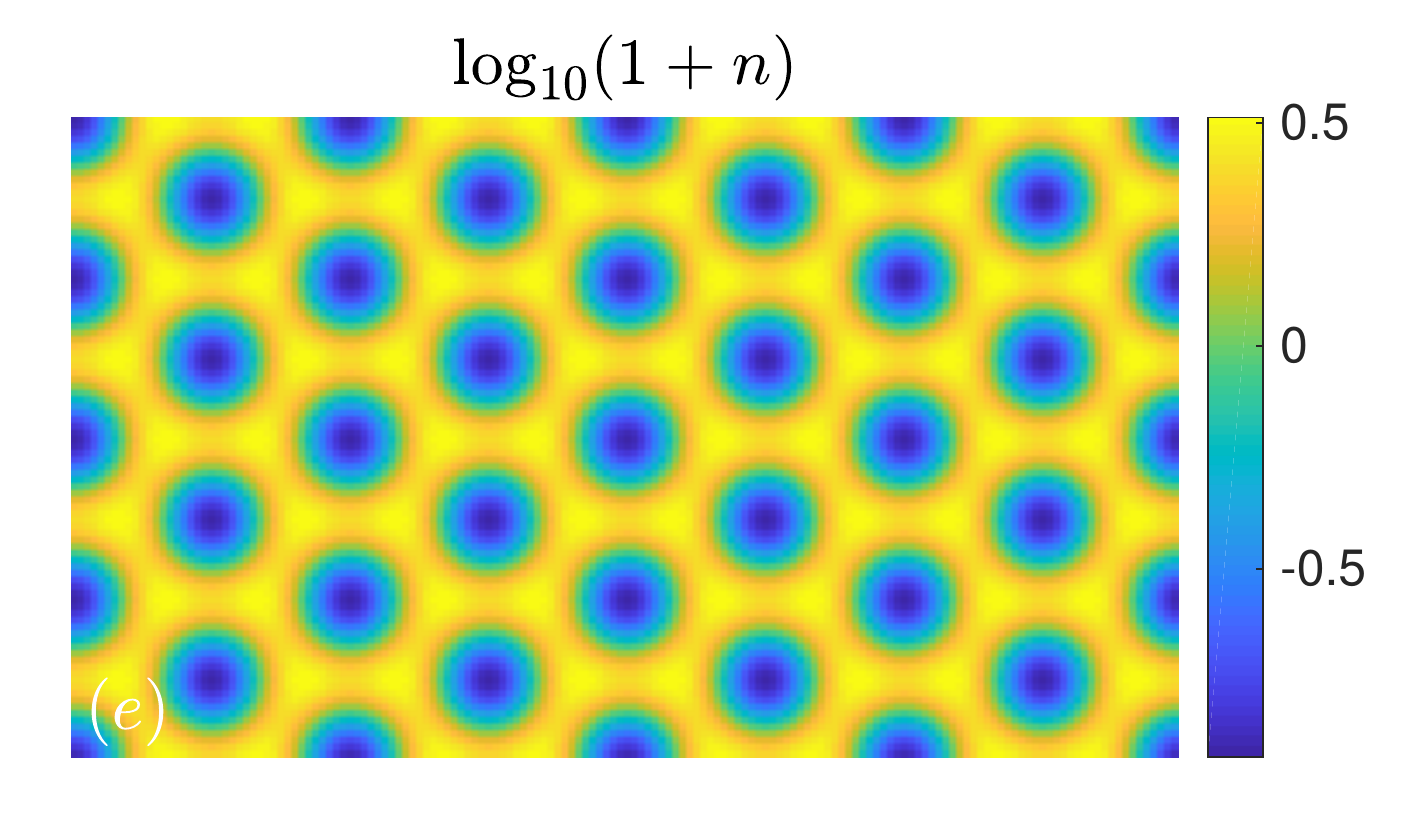}\hfill
                \raisebox{1.5ex}{\includegraphics[width=3.7truecm]{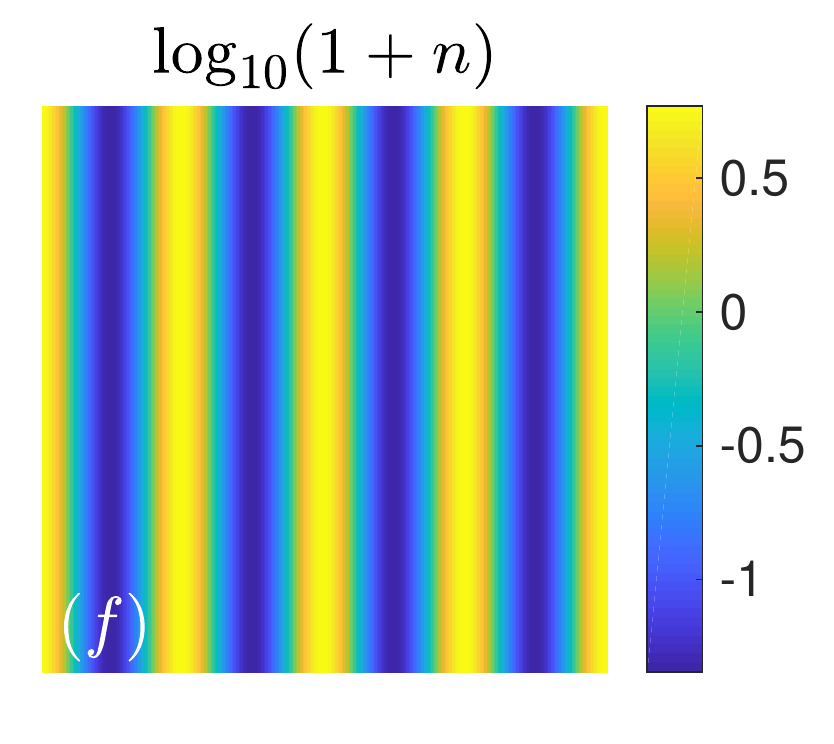}}}
\caption{Examples of solutions of DDFT-3 (\ref{eq:betadFdn3_rewritten}) with $\mu=10$.
(a,d)~Up hexagons (isolated density maxima surrounded by density that is
small but positive) in a $8\pi\times\tfrac{8}{\sqrt{3}}$ wavelength
domain, with $2.3\times10^{-5}<1+n<52.3$. 
(b,e)~Down hexagons (isolated density minima)
in a $8\times\tfrac{8}{\sqrt{3}}$ wavelength domain, with $0.12<1+n<3.3$.
(c,f)~Stripes in a $4\times4$ wavelength domain, with $0.045<1+n<5.9$.
The top row (a--c) shows the density $1+n$ and the bottom row (d--f) shows~$\log_{10}(1+n)$.}
 \label{fig:DDFT3solutions}
\end{center}
 \end{figure*}
  
The solution corresponding to the uniform density liquid state with $n(\bx)=\nliq$
can readily be found. In this case we have $\cL{\nliq}=\sigma(0)\nliq$, and so we
must solve the following algebraic equations for $\nliq$:
 \begin{align}
 &\text{DDFT-3,5:}&
    \log\left(1+\nliq\right)
    - \nliq - \sigma(0) \nliq &= \mu,
 \label{eq:DDFTnliq}\\
 &\text{\PFCgamma,$\epsilon$:}&
    -\tfrac{1}{2}\nliq^2 + \tfrac{1}{3}\nliq^3 - \sigma(0)\nliq &= \mu,
 \label{eq:PFCnliq}
 \end{align}
recalling that the value of~$\sigma(0)$ depends on whether or not the gradient 
expansion is carried out (see Fig.~\ref{fig:dispersionGEM4}). Finding~$\nliq$ 
for a given value of~$\mu$ is done easily using Newton's method, and the 
resulting $\nliq$ and specific~$\Omegaliq$ are shown in 
Fig.~\ref{fig:liquiddensity}. In all cases, we see that $\nliq$ 
is an increasing function of~$\mu$, while $\Omegaliq/A$ is a decreasing function 
of~$\mu$. The figure shows that 
the specific grand potential for the liquid state predicted by all four models
are similar close to~$\mu=0$, but the
predicted liquid state densities are rather different away from $\mu=0$.
This difference originates from the different values of
$\sigma(0)$ ($-18.75$ for DDFT-3 and 
\PFCgamma, in contrast to $-4.37$ for DDFT-5 and \PFCepsilon).
We see from Fig.~\ref{fig:liquiddensity}
that the density of the liquid is erroneously predicted to
increase too rapidly as $\mu$ is increased by the gradient expansion
theories (DDFT-5 and \PFCepsilon). This is because these get the value
of the isothermal compressibility $\chi_T$ to be too large
\cite{Jaatinen2009}. This compressibility is related to $\sigma(0)$ via
$\chi_T=-\beta/[\sigma(0)\rho_0(1+\nliq)]$~\cite{Hansen2013}:
see Eq.~\eqref{eq:sigmactwo} and following discussion. Expanding the logarithm
makes relatively little difference over this range of densities.

Since crystallisation occurs at higher densities, we expect a transition from
the liquid to the crystal to occur as~$\mu$ increases. At the
spinodal the uniform liquid becomes linearly unstable and the patterned state
solution branches bifurcate from the liquid at this point. To find these
states, we seek a solution of the form
 \begin{equation}
 n(\bx) = \nliq + \deltan(\bx),
 \end{equation}
where near the bifurcation point $\deltan\ll1$, and $\deltan$ is
of the form $e^{i\bk\cdot\bx}$, so that
$\cL\deltan=\sigma(k)\deltan$. Expanding Eqs.~(\ref{eq:betadFdn3_rewritten}) 
and~(\ref{eq:betadFdn5_rewritten}--\ref{eq:betadFdnepsilon_rewritten}) 
in powers of $\deltan$
we find that the $\cO(1)$ equations to solve are just
those for finding the liquid state density,
Eqs.~(\ref{eq:DDFTnliq})--(\ref{eq:PFCnliq}). The
$\cO(\deltan)$ equations are
 \begin{align}
 &\text{DDFT-3,5:}&
    \left(\frac{1}{1+\nliq} - 1 - \sigma(k)\right)\deltan &= 0,
 \label{eq:DDFTdeltan}\\
 &\text{\PFCgamma,$\epsilon$:}&
    \left(-\nliq + \nliq^2 - \sigma(k)\right)\deltan  &= 0.
 \label{eq:PFCdeltan}
 \end{align}
The spinodal point for DDFT-3,5 or for \PFCgamma,$\epsilon$ 
is where there are solutions of the equation with $\deltan\neq0$.

Since we are looking for a change in stability, we take the extreme
value of $\sigma(k)$, i.e., $\sigma(k)=0$ (see Fig.~\ref{fig:dispersionGEM4}).
Then, Eq.~(\ref{eq:DDFTdeltan}) is solved (with $\deltan\neq0$) 
only for $\nliq=0$, which leads to $\mu=0$
from~(\ref{eq:DDFTnliq}). In contrast, Eq.~(\ref{eq:PFCdeltan}) with $\sigma(k)=0$ 
has two solutions,
$\nliq=0$ and $\nliq=1$, leading to $\mu=0$ and $\mu=-\tfrac{1}{6}-\sigma(0)$
from~(\ref{eq:PFCnliq}). 
The implication of this
is that the PFC has two
spinodal points: the liquid loses stability at $\nliq=0$ as $\mu$ increases
through~$0$, but it regains stability at $\nliq=1$, which gives $\mu=18.58$ for
\PFCgamma\ and $\mu=4.20$ for
\PFCepsilon. This prediction that the liquid regains stability for higher~$\mu$
is a consequence of expanding the logarithm, or equivalently of Taylor expanding 
the $1/(1+\nliq)$ term in (\ref{eq:DDFTdeltan}) and is confirmed by direct
computation of the crystal solutions below. Of course, this prediction is
erroneous, since the simulation results for the GEM-4 system
\cite{Mladek2006, Prestipino2014} show no sign of a 
second spinodal point or the associated stable second liquid
in the equilibrium system phase diagram.

In Fig.~\ref{fig:DDFT3solutions} we display examples of the three
different types of periodic solutions that can be found for DDFT-3.
These are (i)~the crystal solution, which we refer to as `up hexagons',
which exhibits a triangular array of isolated density maxima surrounded by
hexagonal regions where the
density is close to zero. There are also (ii)~`down hexagons' which
are the opposite, with isolated density minima and hexagonal density maxima. 
Finally, there is (iii)~the stripe state. 
Depending on the state point these solutions are not
necessarily linearly stable. Our naming convention to distinguish
the two different hexagonal solutions originates in the convection literature
\cite{Bodenschatz2000}.
These solutions were initiated at $\mu\approx0$ and then
continued numerically (see Appendix~\ref{app:continuation}) up to $\mu=10$.
For DDFT-3 it is possible to go a bit higher in $\mu$, but with increasing $\mu$
(i.e., increasing average density) the peaks
in the density profile get narrower and higher and so more and more grid points
are required to resolve these peaks correctly.
However, as we show below for some of the other models and for different reasons,
it is not possible to continue the solutions
this far in~$\mu$. The domains on which the profiles
are calculated have periodic boundary conditions, with 4 wavelengths in each
direction (for stripes), or $8\times\tfrac{8}{\sqrt{3}}$ wavelengths (for
hexagons). The wavelength is initially equal to $2\pi$ for $\mu=0$ and
is then adjusted by up to about 2\% in order to minimise the specific 
grand potential as $\mu$ is varied; i.e.,
we minimise~$\Omega/A$ with respect to variations in the size of the crystal
unit cell or, for the stripe phase, we minimise with respect to variations in
the spacing between the stripes -- see Appendix B for details.

\begin{figure*}[t]
\begin{center}
\hbox to \hsize{\includegraphics[width=5.5truecm]{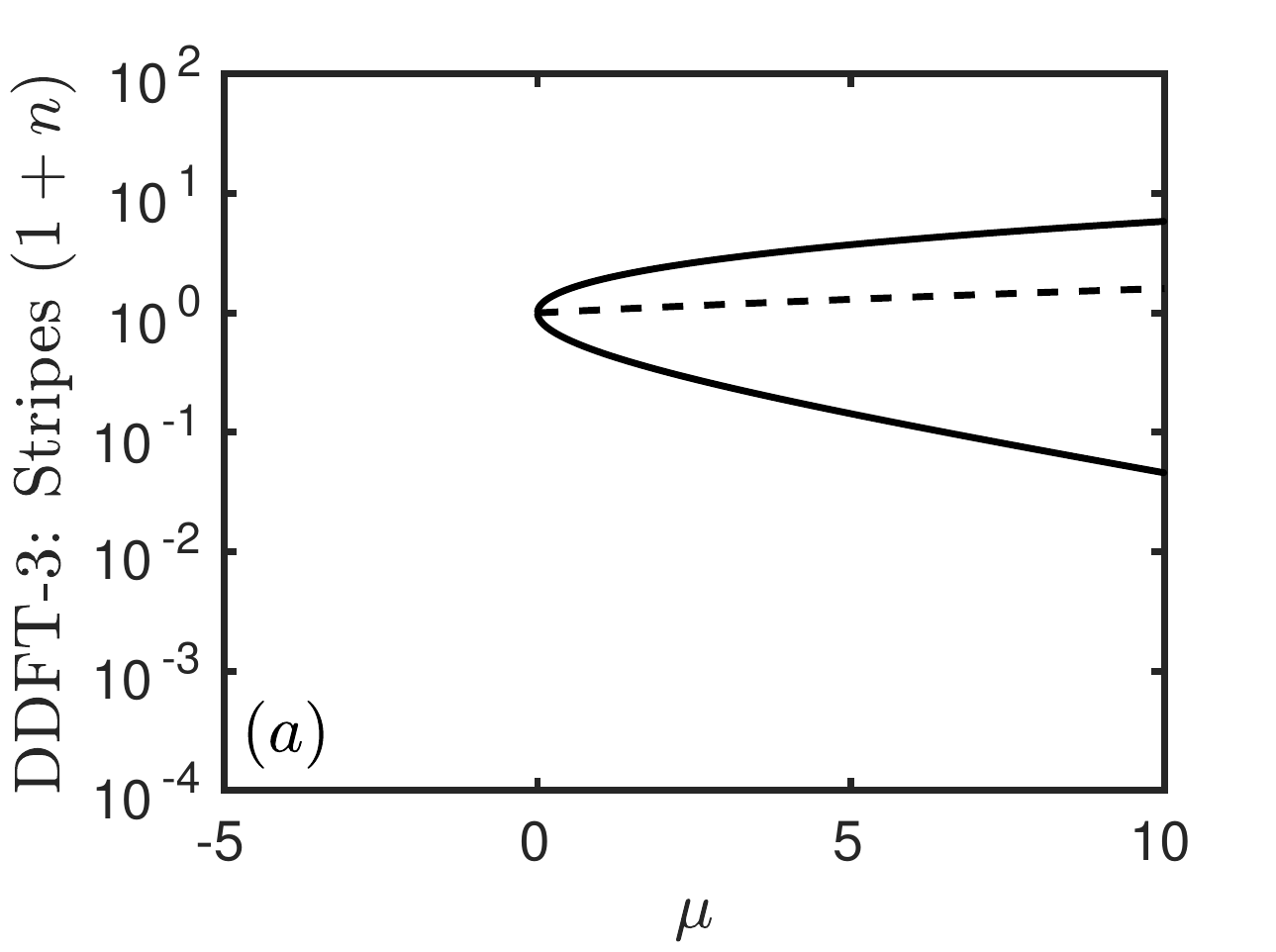}\hfill
                \includegraphics[width=5.5truecm]{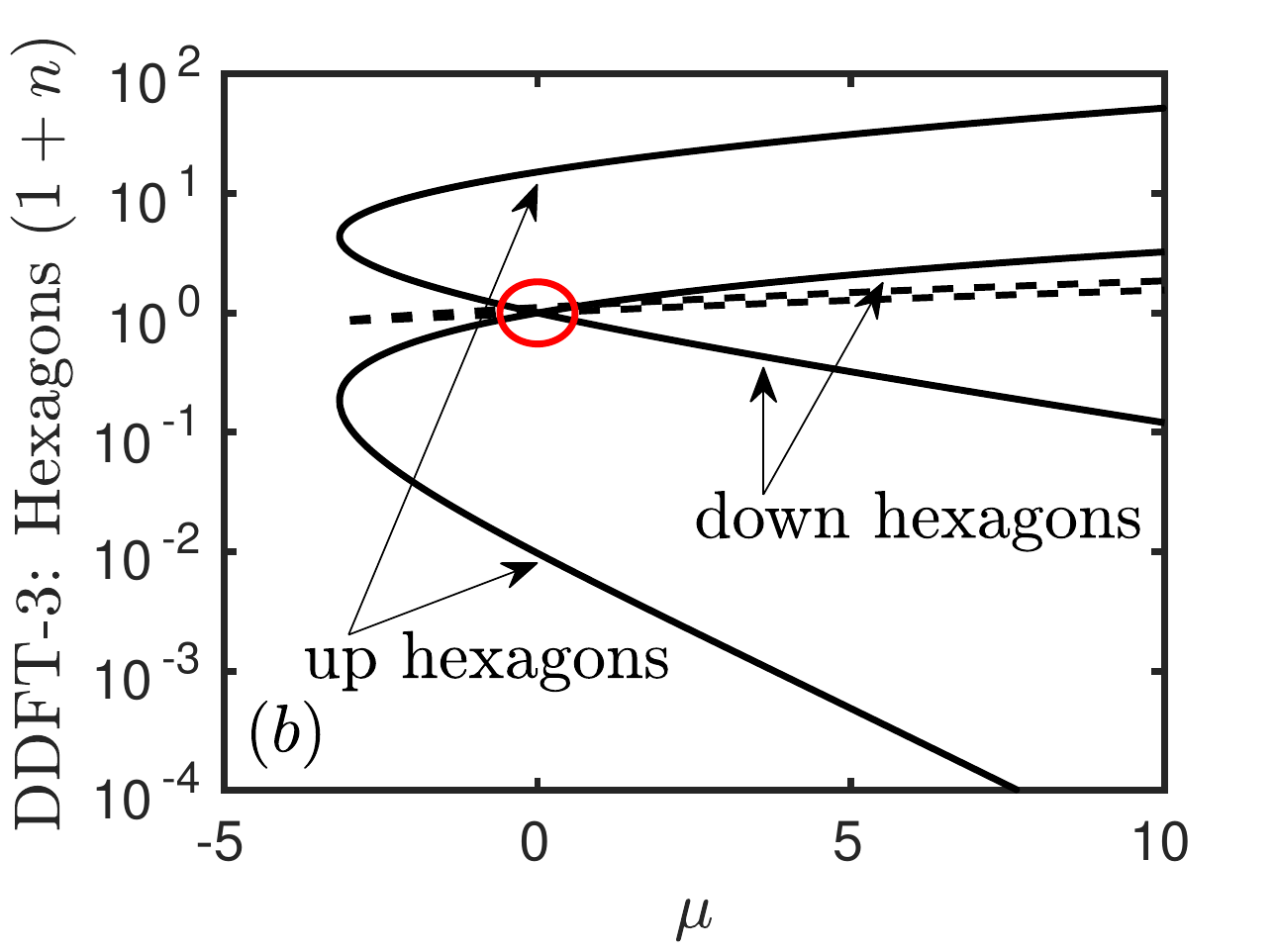}\hfill
                \includegraphics[width=5.5truecm]{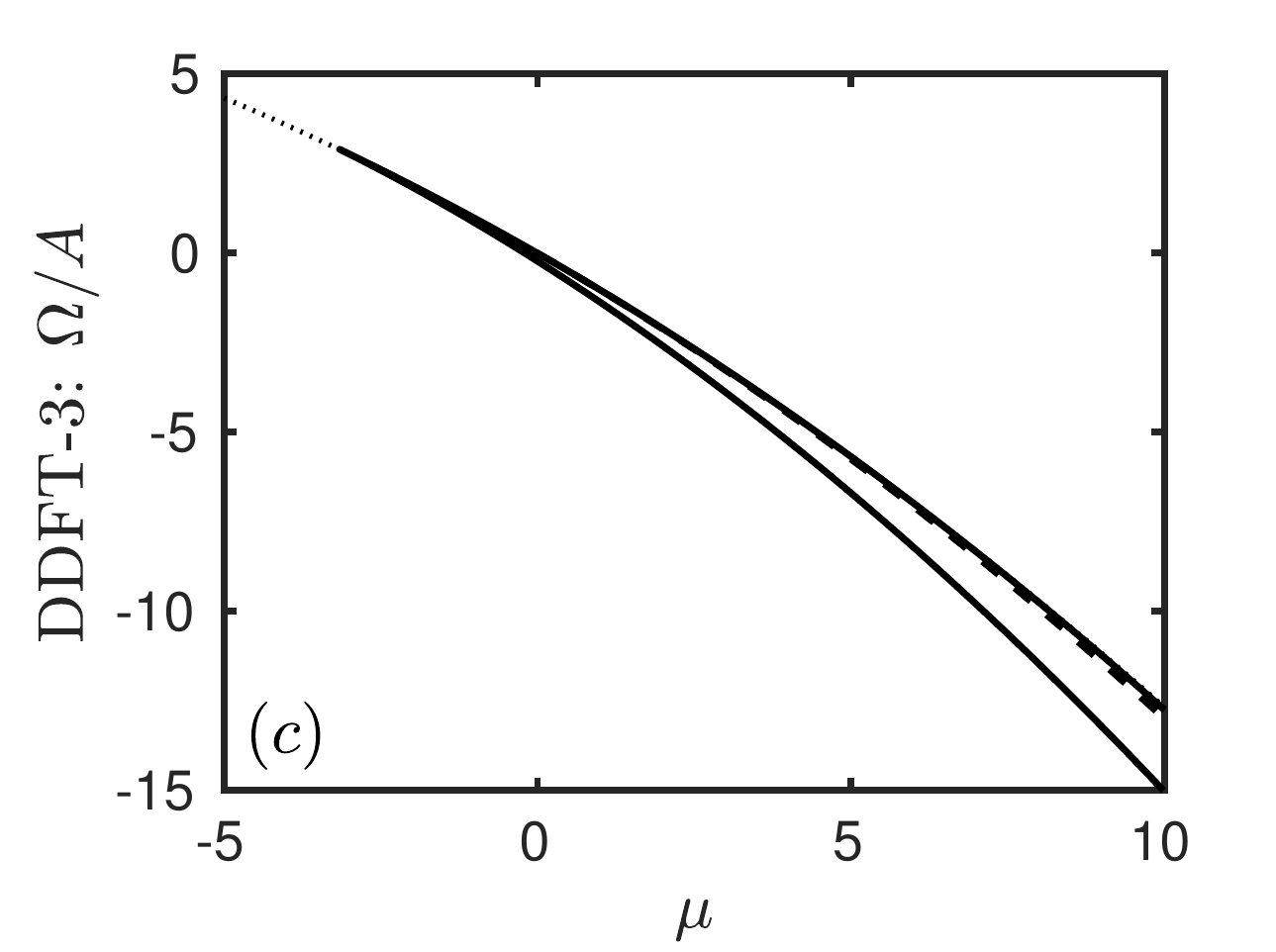}}
\hbox to \hsize{\includegraphics[width=5.5truecm]{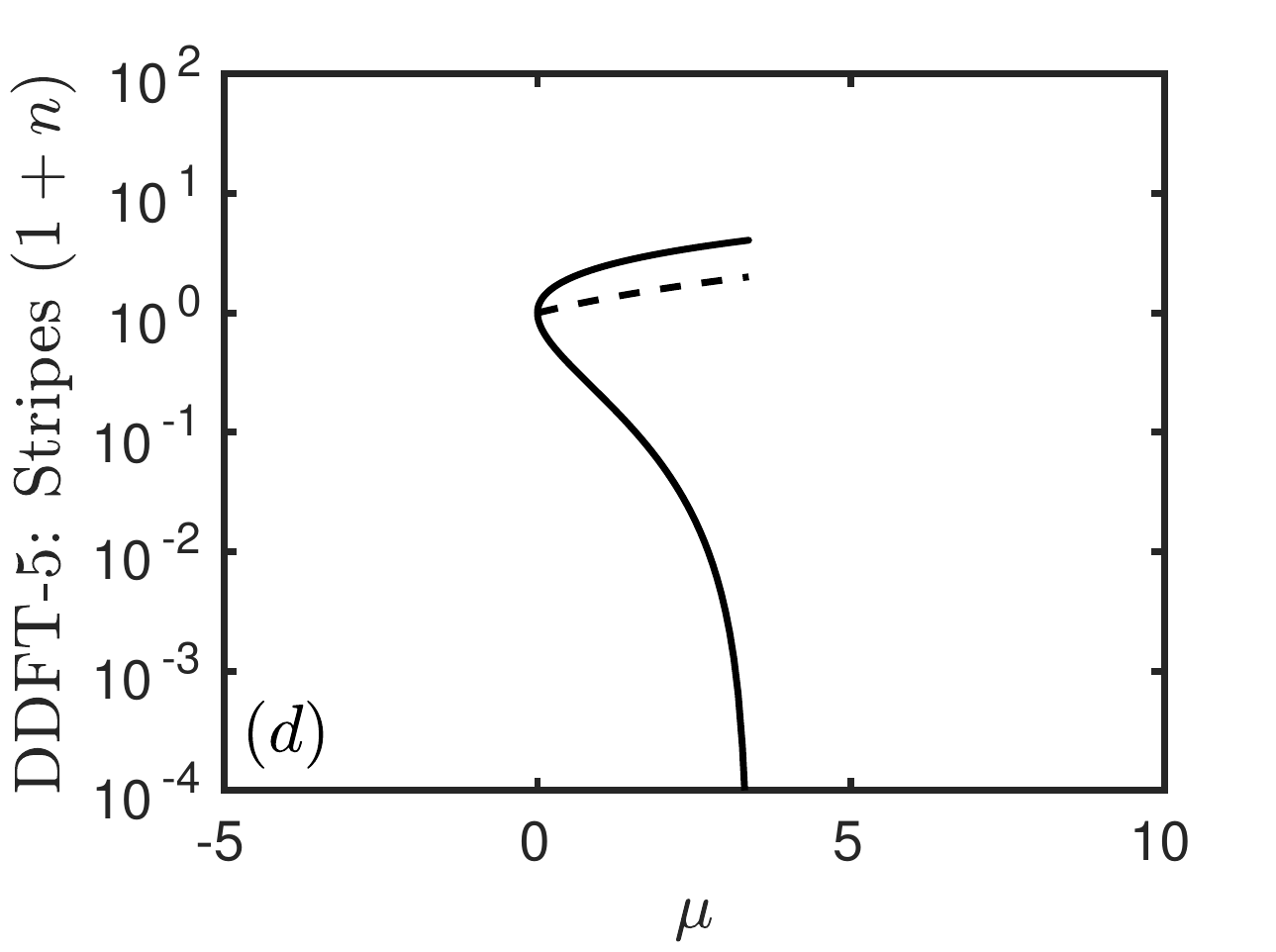}\hfill
                \includegraphics[width=5.5truecm]{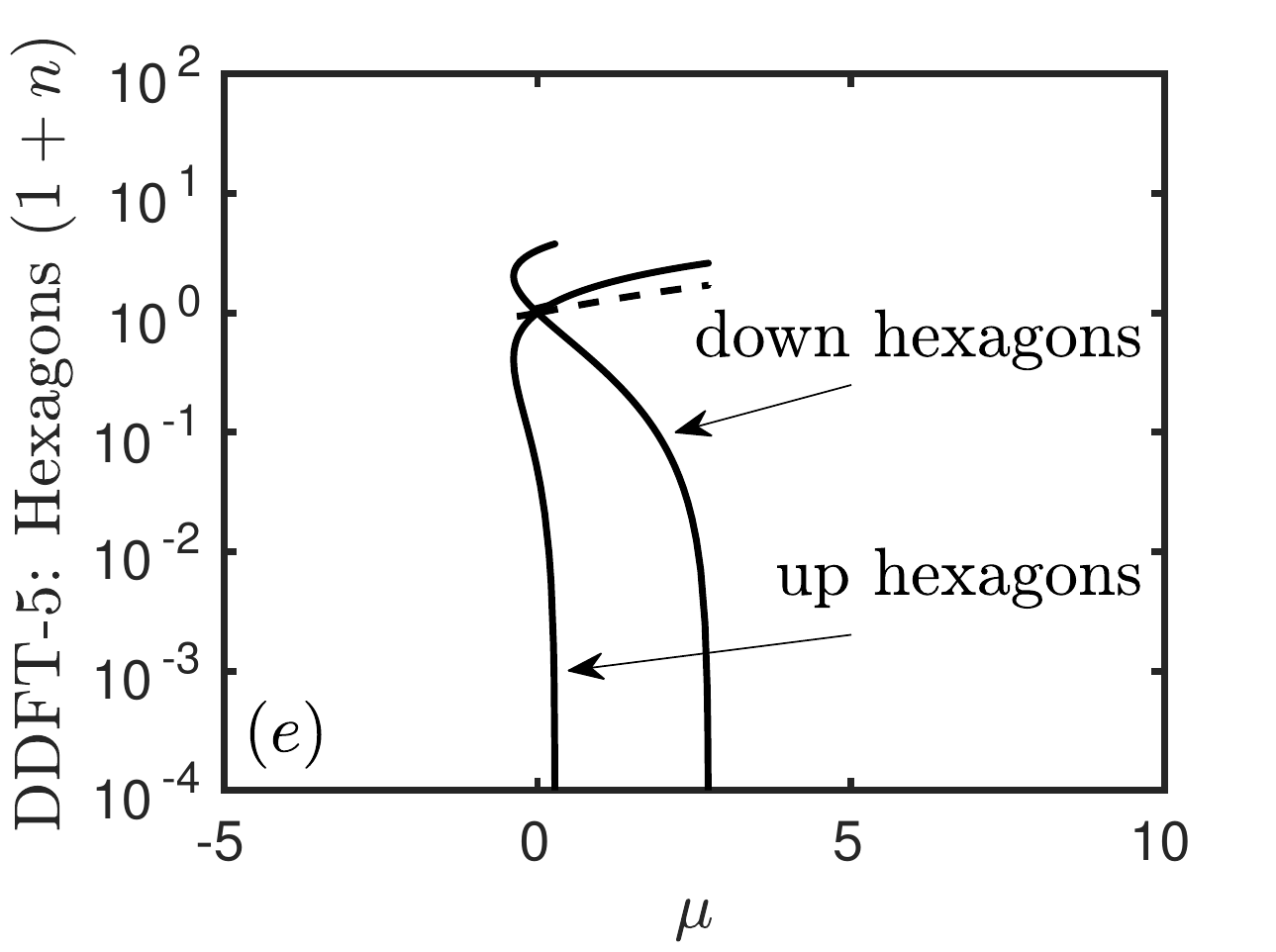}\hfill
                \includegraphics[width=5.5truecm]{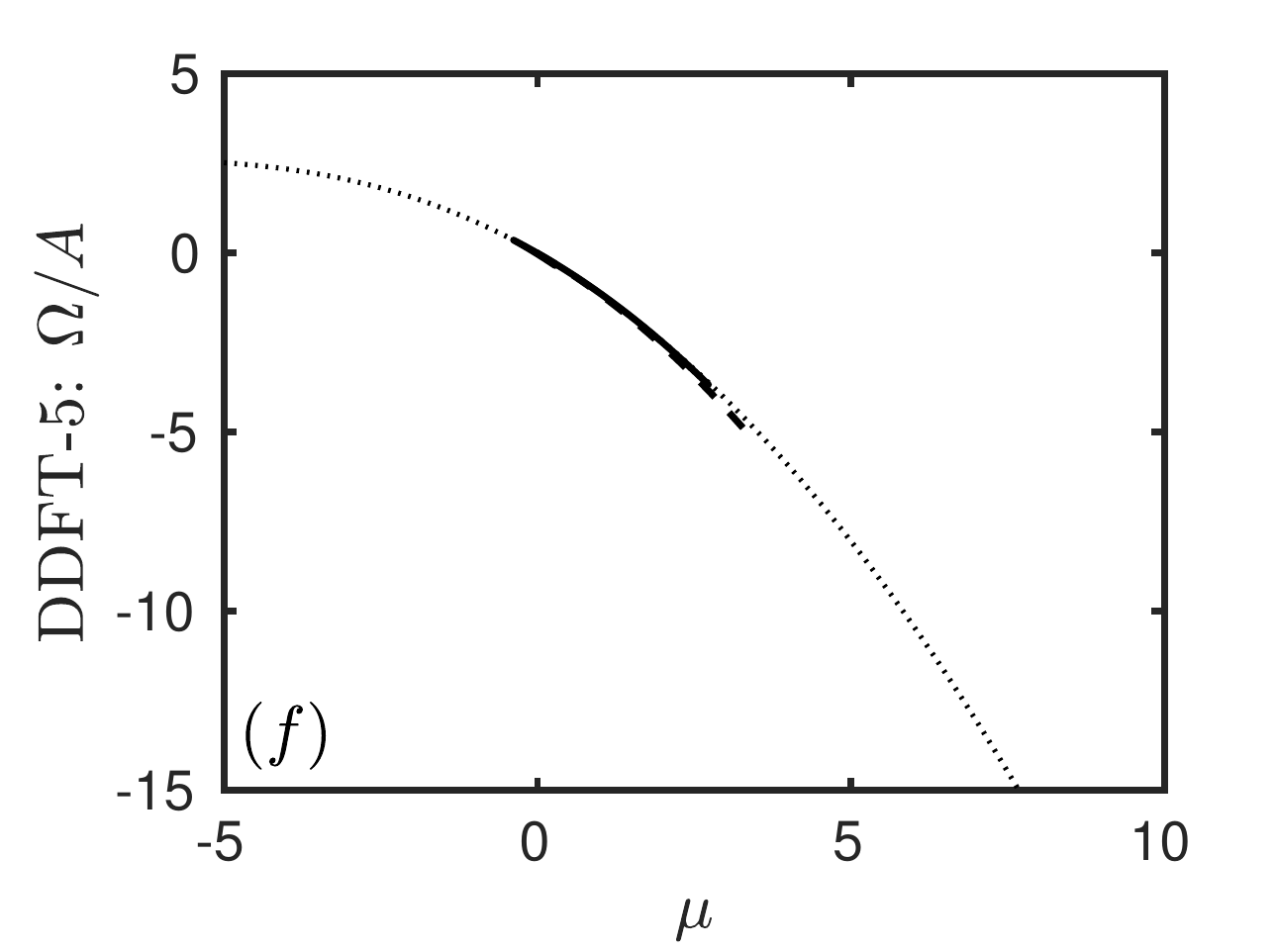}}
\hbox to \hsize{\includegraphics[width=5.5truecm]{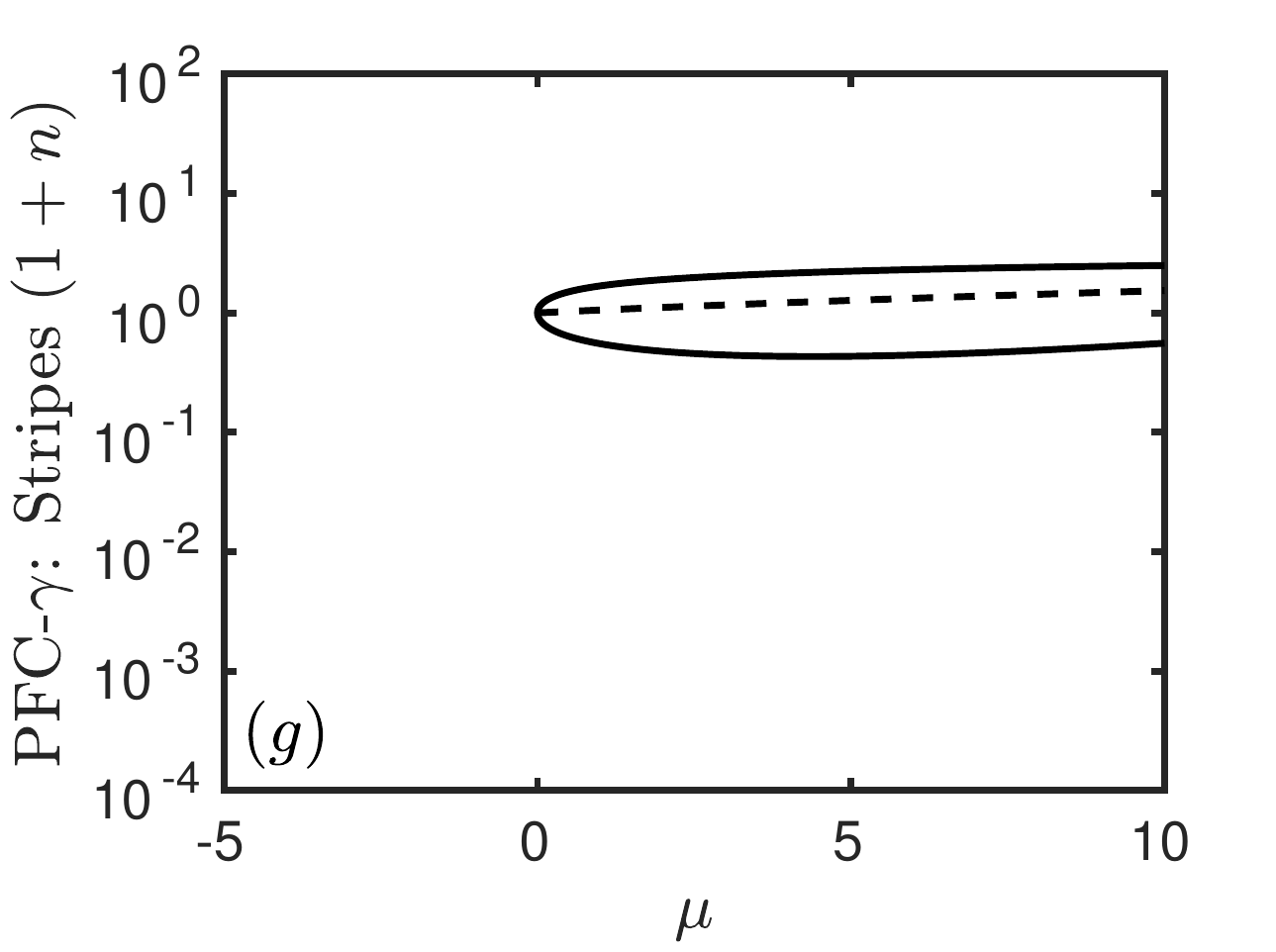}\hfill
                \includegraphics[width=5.5truecm]{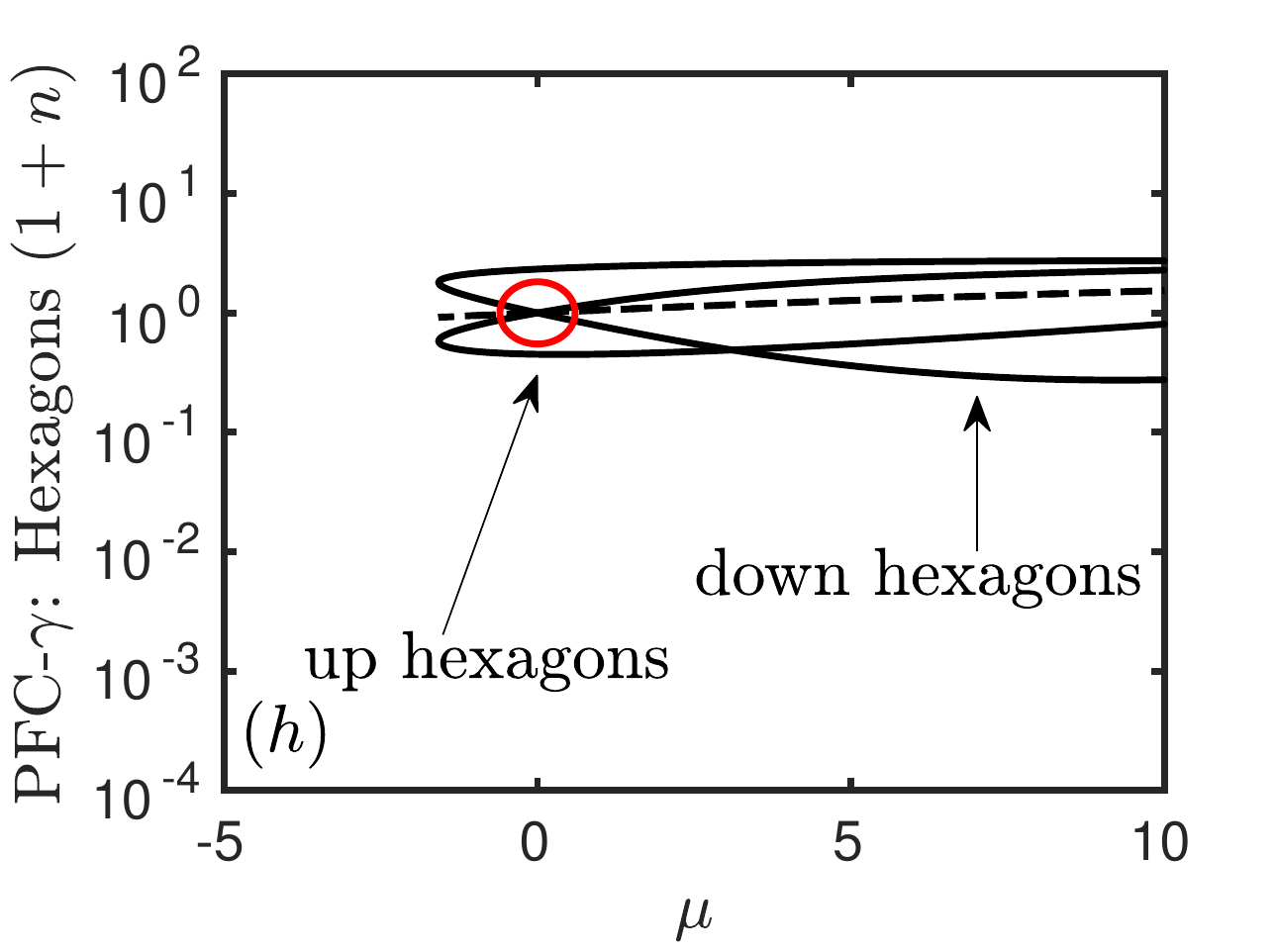}\hfill
                \includegraphics[width=5.5truecm]{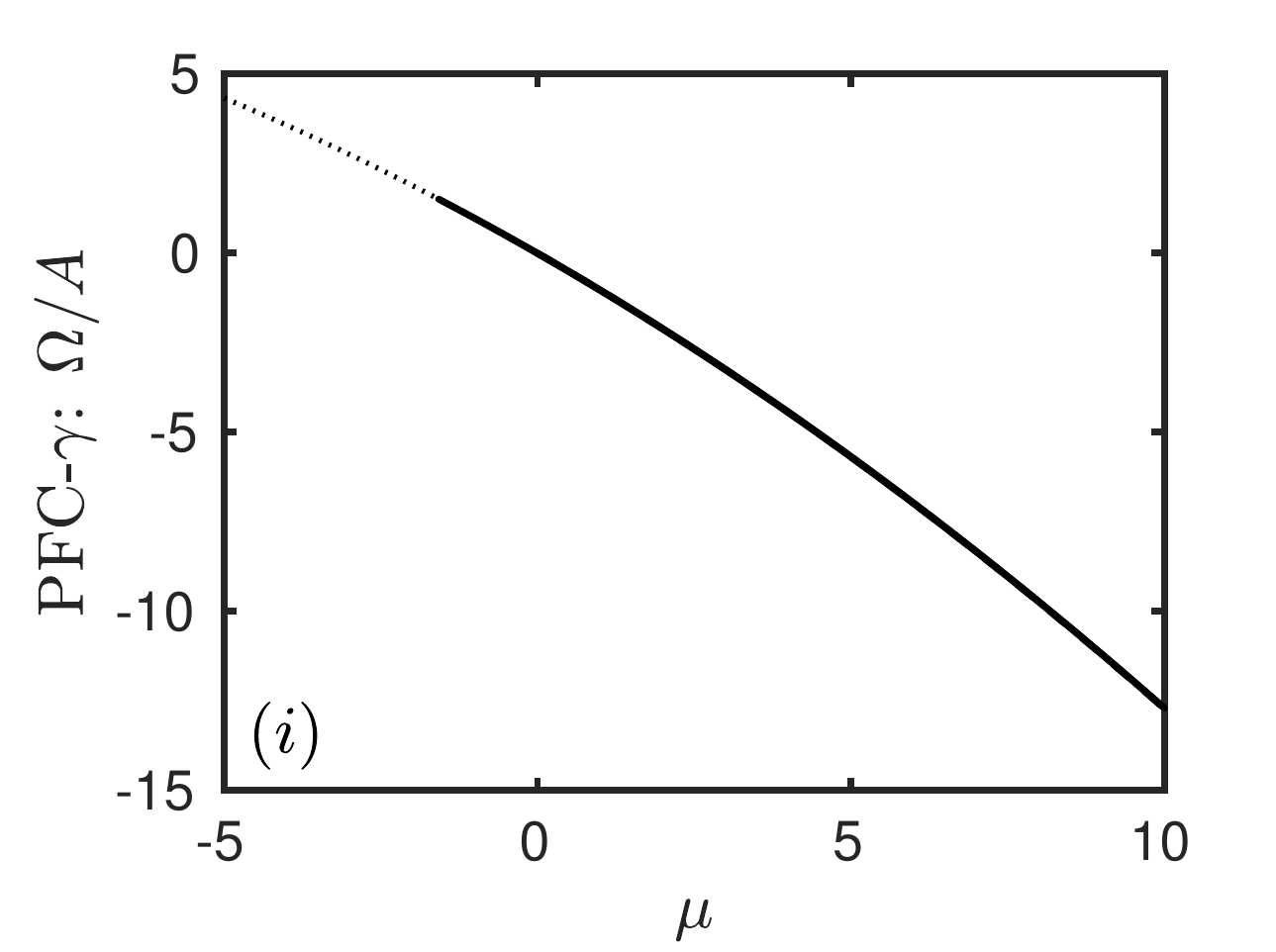}}
\hbox to \hsize{\includegraphics[width=5.5truecm]{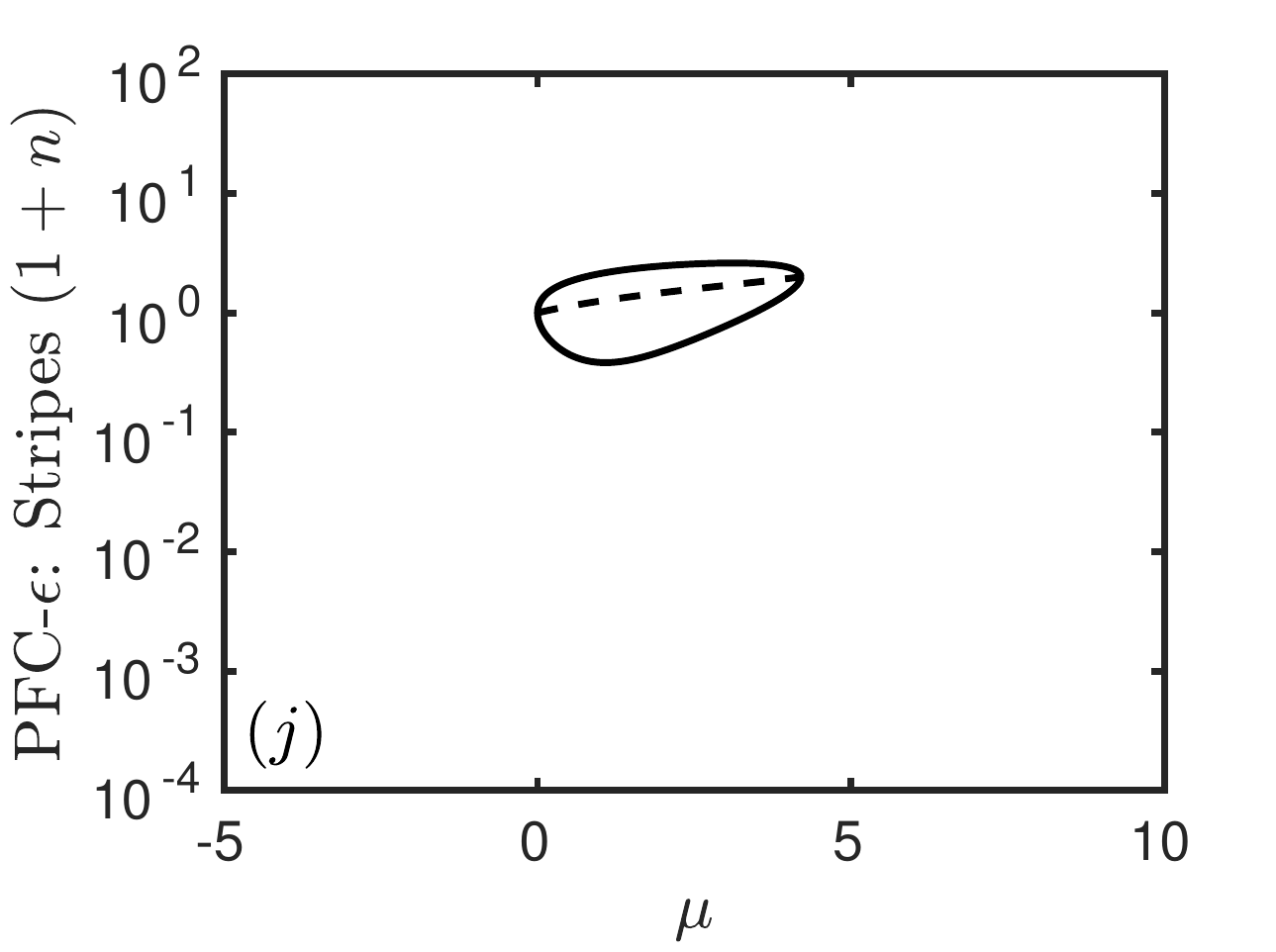}\hfill
                \includegraphics[width=5.5truecm]{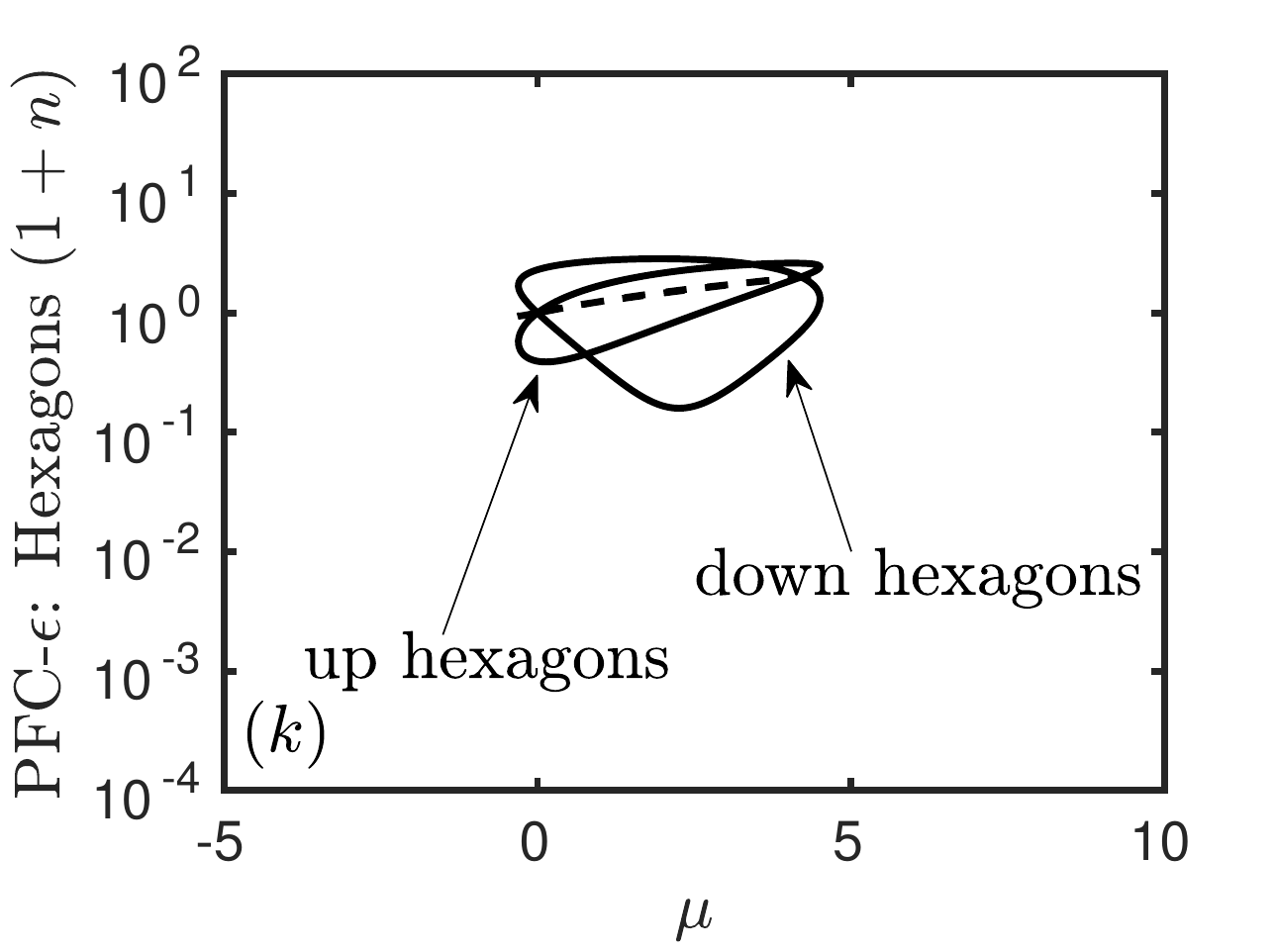}\hfill
                \includegraphics[width=5.5truecm]{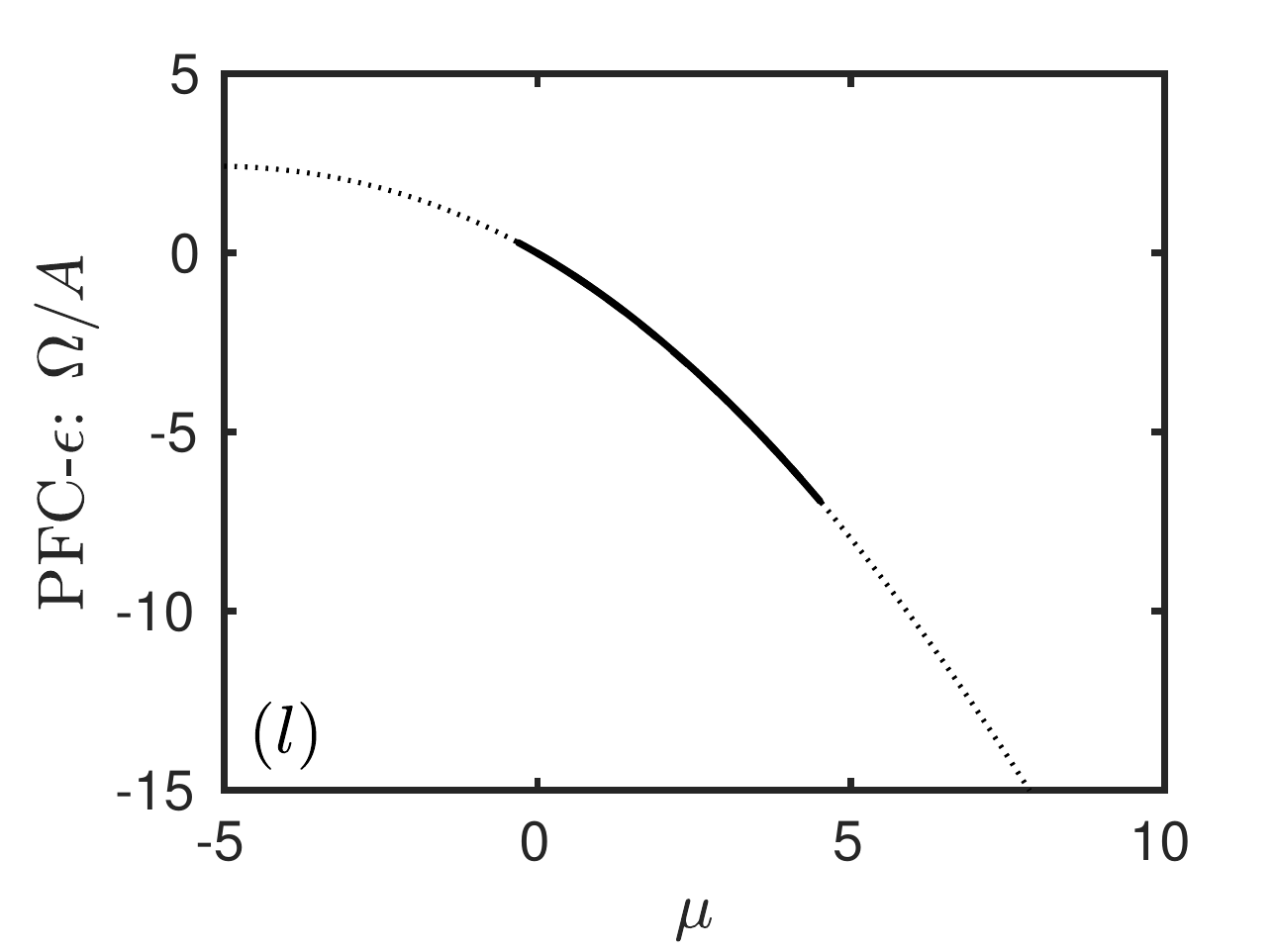}}
\caption{Data summarizing the nature of the stripe and the two hexagonal
solutions in the four cases:
(a--c)~DDFT-3, (d--f)~DDFT-5, (g--i)~\PFCgamma\ and (j--l)~\PFCepsilon. 
The graphs have the same axis limits in order to emphasise similarities
and differences. The left column (a,d,g,j) shows the stripe (lamellar) phase, with
maximum and minimum values of $1+n$ shown as solid lines, and average values of
$1+n$ shown as a dashed line. The center column (b,e,h,k) shows the same quantities for
the two hexagonal branches: the up hexagon branch bifurcates to negative
values of~$\mu$ before turning around in a fold (saddle-node) bifurcation and
continuing to larger~$\mu$. The right column (c,f,i,l -- see also Fig.~\ref{fig:DDFTPFCrelativeOmega}) 
shows $\Omega/A$ for these three
periodic solutions and also for the uniform liquid (dotted line).
In DDFT-3, for $\mu\geq-2.8$ the up hexagons (solid line) have the lowest value of
$\Omega/A$, and of the non-uniform phases the stripes (dashed line) are next, with the
down hexagons (solid line) have the highest value of $\Omega/A$. In the DDFT-5
and the PFC cases, the lines are so close that on this scale they appear overlaid.
{The (approximate) region of quantitative agreement between DDFT-3 and \PFCgamma\ is
circled in red in~(b) and~(h).}}
 \label{fig:DDFTPFCsummary}
\end{center}
 \end{figure*}

In Fig.~\ref{fig:DDFTPFCsummary} we display a series of plots showing the
maximum, minimum and average values of the density profiles $1+n$ for the
stripe and hexagonal structures as a function of~$\mu$. We also plot the specific grand
potential~$\Omega/A$ for the different structures. 
Recall that for a given $\mu$ the thermodynamic equilibrium phase
corresponds to the global minimum of~$\Omega/A$. The results for DDFT-3
are shown in Fig.~\ref{fig:DDFTPFCsummary}(a--c). 
The (a)~stripes originate in a supercritical
pitchfork bifurcation at $\mu=0$, and (b)~hexagons
originate in a transcritical bifurcation at the same value of~$\mu$. The density of
the up hexagons ranges from about $2\times10^{-5}$ up to about 50, for $\mu=10$.
All of these branches can be continued to larger values of~$\mu$.

DDFT-5, in Fig.~\ref{fig:DDFTPFCsummary}(d--f), 
initially behaves in the same way, but all three branches have
their minimum density heading to zero before $\mu$ gets to~10: this happens at
$\mu\approx3.37$ for (d)~stripes, and for $\mu\approx0.28$ and $\mu\approx2.73$ for
(e)~the up and down branches of hexagons, respectively. The numerical method cannot
continue the branches beyond these points. We argue in
Sec.~\ref{sec:4} that this is not an artefact of the numerical
method, rather it is a genuine feature of solutions
of Eq.~(\ref{eq:betadFdn5_rewritten}) that the density $1+n$ can go to zero. In this
limit, $\log(1+n)\rightarrow-\infty$, but this is balanced by a lack of smoothness
in $n(\bx)$: the fourth derivative in $\cLgrad n$ can go to~$+\infty$ and so
balance the singularity in $\log(1+n)$. Therefore, $\mu\approx0.28$ is the limit
of validity of the DDFT-5 model.

The two PFC examples are similar to each other, and it is easier to discuss \PFCepsilon,
in Fig.~\ref{fig:DDFTPFCsummary}(j--l), first. 
Here, (j)~stripes and (k)~hexagons bifurcate from the liquid at
$\mu=0$, but they rejoin the liquid at $\mu=4.20$ as explained in the 
discussion following Eqs.~(\ref{eq:DDFTdeltan}--\ref{eq:PFCdeltan}). The
maximum and minimum densities for the up and down hexagon cross between the two
bifurcations. The behaviour of \PFCgamma, in Fig.~\ref{fig:DDFTPFCsummary}(g--i), is similar, though
the second bifurcation is at $\mu=18.58$, off the scale of the figure.

\begin{figure*}
\begin{center}
\hbox to \hsize{\includegraphics[width=8.6truecm]{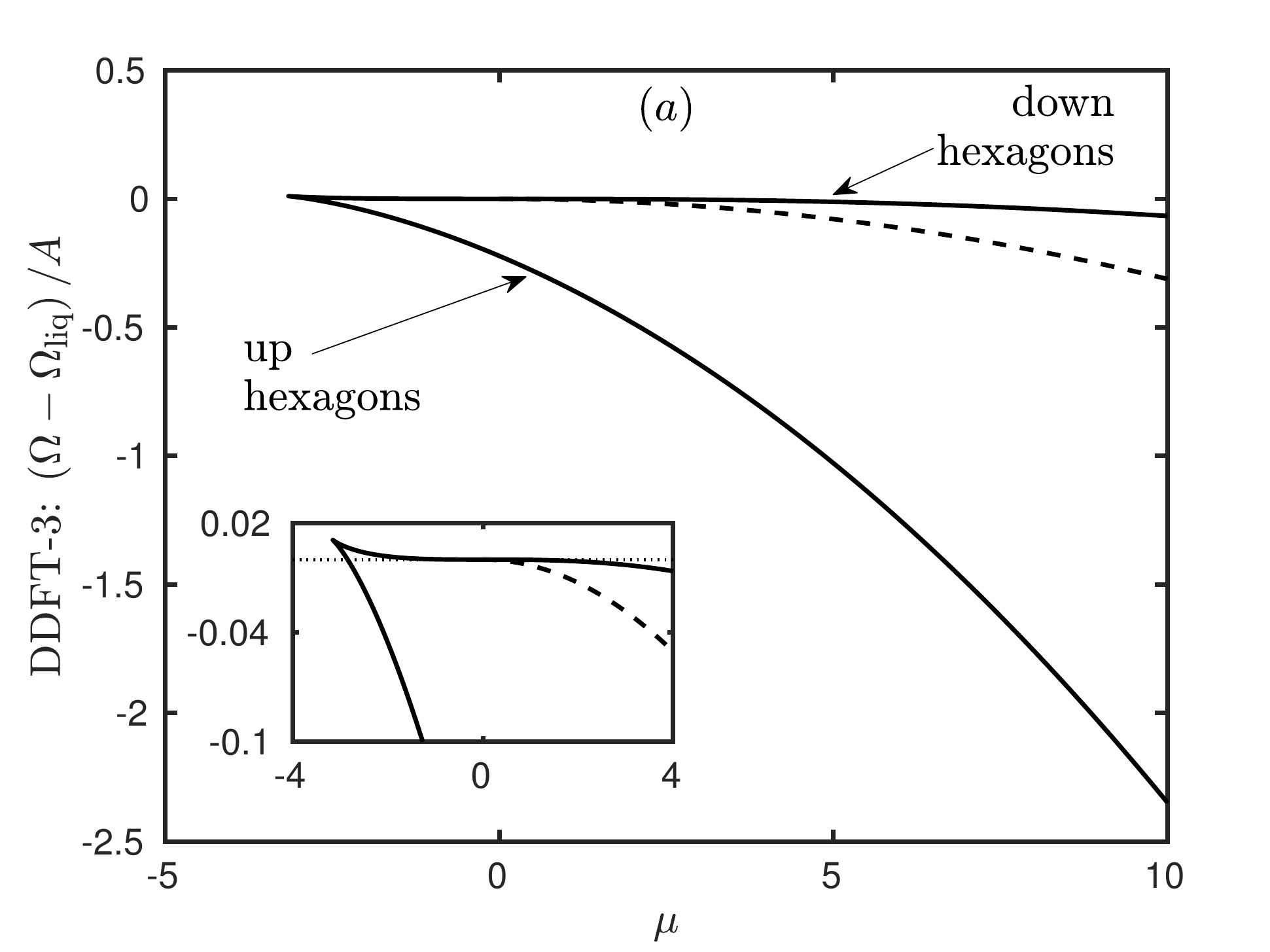}\hfill
                \includegraphics[width=8.6truecm]{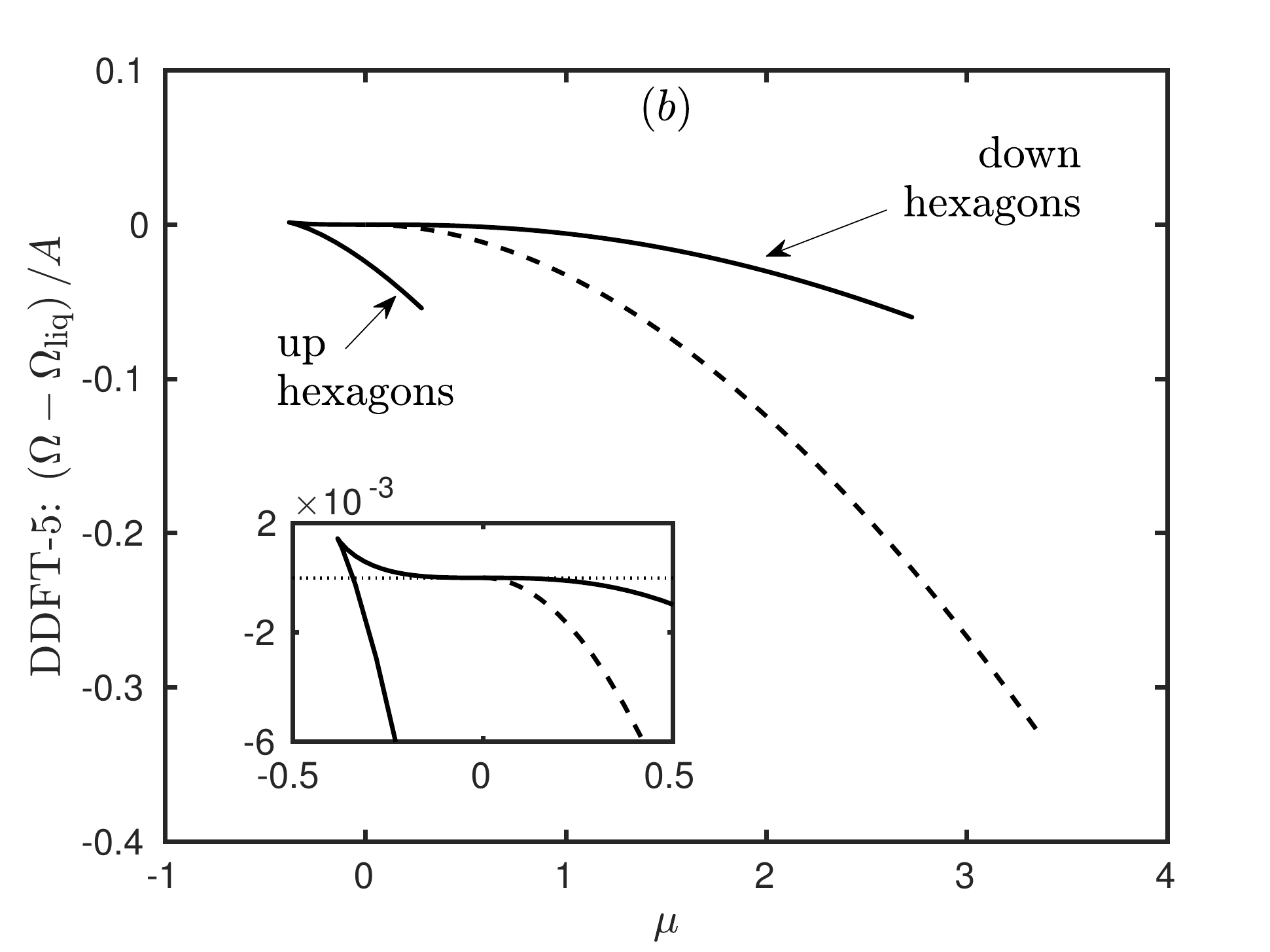}}
\hbox to \hsize{\includegraphics[width=8.6truecm]{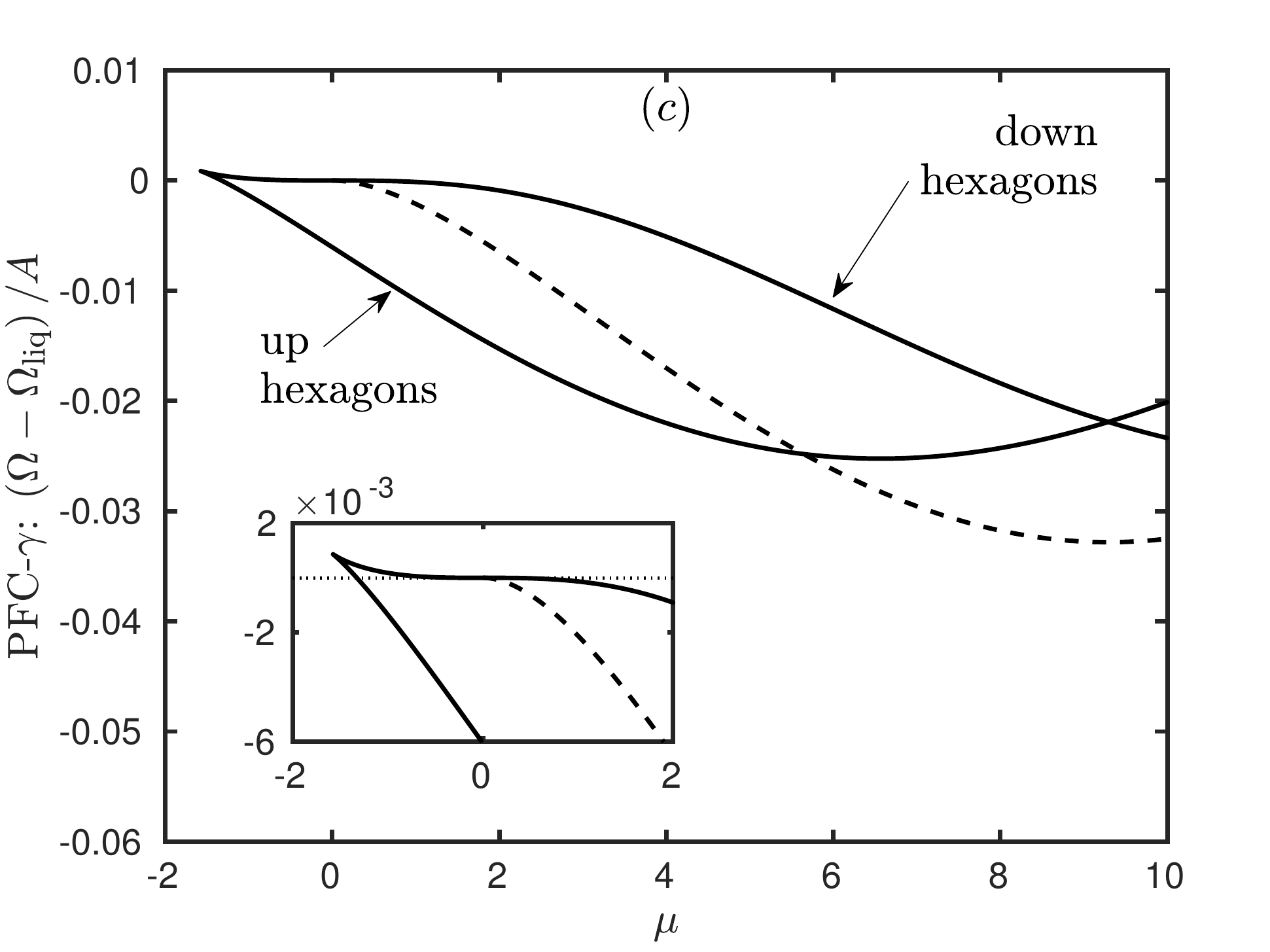}\hfill
                \includegraphics[width=8.6truecm]{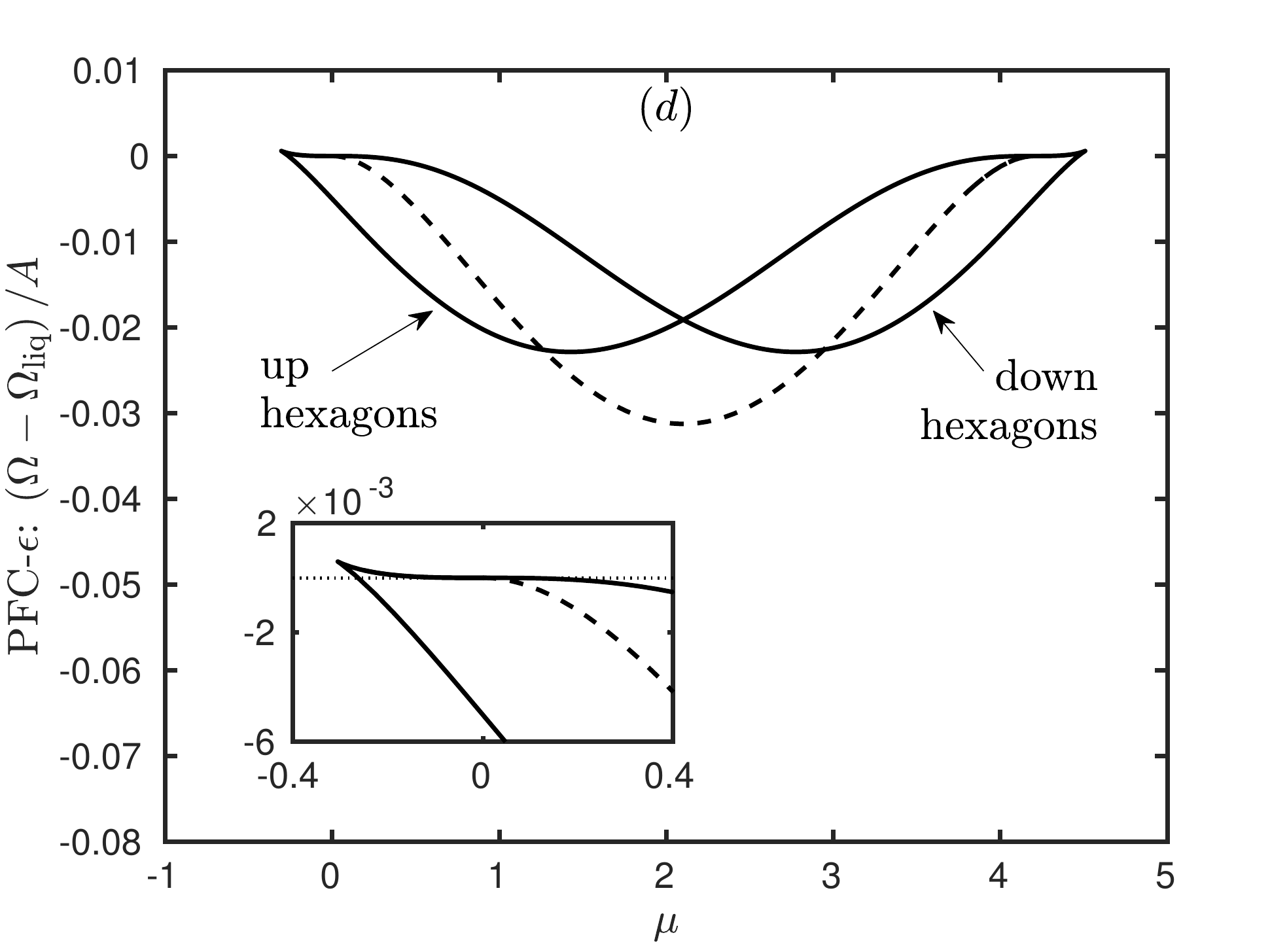}}
\caption{Specific $\Omega$ relative to the value for the liquid at the same 
value of~$\mu$, with hexagons shown as solid lines (the up hexagons
have the lowest $\Omega-\Omegaliq$ for $\mu$ close to zero), and stripes
shown as dashed lines. The insets show details of the region close to $\mu=0$, 
as well as (dotted line) $\Omega-\Omegaliq=0$. The panels are
(a)~DDFT-3, (b)~DDFT-5, (c)~\PFCgamma\ and (d)~\PFCepsilon. 
In the PFC examples, 
the state with lowest grand potential changes from up hexagons to stripes to down 
hexagons as $\mu$ increases. Note the factor of~$10$ difference between the
inset axis scales between DDFT-3 and \PFCepsilon.}
 \label{fig:DDFTPFCrelativeOmega}
\end{center}
 \end{figure*}

Figure~\ref{fig:DDFTPFCsummary}(c,f,i,l) shows that the curves
of the specific grand potential~$\Omega/A$ as functions of~$\mu$ are very close,
so in 
Fig.~\ref{fig:DDFTPFCrelativeOmega}, we plot instead
$(\Omega-\Omegaliq)/A$ versus
$\mu$, where $\Omegaliq$ is the specific grand potential
for the liquid at the same value of~$\mu$. In (a)~the DDFT-3 case,
the up hexagons clearly have the lowest grand potential for $\mu\geq-2.8$
with the uniform liquid being the global minimum for $\mu<-2.8$,
and at no point do stripes come anywhere near, as one should expect from
the particle simulation results~\cite{Prestipino2014}. 
For (b)~DDFT-5, the two hexagon branches stop before
the stripe branch when their minimum densities go to zero (the limit of validity), 
but otherwise the relative values for the hexagon and stripe grand potentials
is qualitatively similar to
DDFT-3. For (c,d)~the PFC examples, once again it is easiest to discuss \PFCepsilon\
first. In Fig.~\ref{fig:DDFTPFCrelativeOmega}(d), 
the hexagon and stripe branches bifurcate from the liquid at $\mu=0$ and
rejoin the liquid at $\mu=4.20$, with stripes having the lowest grand potential
for intermediate values of~$\mu$, and up or down hexagons being the lowest 
grand potential state  for smaller or larger values of~$\mu$. The behaviour of
(c)~\PFCgamma\ is similar, but stretched to larger values of~$\mu$ (off scale).

The insets in the four panels of Fig.~\ref{fig:DDFTPFCrelativeOmega} display
magnifications that show that
the behaviour near the spinodal point at $\mu=0$ is 
qualitatively similar in all four
cases: the up hexagons start with $\Omega>\Omegaliq$ for negative~$\mu$, but
the branch changes direction, forming a cusp, 
close to which is the thermodynamic
coexistence point (Maxwell point), 
where $\Omega=\Omegaliq$. The down hexagons start with
$\Omega<\Omegaliq$ for positive~$\mu$, and the stripes, also with
$\Omega<\Omegaliq$ for positive~$\mu$, have a value of the grand
potential intermediate
between the up and down hexagons. We note that the range of~$\mu$ over which 
this behaviour occurs is about a factor of ten smaller in the DDFT-5 and 
\PFCepsilon\ cases as compared to DDFT-3, also with a roughly ten-fold drop in the overall
range of values of $(\Omega-\Omegaliq)/A$.

\begin{figure*}
\begin{center}
\hbox to \hsize{{\includegraphics[width=8.6truecm]{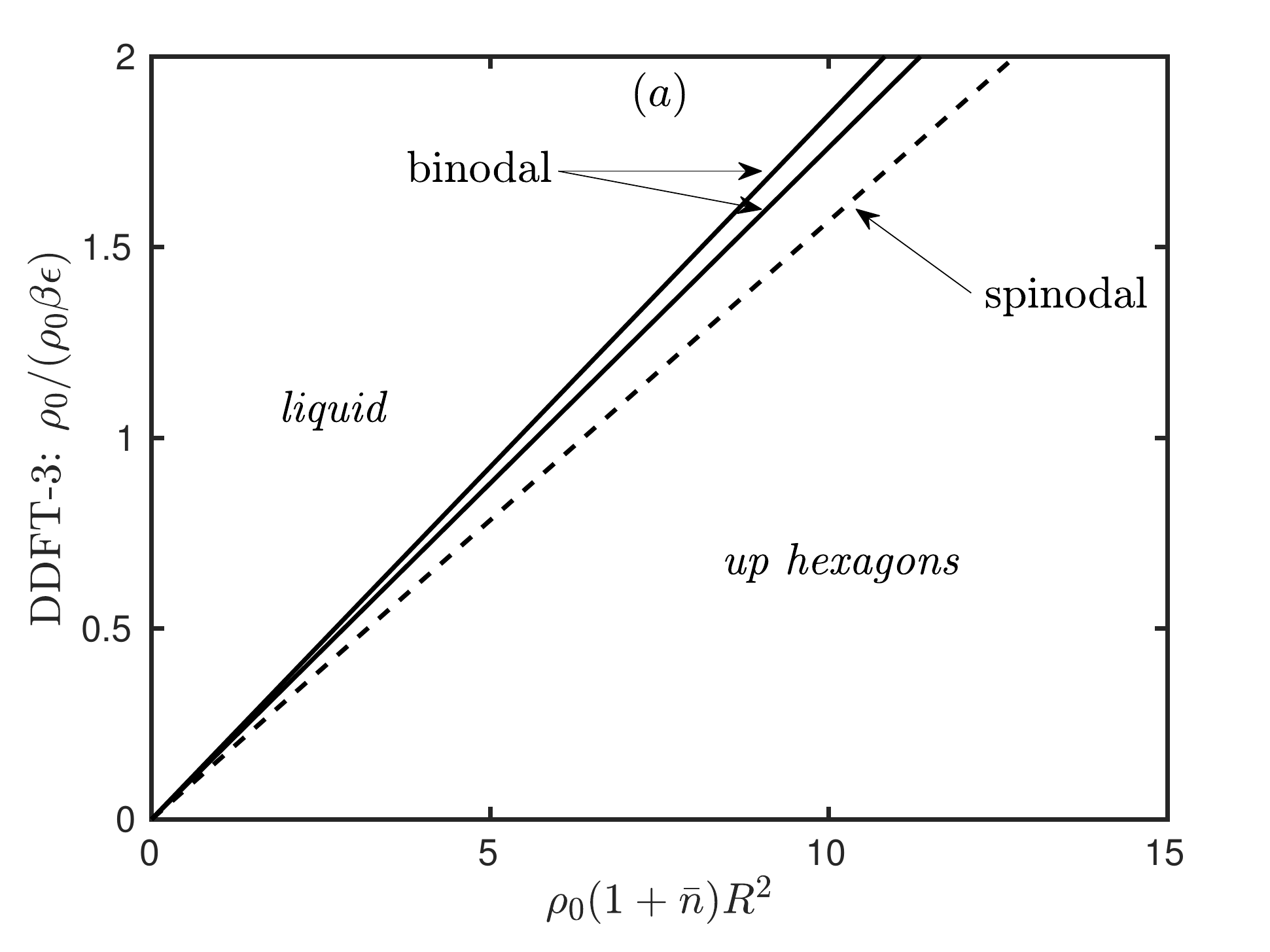}}\hfill
                {\includegraphics[width=8.6truecm]{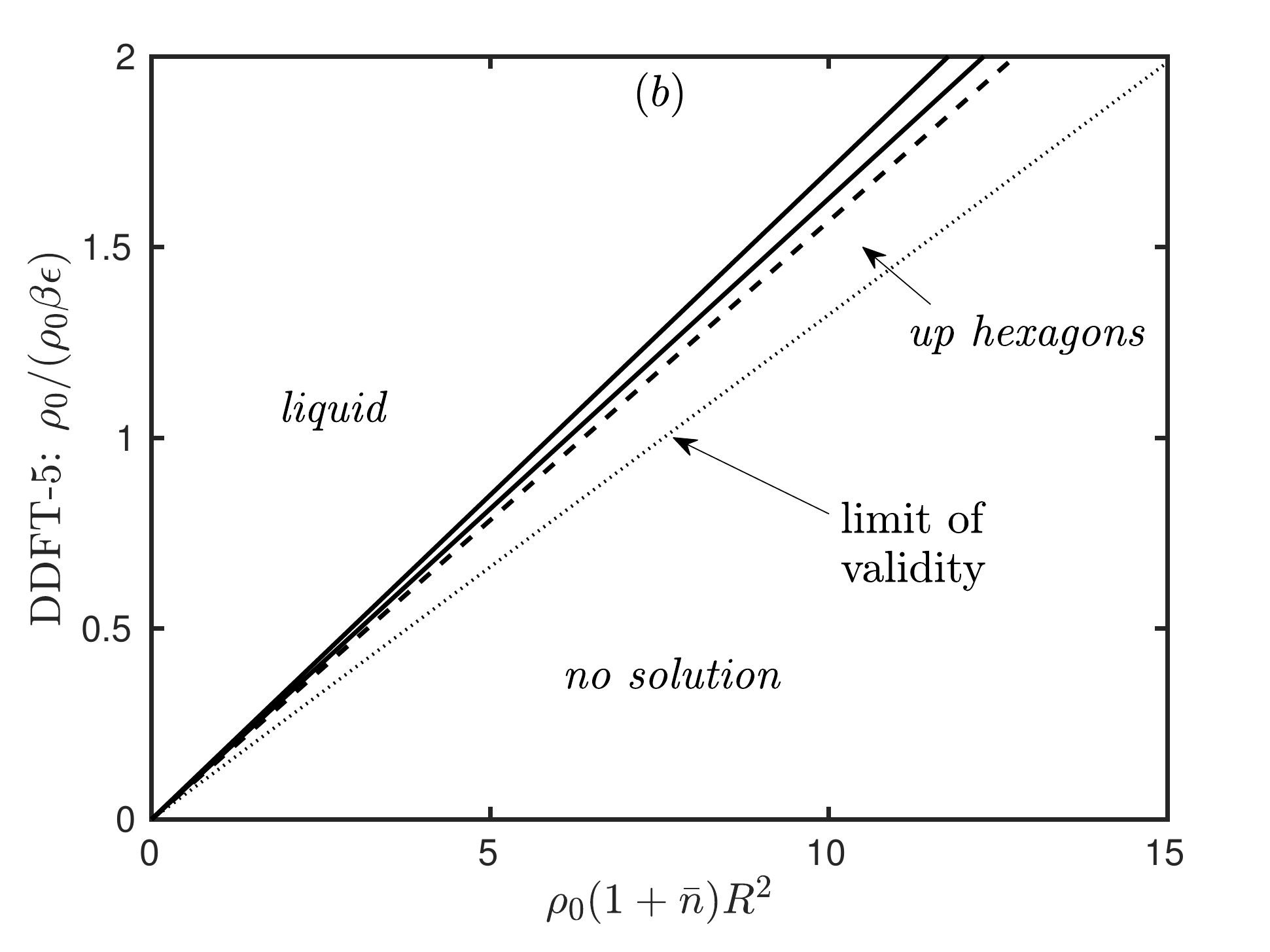}}}
\hbox to \hsize{{\includegraphics[width=8.6truecm]{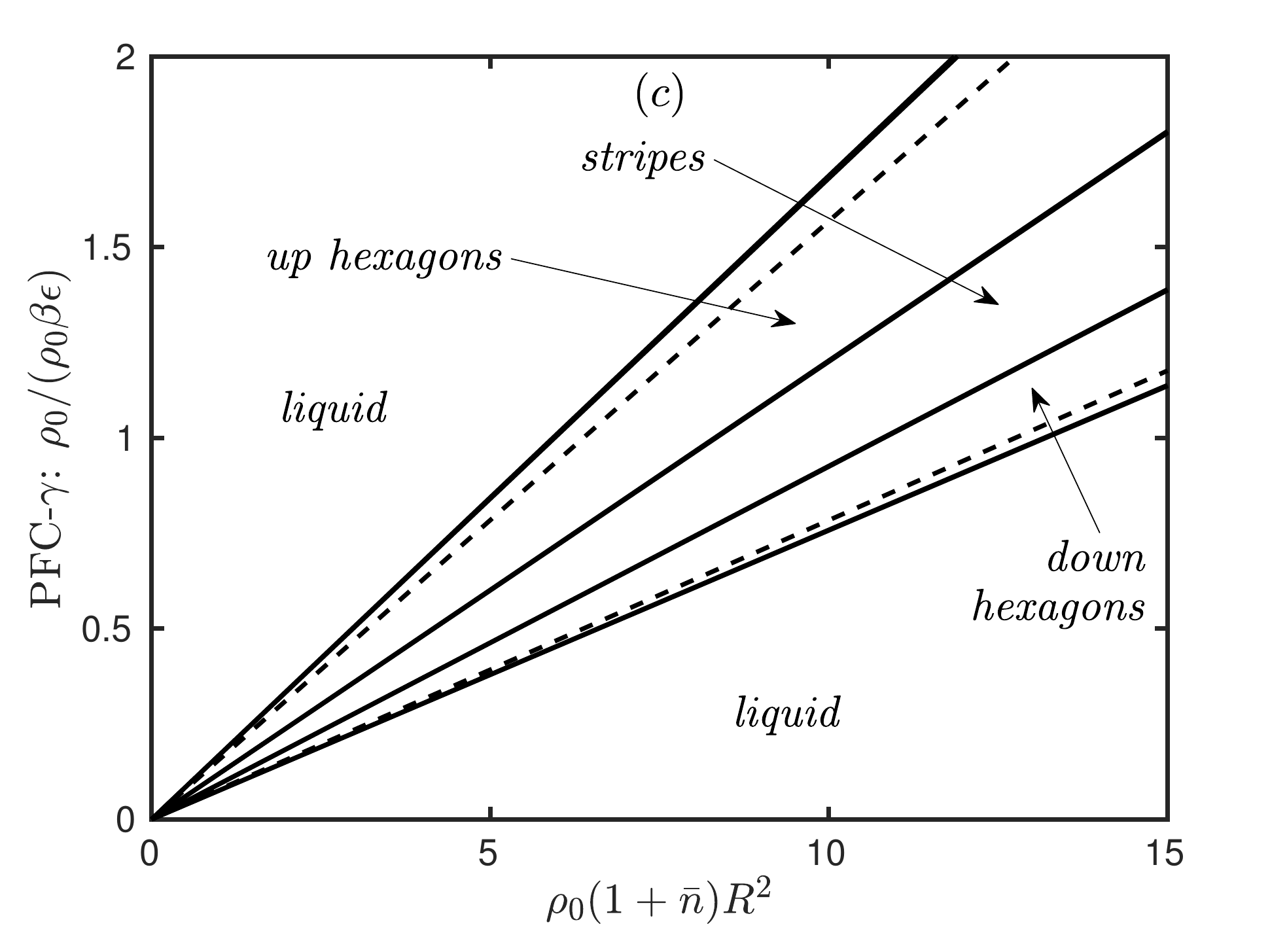}}\hfill
                {\includegraphics[width=8.6truecm]{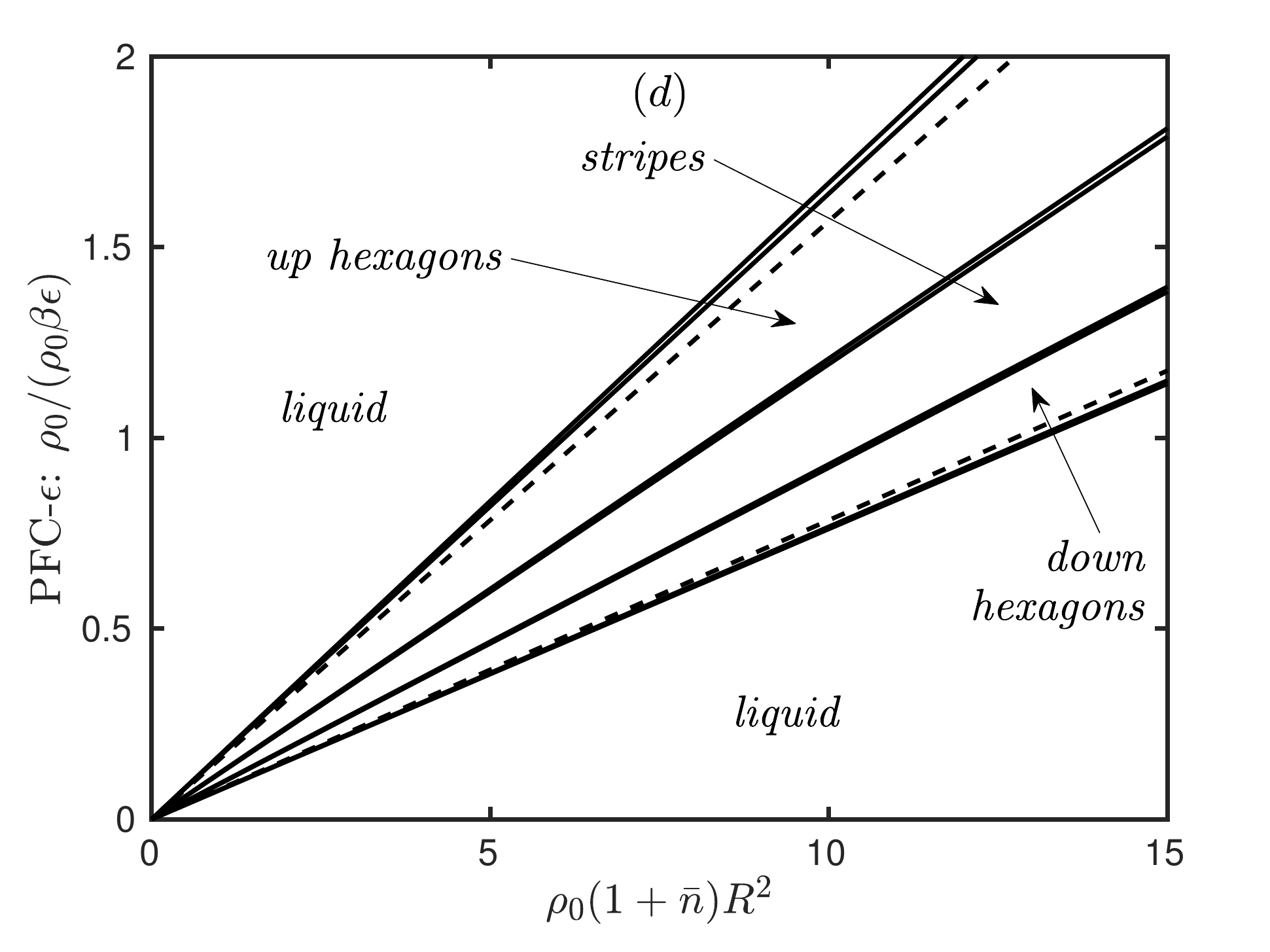}}}
\caption{The phase diagram for the GEM-4 model as predicted by (a)~DDFT-3, (b)~DDFT-5,
(c)~\PFCgamma\ and (d)~\PFCepsilon, plotted in the
average density $\bar{\rho}R^2=\rho_0(1+\bar{n})R^2$ versus temperature
$k_BT/\epsilon=\rho_0/(\rho_0\beta\epsilon)$ plane. In these phase diagrams, 
which show only the states with lowest~$\Omega$, 
lines are
labelled with a roman font and regions are labelled in {\it italics}. For
$k_BT/\epsilon\gtrsim0.1$ the DDFT-3 phase diagram is in good agreement with the true
phase diagram \cite{Prestipino2014}. However, the two PFC phase diagrams wrongly predict
at high densities the occurrence of a stripe phase, down hexagons and a second liquid
phase. In the DDFT-5 phase diagram there is no solution in the bottom right half of the
phase diagram, below the limit of validity.}
 \label{fig:phase_diagrams}
\end{center}
 \end{figure*}
 
The observation that the bulk phase behaviour of the system
depends only on $\mu$ and the value of $\rho_0\beta\epsilon$ if the pair
potential can be written as $u(x_{12})=\epsilon\psi(x_{12})$ -- see the discussion
around Eq.~\eqref{eq:convol_form} -- is true for the GEM-4 system.
As a consequence, having calculated the coexisting densities
for a particular value of $\rho_0\beta\epsilon$, 
the linear stability threshold, these results can be scaled to give the
phase diagram in the full average density $1+\bar{n}$ versus dimensionless
temperature $k_BT/\epsilon$ plane, which is one of the usual ways the GEM-4
phase diagram is displayed \cite{Mladek2006, Mladek2007, Prestipino2014, Archer2014, Archer2016}.
 %The scaling is done by dividing the coexisting density values by the value of $\rho_0$
 %at which they were calculated and then multiplying by the value of
 %$\rho_0$ corresponding to the linear stability threshold density at the desired
 %value of $\beta\epsilon$. 
 The phase diagrams obtained from doing this are
in Fig.~\ref{fig:phase_diagrams}. In Fig.~\ref{fig:phase_diagrams}(a)
we display the phase diagram
obtained from DDFT-3, which is identical (to within the resolution of the calculations)
to that previously calculated in
Refs.~\cite{Archer2014, Archer2016}. For example,
when $\beta\epsilon=1$, the average densities $\rho_0(1+\bar{n})R^2$ of the
coexisting liquid and the crystal are $5.41$ and $5.68$, respectively. 
As a result
of the scaling behaviour, the coexisting densities (binodals) are two straight
lines going from the origin and passing through these two points.

In Fig.~\ref{fig:phase_diagrams}(b) we display the phase
diagram obtained for the DDFT-5. The binodals are a little closer to the linear
stability threshold line than for DDFT-3, but other than that, it looks
similar overall. Note however that the up hexagon branch cannot be continued
beyond~$\mu\approx0.28$ (where the minimum density goes to zero): this line is
indicated as the `limit of validity'. Beyond this line, in the bottom right
region of the phase diagram, there is no up hexagon solution to the equations,
for the reasons discussed in Sec.~\ref{sec:4}.

In Fig.~\ref{fig:phase_diagrams}(c) we display the phase
diagram for \PFCgamma\ and in~(d) for \PFCepsilon. The binodals
almost overlie each other, so the predicted difference between the
average densities of the liquid and the crystal at coexistence
are much smaller than that predicted
by DDFT-3 and DDFT-5. Furthermore, on moving to higher average
densities or to lower temperatures $k_BT/\epsilon$ one encounters the stripe
phase, followed by the down hexagon phase and then finally the uniform liquid
becomes stable again. The prediction of the occurrence of these later phases is
of course wrong, signifying a breakdown in the accuracy of the PFC theory at
even the qualitative level.

Before finishing this section, we note that it is possible to extend the 
gradient expansion in~(\ref{eq:cLgradGEM4}) by including higher powers of the 
Laplacian. For example, Ref.~\cite{Jaatinen2009} proposed an eighth-order 
fitting (EOF), which in our notation is
 \begin{equation}
 \cLgradJ n(\bx) = - \gamma(1+\nabla^2)^2 n(\bx)
                   - E_B (1+\nabla^2)^4 n(\bx), 
 \label{eq:cLgradJaatinen}
 \end{equation}
where $\gamma$ fits the curvature of the dispersion relation as before, and
$E_B$ allows the eigenvalue~$\sigma(0)$ of~$\cL$ to be matched as well, i.e.,
allows the model to match correctly the isothermal compressibility $\chi_T$. An
example of the dispersion relation for this operator is shown as a dotted line
in Fig.~\ref{fig:dispersionGEM4}.  This EOF version of the theory, with~$\cLgradJ$
in~(\ref{eq:cLgradJaatinen}), improves over the standard version,
with~(\ref{eq:cLgradGEM4}), since $\sigma(0)$~for $\cL$ and $\cLgradJ$ are the
same. Therefore, the liquid properties of DDFT-5 match those of DDFT-3, and the liquid
properties of \PFCgamma\ match those of \PFCepsilon, once $\cLgrad$ is replaced by 
$\cLgradJ$.

However, the drawbacks of the gradient expansion are still present. 
With~$\cLgradJ$, the values of~$\mu$ at which the DDFT-5 stripe and hexagon 
densities go to zero are larger, but 
this undesirable feature is only deferred, not eliminated. The reason is that 
the singularity in the logarithm is now balanced against an eighth-order 
derivative. Note too that introducing even higher derivatives does not cure
this problem, it just pushes the singularity to higher order. In 
addition, with~$\cLgradJ$ the second liquid spinodal in the \PFCepsilon\
model is still present, it is just pushed to higher values of~$\mu$ (similar
to the value for \PFCgamma) and since this second spinodal is
present, there is still 
a range of values of~$\mu$ for which stripes have the lowest specific~$\Omega$.

\section{Effect of the approximations} \label{sec:4}

The qualitative change in going from a DDFT to a PFC model (dropping the 
$\nabla\cdot\left[n\nabla\cL{n}\right]$ term, assuming constant mobility and expanding 
the logarithm) is apparent in the phase diagrams shown in Fig.~\ref{fig:phase_diagrams},
comparing (a,b) to~(c,d). Here we discuss in detail additional
effects of the approximations.

\subsection{Expanding the logarithm}

The effect of expanding the logarithm as in Eq.~(\ref{eq:expandlogarithm}) is
very significant. In Sec.~\ref{sec:3} we demonstrated that this expansion
leads to the liquid having a second spinodal point,
illustrated in the
phase diagrams in Fig.~\ref{fig:phase_diagrams}(c,d), with the crystal re-melting as
the density is increased. The reason for this is that Eq.~(\ref{eq:PFCdeltan})
can be solved with $n=0$ and $n=1$, while Eq.~(\ref{eq:DDFTdeltan}) is only
solved by $n=0$, with $\sigma(k)=0$ in both cases.

Intimately connected to the existence of
this second spinodal is the transition from stable up
hexagons at the $\mu=0$ spinodal to stable down hexagons at the higher
density spinodal. These connections to the spinodals mean that the free energy
of the up hexagons increases again compared to the liquid state free energy
as the chemical potential is increased, in
order to reconnect to the liquid state at the upper spinodal.
An intermediate region of stable stripes is not inevitable, but is
evident in both PFC examples in
Fig.~\ref{fig:DDFTPFCrelativeOmega}(c,d).

Taking more terms in the expansion in Eq.~(\ref{eq:expandlogarithm}) does not 
help. The highest power should be even (otherwise the free energy is not 
bounded below), and the improved versions of Eq.~(\ref{eq:PFCdeltan}), which 
involves the second derivative of Eq.~(\ref{eq:expandlogarithm}) with respect 
to~$n$, also have $n=0$ and $n=1$ as (the only) real roots, regardless of how
many terms are kept in the expansion of the logarithm. {The exception is if
only terms up to~$n^2$ are kept in~(\ref{eq:expandlogarithm}); in this case, 
$c^{(3)}$ and $c^{(4)}$ (if they are non-zero) provide the stabilizing nonlinearities
and may also lead to a spurious spinodal.}

\begin{figure*}
\begin{center}
\hbox to \hsize{{\includegraphics[width=8.6truecm]{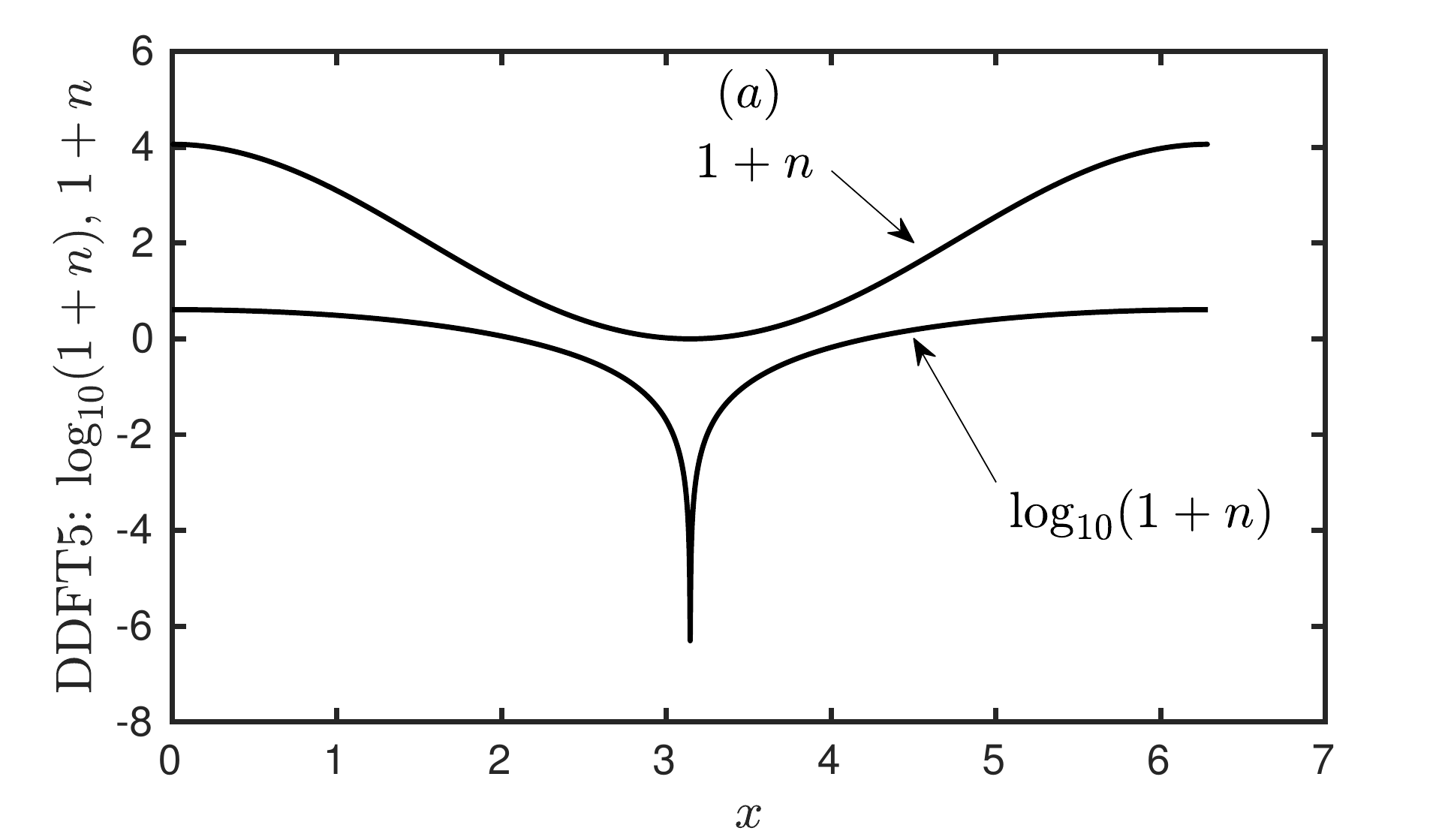}}\hfill
                {\includegraphics[width=8.6truecm]{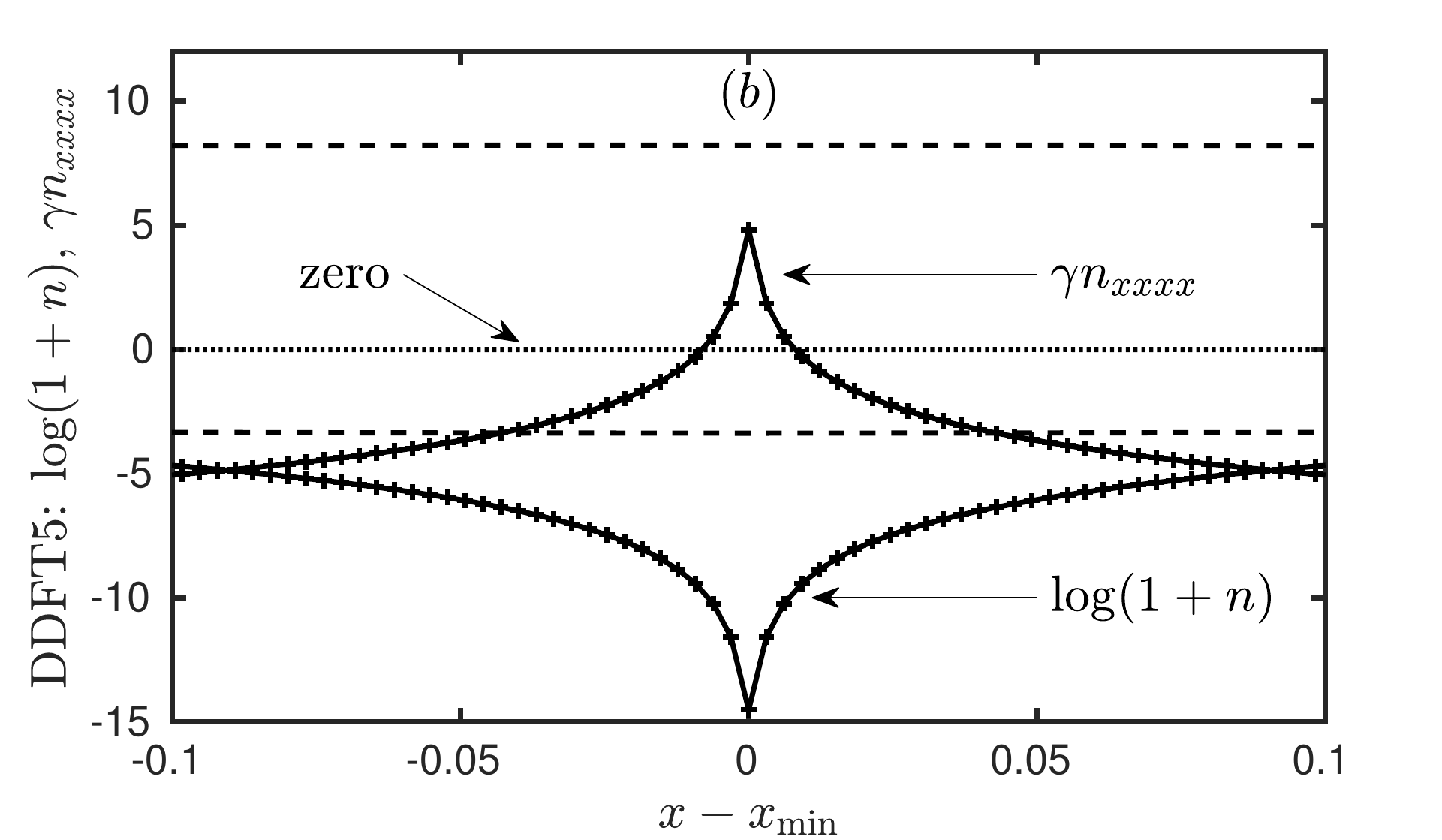}}}
\hbox to \hsize{{\includegraphics[width=8.6truecm]{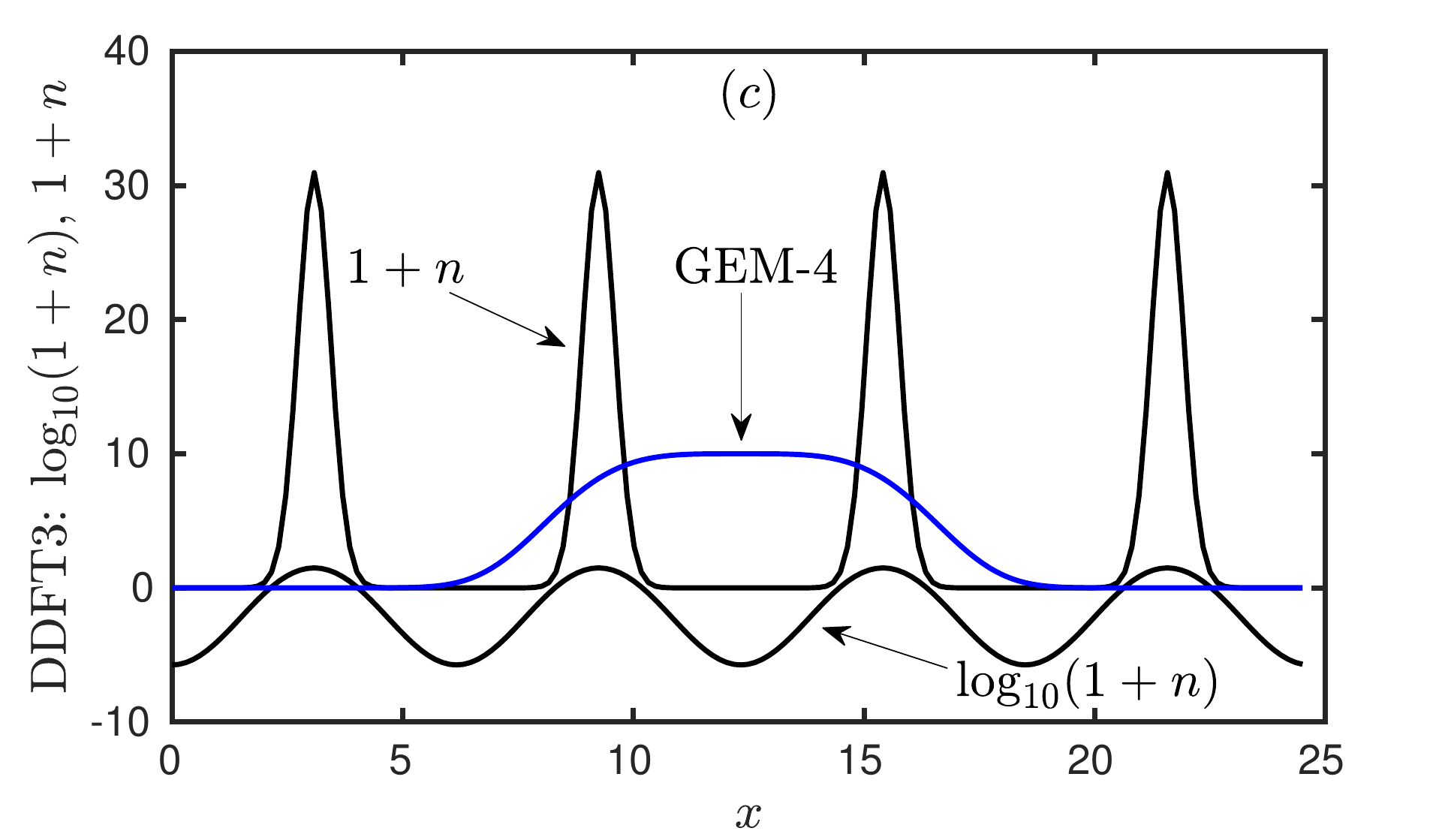}}\hfill
                {\includegraphics[width=8.6truecm]{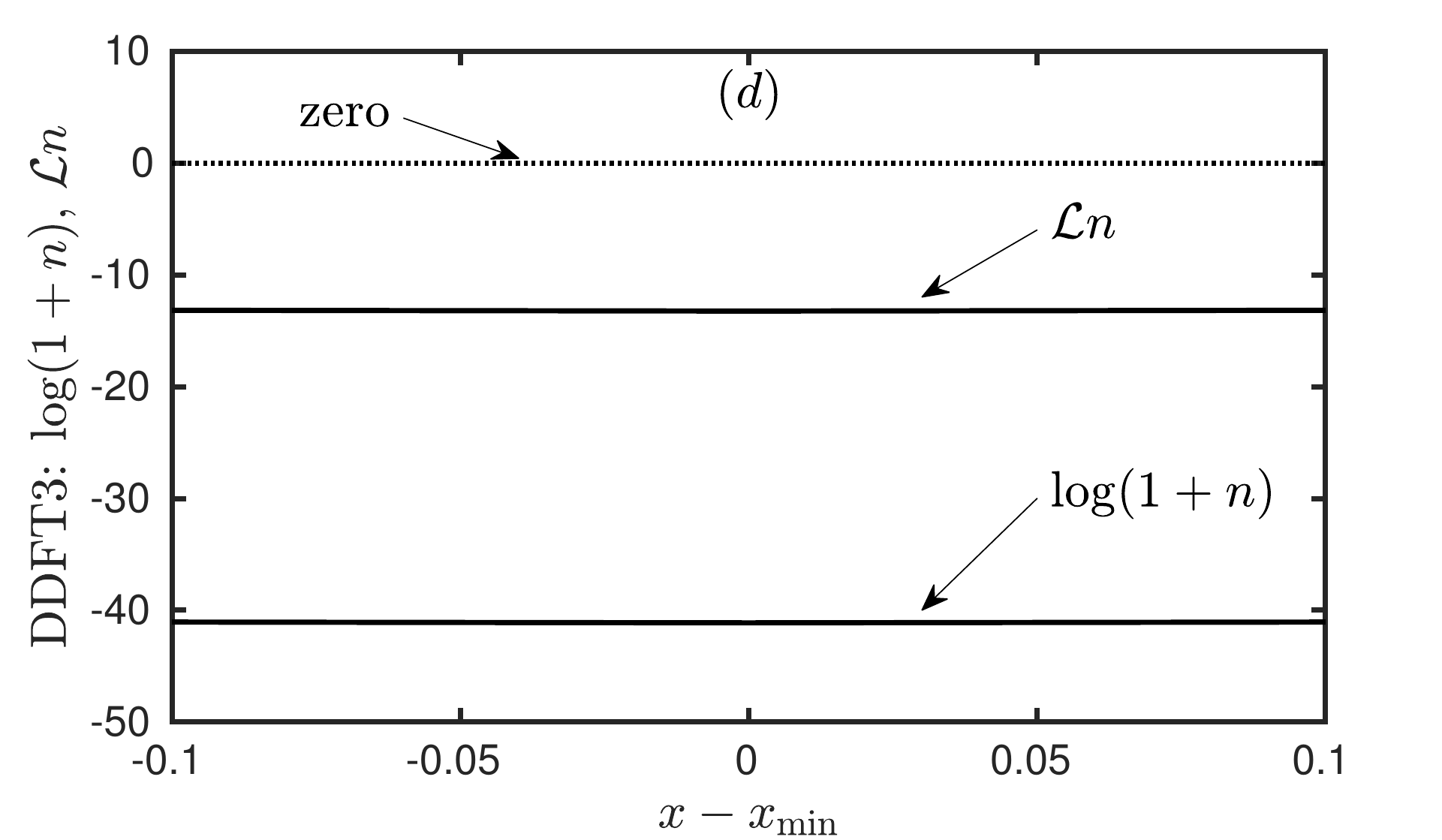}}}
\caption{(a)~DDFT-5 stripe density profile $1+n$, its logarithm and other terms
in~(\ref{eq:betadFdn5_rewritten}), for $\mu=3.3688$, with
$1+n(x_{\text{min}})\approx5\times10^{-7}$. The resolution was $N_x=2048$
grid points in a
domain with one wavelength. (b)~Detail around the
location of the density minimum at~$x_{\text{min}}$, showing various terms in
(\ref{eq:betadFdn5_rewrittenagain}): $\log(1+n)$ and $\gamma{n_{xxxx}}$ as
solid lines, with $+$~symbols showing the locations of the grid points, and
$(\gamma-1)n$ and $\gamma n_{xx}$ as the lower and upper dashed lines. The
total of all terms in~(\ref{eq:betadFdn5_rewrittenagain}) is shown as a dotted
line at zero, showing that the equation is satisfied at every grid point. The
balance between $\log(1+n)$ and $\gamma{n_{xxxx}}$ is clear. (c,d)~Similar data for
1D DDFT-3 stripes, with $\mu=28.9$ in a domain
containing four wavelengths, with other parameters as in
Table~\ref{tabApp:GEM4parameters}. A (scaled) plot of the GEM-4 potential
is shown in
blue. The density is sharply peaked, and has a minimum value
$\approx1.8\times10^{-6}$. In~(d), close to one of the density
minima, the $\log(1+n)$ and the $\cL{n}$ terms (solid lines) are well behaved,
and the DDFT-3 equation~(\ref{eq:betadFdn3_rewritten}) as a whole (dotted line)
is satisfied. Note that (a,c) use $\log_{10}(1+n)$ while (b,d) use the natural logarithm.}
 \label{figApp:DDFT5singularity}
\end{center}
 \end{figure*}

\subsection{Gradient expansion of~$\cL$}

As discussed in Sec.~\ref{sec:3} above, the results in
Fig.~\ref{fig:DDFTPFCsummary}(d,e) suggest that for certain values
of~$\mu$, the DDFT-5 equation~(\ref{eq:betadFdn5_rewritten}) has solutions
for the density $1+n$ which go to zero at certain places.
In contrast, the density
in the PFC models appears to stay away from zero
(although there is no reason
for it to do so and there would be no singularity if it did),
and in DDFT-3, the density minimum gets smaller and
smaller as $\mu$ increases, but remains positive,
without the sharp cutoff seen in DDFT-5. In
this section we argue that the density reaching zero is not an artefact of
numerical difficulties, rather it is a feature of the DDFT-5
equation~(\ref{eq:betadFdn5_rewritten}). 
{Here, we focus on singularities in the solution, not on stability.
Our discussion in this section is mainly framed in 
terms of the stripe solution, which is unstable, but the stable up-hexagon
branch has similar issues, as is illustrated in Fig.~\ref{fig:DDFTPFCsummary}(e).}

Figure~\ref{figApp:DDFT5singularity}(a) shows that even close to
the end of the branch of DDFT-5 stripes, the density profile
$1+n$ remains smooth, 
but since its minimum at $x=x_{\rm min}$ is very close to zero
($1+n(x_{\rm min})\approx5\times10^{-7}$), therefore
the logarithm $\log(1+n)$ is sharply spiked towards large negative
values at $x_{\rm min}$. Writing out the terms
in~(\ref{eq:betadFdn5_rewritten}) for a density profile only varying in the
$x$-direction, we have:
 \begin{equation}
 \log\left(1+n\right)
    + (\gamma-1) n + \gamma n_{xx} + \gamma n_{xxxx} - \mu = 0,
 \label{eq:betadFdn5_rewrittenagain}
 \end{equation}
which suggests that the only way to balance a large negative contribution from
$\log(1+n)$ is to have a large positive~$\gamma{n_{xxxx}}$. 
Figure~\ref{figApp:DDFT5singularity}(b) shows that these two terms (solid lines
with markers at the grid points) do indeed balance each other. The figure also
shows that the other terms in (\ref{eq:betadFdn5_rewrittenagain}) are well
behaved and that the equation is satisfied at each grid point. Therefore this
singularity is not a numerical artefact, but rather a genuine feature of DDFT-5
stripes: the minimum density goes to zero at a certain finite value of~$\mu$.
In Fig.~\ref{figApp:DDFT5singularity}(c,d) we show for comparison
results from DDFT-3: the stripes have sharply peaked
density maxima and density minima just as small as in DDFT-5, but all terms in
Eq.~(\ref{eq:betadFdn3_rewritten}) (Fig.~\ref{figApp:DDFT5singularity}d) are well behaved.

As~$\mu$ is further increased,
the minimum of the density in DDFT-5 gets closer to zero,
so the logarithm of the density goes further towards~$-\infty$ and correspondingly
$\gamma{n_{xxxx}}$ goes towards~$+\infty$.
Fig.~\ref{fig:DDFTPFCsummary}(d) shows that the density
minimum gets
to zero at a finite value of~$\mu$. We have not been able to develop a
consistent asymptotic approximation for this limit in the DDFT-5 equation.
However, to illustrate that apparently smooth solutions with logarithmic
singularities in their fourth derivatives can easily be found, consider for
example taking $\gamma=1$ in~(\ref{eq:betadFdn5_rewrittenagain}) and
taking a density profile that has a quadratic minimum at $x=x_{\rm min}=0$:
 \begin{equation}
 1 + n(x) = Ax^2 + Bx^4 + Cx^4\log(x^2),
 \label{eqApp:densityexpansion}
 \end{equation}
where $A$, $B$ and $C$ are constants. For small~$x$, the largest of these three 
terms is $Ax^2$, so $\log(1+n)\approx\log(Ax^2)$, which goes to~$-\infty$ as 
$x\rightarrow0$. The other terms are  
$n_{xx}\approx2A+\cO(x^2,x^2\log(x^2))$ and 
$n_{xxxx}\approx24C\log(x^2) + 24B + 100C$.
Adding these three together requires $1+24C=0$ to cancel the logarithmic 
singularity at~$x=0$, and the remaining terms are constants or go to zero as 
$x\rightarrow0$.

As can be seen from Fig.~\ref{figApp:DDFT5singularity},
having an adequate resolution for our numerical
calculations was a challenge but for different reasons for
the different models. In
DDFT-3, the density maxima can be sharply peaked while the logarithm of the
density is smooth, so inadequate resolution in the density field prompts
an increase in the number of grid points (we implement
automatic regridding, as discussed in
Appendix~\ref{app:continuation}). In contrast, in DDFT-5, the density field can
be smooth but with minima very close to zero, so its logarithm has very sharp
negative peaks. In this case, inadequate resolution in the logarithm of the
density prompts regridding. The difference is that in DDFT-5, the equation 
involves derivatives so any problem is magnified, while in DDFT-3, the equation 
involves convolutions that smooth out any problems.

These arguments indicate that a singular solution to the DDFT-5
equation~(\ref{eq:betadFdn5_rewritten}) of the type seen in
Fig.~\ref{figApp:DDFT5singularity} is possible, with the density going to zero,
and that this is not a problem of inadequate numerical resolution, but rather a
consequence of replacing the convolution in DDFT-3 with derivatives. A full
asymptotic theory should result in a prediction for the value of~$\mu$ at which
the branch terminates. The stripe solutions of the DDFT-3
equation~(\ref{eq:betadFdn3_rewritten}) can also have small density but without
any singularity in the solution.

\section{One mode approximation for DDFT}\label{sec:5}

The data displayed
in Fig.~\ref{figApp:DDFT5singularity}(a,c) lead to
an interesting observation: in (a)~DDFT-5, the logarithm of the
density is sharply negatively peaked, while the density is smooth (at least up
to its second derivative), slowly varying and resembles a cosine.
In contrast, in (c)~DDFT-3, the density is sharply peaked, while the
logarithm of the density is slowly varying and resembles a cosine. One of the
attractions of PFC theory is that it has slowly varying solutions that are well
represented by a few Fourier
modes~\cite{Elder2002, Elder2004, Emmerich2012, Wu2010}, and this carries
over to some extent to DDFT-5. Such Fourier representations of the density
profiles in
DDFT-3 are unsatisfactory, apart from for the unstable solutions very close to the
spinodal point, since any solution of reasonable amplitude is sharply peaked.

\begin{figure*}
\begin{center}
\hbox to \hsize{{\includegraphics[width=5.5truecm]{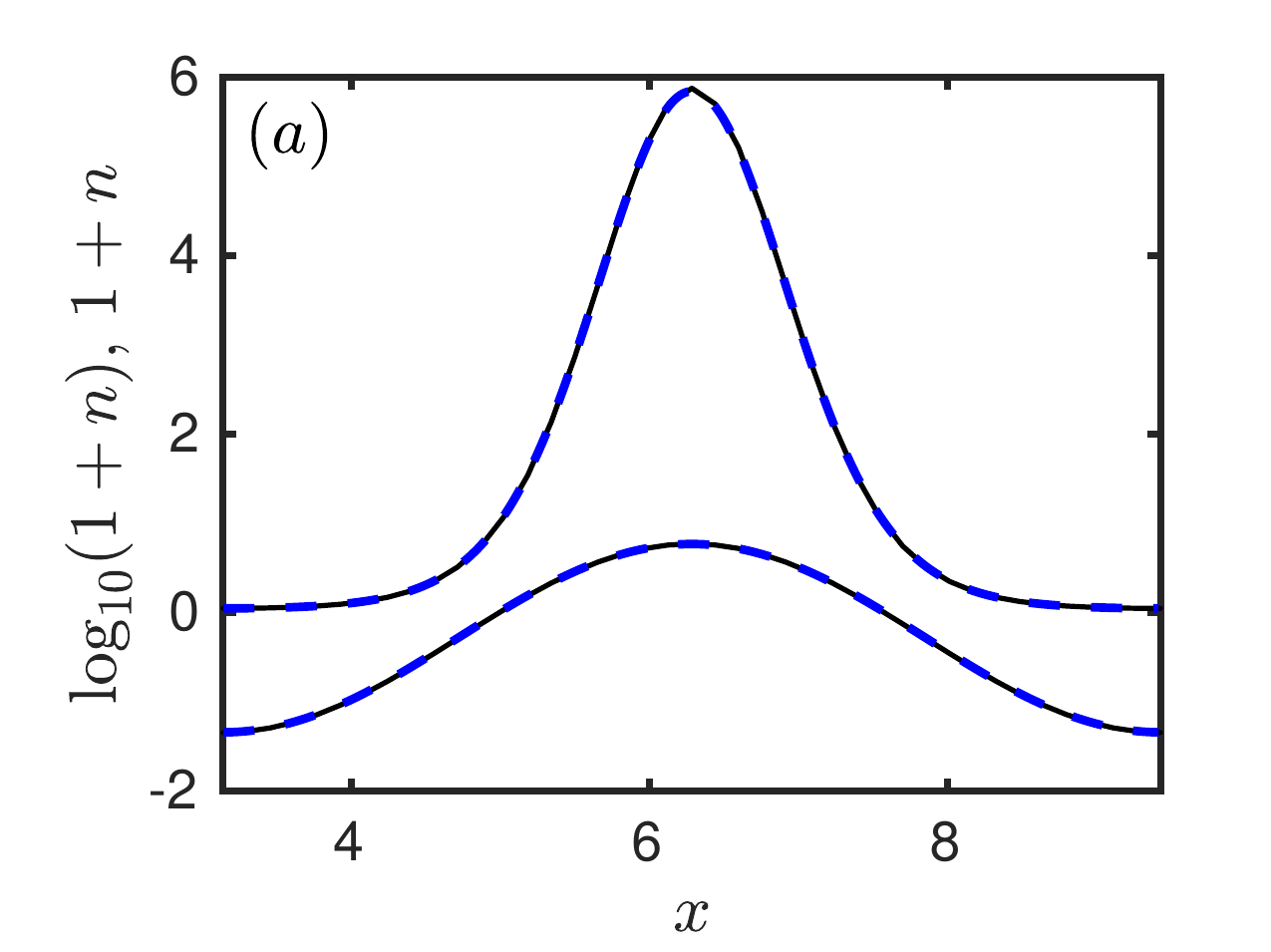}}\hfill
                {\includegraphics[width=5.5truecm]{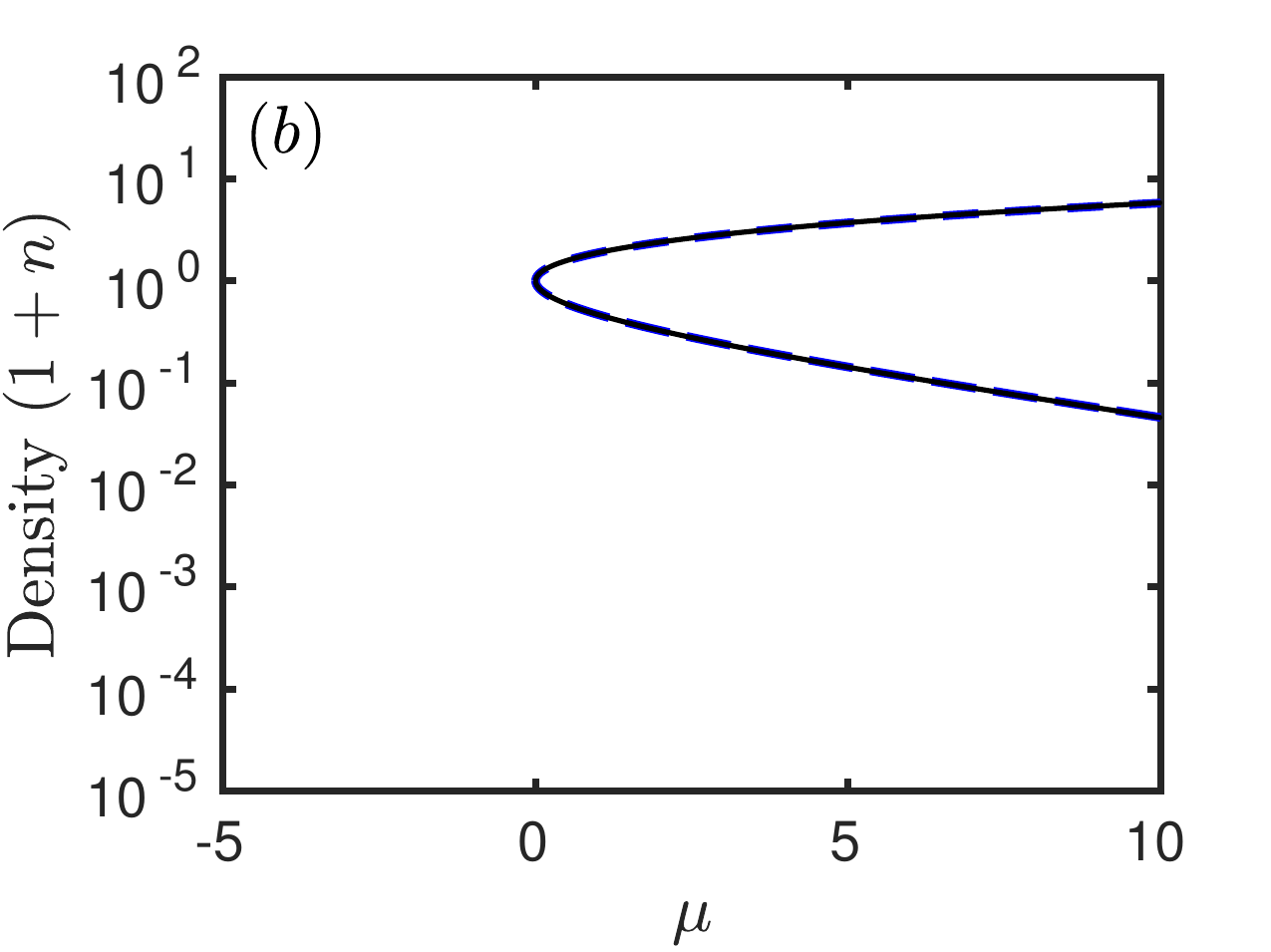}}\hfill
                {\includegraphics[width=5.5truecm]{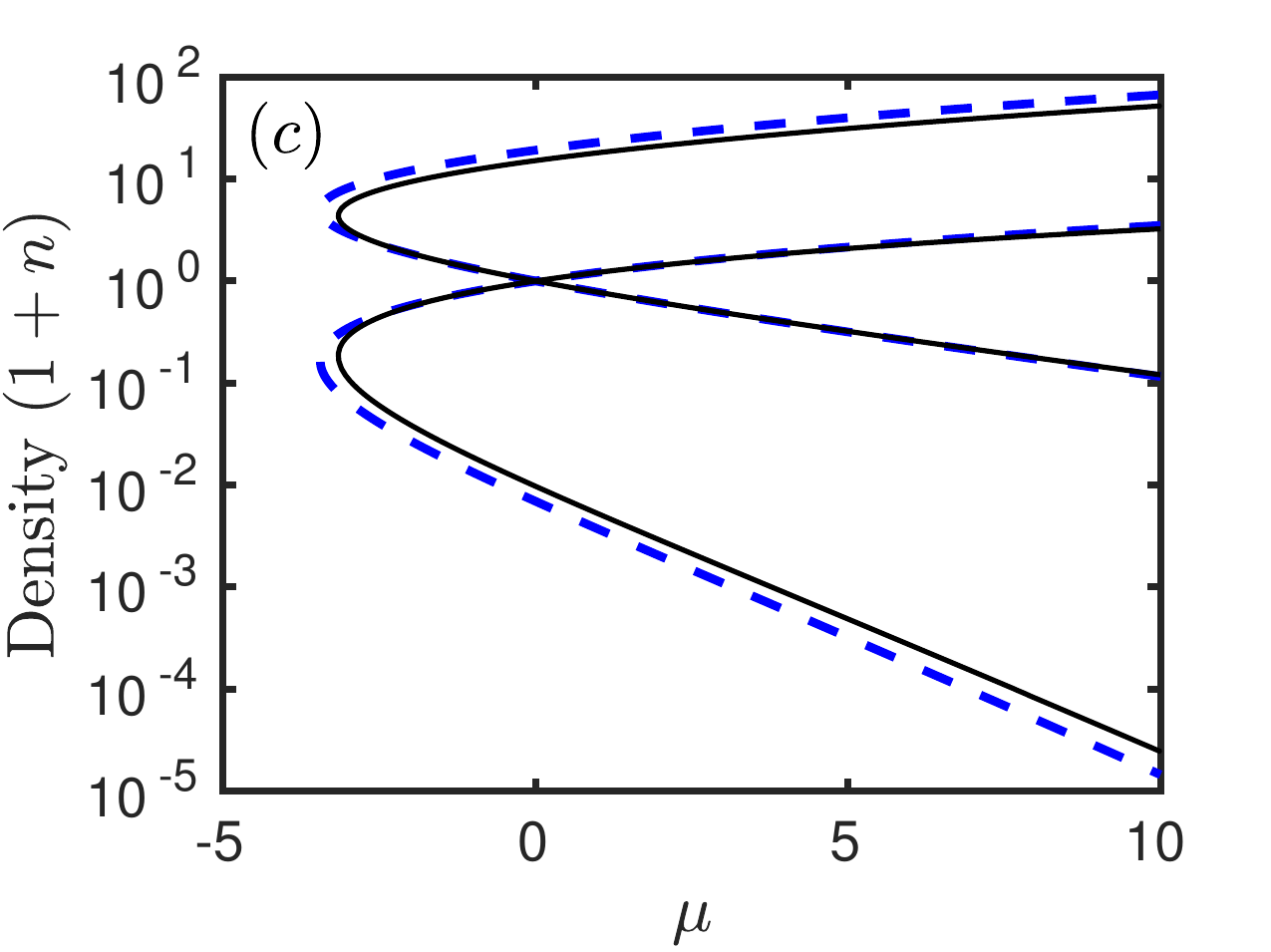}}}
\caption{Full numerical solutions of DDFT-3 plotted together with
the one-mode approximation. 
(a)~Stripes in 1D DDFT-3, showing the
density profile itself and its logarithm from solving
Eq.~(\ref{eq:betadFdn3_rewritten}) at $\mu=28.9$ (solid black line), with the
one-mode approximation overlaid (dashed blue line). 
Replotting the 2D DDFT-3 (b)~stripe and (c)~hexagon data from
Fig.~\ref{fig:DDFTPFCsummary}(a,b) (solid black lines), with the one-mode
approximation overlaid (dashed blue lines).}
 \label{figApp:DDFT3onemodeapprox}
\end{center}
 \end{figure*}
 
However, the data in Fig.~\ref{figApp:DDFT5singularity} suggest that 
representing the \emph{logarithm} of the density with a few Fourier modes should 
work well in DDFT-3. In this section, we elaborate how such a theory can be 
developed.

The key is to write Eq.~(\ref{eq:betadFdn3_rewritten}) in terms of
$\phi(\bx)\equiv\log(1+n(\bx))$, and approximate~$\phi$ by a few Fourier modes. As a first
step, we write out $\cL$ using Eq.~(\ref{eq:defnL_GEM4}) and
re-write~(\ref{eq:betadFdn3_rewritten}) as:
 \begin{equation}
    \log(1+n(\bx))
    + \rho_0\beta\epsilon \!\!\int\! 
                         \psi(|\bx-\bx_2|) n(\bx_2)\rdbx_2
    - \mu = 0,
 \label{eq:betadFdn3_rewrittenagain}
 \end{equation}
where $\psi(|\bx-\bx_2|)=e^{-|\bx-\bx_2|^4/R^4}$. The convolution term (including the $\rho_0\beta\epsilon$ prefactor) in this
equation is $-n(\bx)-\cL n(\bx)$. We know the eigenvalues of $\cL$:
$\cL\exp(ikx)=\sigma(k)\exp(ikx)$, which means that the convolution term
acting on
a Fourier mode $\exp(ikx)$ has eigenvalue~$-(1+\sigma(k))$. We also know that
for high wavenumbers, the convolution averages to zero, and indeed
$\sigma(k)\rightarrow-1$, as can be seen in Fig.~\ref{fig:dispersionGEM4}.

We focus first on stripes, which have Fourier components only at integer
wavenumbers, and notice that $1+\sigma(2)$ is already very small (less
than~$0.01$), because $\hat{\psi}(2)$ is small. This implies that the Fourier components of the convolution term
at $k=2$ and higher will be much smaller than the Fourier components at $k=0$
and $k=1$ -- regardless of the spectrum of~$n$ itself. The other two terms in
Eq.~(\ref{eq:betadFdn3_rewrittenagain}) are~$\mu$, which is constant ($k=0$
only), and~$\log(1+n)$, so $\log(1+n)$ can only have significant Fourier
components at $k=0$ and $k=1$: there is nothing to balance modes with
$|k|\geq2$. This explains why $\log(1+n)$ in the lower left panel of
Fig.~\ref{figApp:DDFT5singularity} is smooth, and it is also why approximating
the logarithm of the density by a few Fourier modes must work regardless of 
the amplitude of the modulations in the density,
or how sharply they are spiked, or of the value of~$\mu$.

For the stripe phase we write 
 \begin{equation}
 \log(1+n(x)) = \phi(x) = \phi_0 + \phi_1 e^{ix} + {\bar \phi}_1 e^{-ix},
 \label{eq:logdensityFourier}
 \end{equation}
where $\phi_0$ and $\phi_1$ are constants (real and complex, respectively)
that we need to find. This can easily be generalised for hexagons and
other periodic phases by adding more modes in \eqref{eq:logdensityFourier}.
The $k=0$ and $k=1$ components of
$\exp\left(\phi_1e^{ix}+\text{c.c.}\right)$, where $\text{c.c.}$ denotes the
complex conjugate, can be expressed in terms of
integrals, defining two functions $f_0(\phi_1)$ and $f_1(\phi_1)$ given~by
 \begin{equation}
 \begin{split}
 f_0(\phi_1) &= \frac{1}{2\pi}\int_0^{2\pi} \exp\left(\phi_1e^{ix}+\text{c.c.}\right)\rd{}x, \\
 f_1(\phi_1) &= \frac{1}{2\pi}\int_0^{2\pi} e^{-ix}\exp\left(\phi_1e^{ix}+\text{c.c.}\right)\rd{}x,
 \label{eq:f0f1defn}
 \end{split}
 \end{equation}
i.e., $f_0$~is modified Bessel function of the first kind of order zero and 
$f_1$~is a Fourier transform generalisation of~$f_0$. Using these functions,
$n(x)=e^{\phi(x)}-1$ can be written in terms of its Fourier
components as
 \begin{equation}
 \begin{split}
 e^{\phi(x)}-1 &= e^{\phi_0}f_0(\phi_1) - 1\\
            &\quad{} + \left(e^{\phi_0}f_1(\phi_1)e^{ix} + \text{c.c.}\right)\\
            &\quad{} + \text{modes with $|k|\geq2$.}
 \label{eq:densityFourier}
 \end{split}
 \end{equation}
The modes with $|k|\geq2$ in~(\ref{eq:densityFourier}) are large in amplitude,
but they are reduced in significance in Eq.~(\ref{eq:betadFdn3_rewrittenagain})
by the convolution, as explained above. 
The action of the convolution on modes with $|k|<2$ can be written in terms 
of~$\sigma(k)$.
Retaining only these terms, we are left with the
$k=0$ and $k=1$ components of~(\ref{eq:betadFdn3_rewrittenagain}):
 \begin{equation}                                                              
 \begin{split}
 \phi_0 + (1+\sigma(0))\left(1-e^{\phi_0}f_0(\phi_1)\right) - \mu &= 0,\\
 \phi_1 - e^{\phi_0}f_1(\phi_1)(1+\sigma(1)) &= 0.
 \label{eq:DDFT3onewaveeqns}
 \end{split}                                                        
 \end{equation}
Notice that the only information remaining from the GEM-4 potential is
the values of~$\sigma(0)$ and $\sigma(1)$, i.e., the values of $\hat{u}(0)$
and $\hat{u}(1)$. Recall also that if the reference
density $\rho_0$ is chosen to be the value at the spinodal, then we have
$\sigma(1)=0$. These equations can also be written in terms of the pair
potential as
\begin{equation}                                                              
 \begin{split}
 \phi_0 - \rho_0\beta\hat{u}(0)\left(1-e^{\phi_0}f_0(\phi_1)\right) - \mu &= 0,\\
 \phi_1 + e^{\phi_0}f_1(\phi_1)\rho_0\beta\hat{u}(1) &= 0.
 \label{eq:DDFT3onewaveeqns_u}
 \end{split}                                                        
 \end{equation}
The two equations in~(\ref{eq:DDFT3onewaveeqns}) can easily be
solved for $\phi_0$ and~$\mu$ in terms of~$\phi_1$, from which the density
can be reconstructed.
The agreement between this and the full solutions of DDFT-3 is astonishing.
Figure.~\ref{figApp:DDFT3onemodeapprox}(a) shows a 1D example at
$\mu=28.9$, with the full solution as a black line and the approximate solution
as a dashed line. Even though the density varies by two orders of magnitude,
the two are almost indistinguishable. There is similar excellent agreement
with the branch of stripe solutions (Fig.~\ref{figApp:DDFT3onemodeapprox}b) in 2D DDFT-3 (recall that
the GEM-4 potential has different values of~$\sigma(0)$ in 1D and 2D).

For 2D hexagons, the approach is similar, with the two $e^{\pm{ix}}$ terms in
Eq.~(\ref{eq:f0f1defn}) replaced by six similar terms, with wavenumbers $\bk$
that are uniformly spaced around a circle of radius 1 in $\bk$-space (c.f.~Eq.~\eqref{eqApp:weaklynonlinearhexagons} in Appendix B).
The agreement in this
case (Fig.~\ref{figApp:DDFT3onemodeapprox}c) is also very good.
If one adds a further six modes with $|\bk|=\sqrt{3}$, then
the agreement is as good as that for
stripes. These approximate solutions can easily be continued up to $\mu=100$
without difficulties, where we observe 
very sharply peaked density maxima and 
extremely small but non-zero
values for the density minimum.

\section{Discussion and conclusions}\label{sec:6}

In this paper, starting from DDFT, we have presented a step-by-step derivation of PFC
theory, at each stage explaining the consequences of the approximations. The
approximations can be listed under three main groupings: (i)~making a truncated
functional Taylor expansion approximation for the excess Helmholtz free energy, and then
making the RPA/RY approximation. This leads to DDFT-3 in our classification.
(ii)~Neglecting the $\nabla\cdot[n\nabla\cL n]$ term, which effectively also forces
making a Taylor expansion of the logarithmic ideal gas term and assuming constant 
mobility. (iii)~Replacing the nonlocal convolution in $\cL$ with a local gradient
expansion. The consequence of~(ii) is to introduce a second spinodal into the phase
diagram and to significantly alter the relative stabilities of the different periodic
states, in particular making striped states to become an equilibrium phase for some state
points, which is contrary to the physics. The consequence of making (iii) without first
making (ii), is to generate a theory (DDFT-5) that has a no-solution region in the phase
diagram, such as that displayed for the GEM-4 model in
Fig.~\ref{fig:phase_diagrams}(b). All these consequences have been illustrated for the GEM-4
system, chosen because DDFT-3 is fairly accurate for this model for temperatures
$k_BT/\epsilon>0.1$, allowing us to see the influence of the subsequent approximations.

{Throughout, there is good quantitative agreement between DDFT-3, \PFCgamma\ and
the EOF versions of DDFT-5 and \PFCepsilon\ (data not shown) \emph{only} for unstable
small amplitude solutions close to the spinodal point. The region of quantitative agreement
agreement between DDFT-3 and \PFCgamma\ is circled in red in Fig.~\ref{fig:DDFTPFCsummary}(b,h). 
Beyond this region, the agreement between the four theories is at best qualitative.}

Given all these problematic consequences for PFC theory, especially the issues related to
Taylor expanding the logarithmic ideal gas term, it raises the question of why then is
PFC theory so successful? In our view, there are several reasons for this. The first
reason is that PFC theory is qualitatively correct near to the spinodal. Therefore, it
can satisfactorily describe the coexistence between the liquid and the crystal phase,
which is often an important aspect in applications of the theory. Second, despite the
approximations, PFC theory still incorporates some very important physics: (i)~the free
energy satisfies the correct symmetries, (ii)~the dynamical equation gives a time
evolution that decreases the free energy monotonically over time and (iii)~the current is
proportional to the gradient of the chemical potential. These are all important features
for describing many phenomena. Also, many of the features that PFC theory is used to describe are generic, 
and the model parameters can be scaled to fit (for example) iron~\cite{Jaatinen2009}
and graphene~\cite{Fan2017}, but 
could equally well be scaled to match other materials, with similar good 
agreement. This universality (having the correct symmetries etc.) underlines 
the importance of PFC theory as a powerful model of generic features of 
crystalization, but it means that PFC theory will in general not be able to 
predict any unusual (non-generic) behaviour. 

Our results in Sec.~\ref{sec:5} for the GEM-4 system
showing that one can derive a very accurate one-mode amplitude equation approximation for
the field $\phi(\bx)=\log(\rho(\bx)/\rho_0)$, rather than the density itself, gives a
tantalising hint as to how PFC-type theories may more properly be derived and what the
order parameter field in PFC theory really represents: Should we consider the PFC order
parameter to be a scaled logarithm of the density distribution or some other similar
function of the density, rather than being proportional to the density profile itself? On
the basis of the work presented here, the answer to this question is `probably yes', but
clearly more work is required to fully address this.

Returning to theories for the density profile, in our view it is preferable to retain the
logarithmic ideal gas term in the approximation used for the Helmholtz free energy
functional, since this is required to have physical (i.e., positive) density profiles,
and also because with this the DDFT dynamics leads to the (correct physics) linear
diffusion equation term -- see Eq.~\eqref{eq:rhodiffusion}. The difficult consequence of
retaining this in the free energy is that one then has to deal with terms in the
dynamical equation of the form $\nabla\cdot[n\nabla\cL n]$, which makes solving
numerically far more difficult. However, this term also contributes to making the
crystalline phase more stable than the stripe phase, which also makes it important. The
crucial contribution of this term can especially be seen in DDFT-3, since in this version
of the theory it is clearly required for stabilizing the crystal structure.
{In general, we would advocate using one of DDFT-1, DDFT-2 or DDFT-3 (depending on how
$\cFex$~is treated) over all existing PFC theories, for studying the properties of real materials.}

It is worth noting that in all the approximations made here, we only consider those that
retain the form of a dynamics that decreases the free energy monotonically over time --
see Eq.~\eqref{eq:Fdecreases}. This is a feature of both DDFT and PFC theory. In our view
this structure is important and should not be broken by any approximations made, i.e.,
any that might be made in future in attempting to avoid any of the above mentioned
issues.

It is also worth noting that, while we have not discussed the consequences
of the LDA in going from DDFT-1 to DDFT-2 (see Eq.~(\ref{eq:definecs})), the
polynomial terms in the chemical potential~(\ref{eq:betadFdn2}) can
potentially lead to the same problem of having a second spinodal, even
while retaining the logarithm term. This applies to DDFT-4 as well.

In this paper we have largely focussed on making our arguments in two dimensions,
in order to keep the presentation as simple and comprehensible as possible. However, we
should emphasis that all of our arguments apply for three dimensional (3D) systems. For
example, at the higher temperatures we have focussed on here, the 3D GEM-4 model exhibits
at equilibrium a single fluid phase and two crystalline phases: the body-centered cubic
phase at lower densities and the face-centered cubic structure at higher densities.
These are all accurately predicted by the 3D~version of DDFT-3 \cite{Mladek2006}. There
are no columnar or lamellar phases, which are the 3D equivalents of the stripe phase. On
the other hand, the 3D versions of the PFC theories presented here all predict a lamellar
phase at some state points~\cite{Thiele2013}, the 3D generalisations of the down hexagons and a second
spinodal with the uniform liquid becoming the equilibrium phase at higher densities.

{It would be interesting to explore how the dynamics of defects, the elasticity 
and the plasticity of crystals differs between DDFT and~\hbox{PFC}}, and 
our results are also relevant to binary systems. In the derivation in
Ref.~\cite{Huang2010a} of a PFC theory for binary mixtures, the generalisation of the
$\nabla\cdot[n\nabla\cL n]$ term is retained until the last moment in the derivation, but
then dropped for the same reasons that is is neglected for one-component systems. Given
the importance of this term for one-component systems, it is surely also important for
stabilizing crystal structures in binary systems. {Note also that when determining
mechanical properties such as elastic constants, the terms in the free energy that are
linear in $n$ can be important \cite{Wang2018b}. These have been neglected here throughout
since such terms do not contribute to determining density profiles.}

The singularity observed for DDFT-5 as the chemical potential $\mu$ (or equivalently, the 
average density $1+\bar{n}$) is increased was found by continuing equilibrium solutions
determined at lower values of $\mu$. One aspect that needs further investigation relates
to determining the influence of this when DDFT-5 is solved for state points where the
final equilibrium crystal (or stripe) solution for the density profile does not exist.
For example, a situation we have in mind is that studied by van Teeffelen
\etal~\cite{Teeffelen2009} consisting of a solidification front propagating into the
unstable liquid. These authors compared results for this situation between (in our
terminology) DDFT-3, DDFT-5 and \PFCepsilon. Their results are for two-dimensional
dipolar colloidal particles. From the DDFT-5 results (PFC1 in their terminology)
displayed in Figs.~4 and 5 of Ref.~\cite{Teeffelen2009}, it can be seen that they did not
consider values of the coupling parameter large enough to encounter any of the
singularities; the density profiles stay well away from zero. It would be interesting to
quench deeper into the crystal phase to study the evolution of the density distribution
towards the singular state. However, the numerics to resolve this accurately would surely
be difficult. 

One aspect of PFC theory that the derivation from DDFT highlights is that in general one
is not free to independently vary the parameters $r$ and $\bar{n}$ in
Eqs.~\eqref{eq:PFCFalpha} and~\eqref{eq:PFCalphadynamics}. For example, for the GEM-4
model there are certain values of $r$ that are not generated by the mapping from DDFT. Of
course, by changing the pair potential, different combinations of the PFC model
parameters can become accessible. We should also recall that although we have illustrated
many of our conclusions by considering the soft-core GEM-4 model, PFC theory can be
derived for systems of particles with hard potentials since it is the pair direct
correlation function $c^{(2)}(\bx_1,\bx_2)$ that enters the theory; this quantity is
finite for all values of $\bx_1$ and $\bx_2$.

As a final point, we mention that our results will also be of interest to the pure
mathematics community. DDFT-3 is also referred to as the McKean--Vlasov equation and in
this context there are a number of recent interesting rigorous results
\cite{Carrillo2018, Gomes2019}. Our results for DDFT-5, showing that for a finite value
of $\mu$ there is a singularity with the density profile going to zero, may well be of
interest to those who study the mathematics of solutions to partial differential
equations with compact support -- see for example Ref.~\cite{Bernis1992}. For values of
$\mu$ beyond the singular point where $1+n(\bx)\to0$ it may be that the solutions become
complex. If one were interested to find these solutions, we believe it might require
treating $\mu$ as a complex variable. Of course, all of this is out of the realm where
the model represents a theory for matter.

\begin{acknowledgments}
This work was supported in part by a L'Or{\'e}al UK and Ireland Fellowship 
for Women in Science (PS), by the EPSRC under grants EP/P015689/1 (AJA, DR) and 
\hbox{EP/P015611/1 (AMR)}, and by the Leverhulme Trust (RF-2018-449/9, AMR).
We are grateful for conversations with Tapio Ala-Nissil\"a, Daniele Avitabile,
Ken Elder, Zhi-Feng Huang, Kai Jiang, Edgar Knobloch, Ron Lifshitz, Chris Malcotte, Daniel Read, 
Uwe Thiele, Steve Tobias, Gyula T\'oth, Laurette Tuckerman and Joanna Tumelty.
We are grateful also for constructive comments from two anonymous referees.
\end{acknowledgments}

\appendix

\section{Linear theory for GEM-4}

\label{app:LinearGEM4}

In this appendix, we discuss how we compute the linear theory for the GEM-4 
potential in~(\ref{eq:GEM4Potential}). 
To be specific, in a two-dimensional periodic domain, the 
eigenvalue~$\sigma(k)$ is defined by $\cL e^{i\bk\cdot\bx} = \sigma(k) 
e^{i\bk\cdot\bx}$ and~(\ref{eq:defnL_GEM4}), which can be written as
 \begin{equation}
 \sigma(k) = -  1 - \rho_0\beta\epsilon \!\!\int\! 
                         e^{-|\bx-\bx_2|^4/R^4} e^{i\bk\cdot(\bx_2-\bx)}\rdbx_2,
 \label{eqApp:defn_sigmak_GEM4}
 \end{equation}
where the integral is taken over the periodic domain (the GEM-4
potential is replaced by its periodic extension). We set $\bk=(k,0)$, we
integrate from $[-N\pi/k,N\pi/k]$ in each dimension, and we choose the integer~$N$
large enough that the GEM-4 exponential is effectively zero at the boundaries;
$N=4$ suffices. We then scale $\bx$ by a factor of~$k$, replacing $\bx_2-\bx$ by 
$\bx/k$, so that the integral becomes:
 \begin{equation}
 \sigma(k) = -  1 - \rho_0\beta\epsilon 
                    \!\!\iint_{-N\pi}^{N\pi}\!
                         e^{-|\bx|^4/(k^4R^4)} e^{ix}\frac{\rdbx}{k^2},
 \label{eqApp:defn_sigmak_GEM4_scaled}
 \end{equation}
where $x$ is the first component of~$\bx$. With this scaling, the limits of the 
integral do not depend on~$k$.

\begingroup
 % \squeezetable
\begin{table}
\begin{center}
\begin{ruledtabular}
\begin{tabular}{cccccc}
Dimension  & $R$ & $\rho_0\beta\epsilon$ & $\gamma$ & $\sigma(0)$ & $E_B$ \\[0.5ex]
\hline\\[-1.5ex]
 % 1 & 4.591814091503938 & 1.16285266731731c & 3.205086306313718 & -10.679659250202967 \\
 % 2 & 5.096178801151792 & 0.245503483070306 & 4.369210135662223 & -18.751775027487930 \\
 % 3 & 5.571858062552498 & 0.045506774926377 & 5.688860421626603 & -31.304691012062225 \\ [3.0ex]
1 & 4.5918 & 1.1629 & 3.2051 & $-10.680$ & $\phantom{0}7.475$ \\
2 & 5.0962 & 0.2455 & 4.3692 & $-18.752$ & $14.383$  \\
3 & 5.5719 & 0.0455 & 5.6889 & $-31.305$ & $25.616$ \\ 
\end{tabular}
\end{ruledtabular}
\end{center}
\caption{Linear theory for the GEM-4 potential in one, two and three
dimensions. Solving $\frac{\rd\sigma}{\rd{k}}(1)=0$ and $\sigma(1)=0$ gives $R$
and~$\rho_0\beta\epsilon$, while $\gamma$ and $\sigma(0)$ are computed from
$\frac{\rd^2\sigma}{\rd{k^2}}(1)$ and (\ref{eqApp:defn_sigmak_GEM4})
respectively. We also give the values of $E_B$ for the EOF in Eq.~\eqref{eq:cLgradJaatinen}.}
 \label{tabApp:GEM4parameters}
\end{table}
\endgroup

We choose $R$ and $\rho_0\beta\epsilon$ so that $\sigma(1)=0$ and
$\frac{\rd\sigma}{\rd{k}}(1)=0$. The derivative of $\sigma$ with respect to~$k$
is
 \begin{equation}
 \frac{\rd\sigma}{\rd{k}} = - \rho_0\beta\epsilon 
                    \!\!\iint_{-N\pi}^{N\pi}\! 
                         \left(\frac{4|\bx|^4 - 2k^4R^4}{k^7R^4}\right)
                         e^{-|\bx|^4/(k^4R^4)} e^{ix}\rdbx.
 \label{eqApp:defn_sigmak_GEM4_derivative}
 \end{equation}
Evaluating this at $k=1$ and removing the constant factor outside the integral
gives a function $F(R)$:
 \begin{equation}
 F(R) =  \iint_{-N\pi}^{N\pi}\! 
                         \left(\frac{4|\bx|^4 - 2R^4}{R^4}\right)
                         e^{-|\bx|^4/R^4} e^{ix}\rdbx.
 \label{eqApp:GEM4_FRs}
 \end{equation}
We solve the equation $F(R)=0$ using Newton's method to give~$R$. We then
calculate $\rho_0\beta\epsilon$ by requiring that $\sigma(1)=0$
in~(\ref{eqApp:defn_sigmak_GEM4_scaled}). Values for $R$ and
$\rho_0\beta\epsilon$ in one, two and three dimensions are given in
table~\ref{tabApp:GEM4parameters}. We compute the GEM-4 dispersion relation in
Fig.~\ref{fig:dispersionGEM4} in a similar way, and use the second derivative
of~(\ref{eqApp:defn_sigmak_GEM4_scaled}) with respect to~$k$, at~$k=1$, to find
the~$\gamma$ parameter (also given in table~\ref{tabApp:GEM4parameters})
in~$\cLgrad$. The table also gives~$\sigma(0)$, since this is useful for
computing properties of the liquid, as well as $E_B$, the coefficient in the
eighth-order model of~\cite{Jaatinen2009}, given in Eq.~(\ref{eq:cLgradJaatinen}).

\section{Numerical method: continuation}

\label{app:continuation}

We use numerical continuation to solve the four
equations~(\ref{eq:betadFdn3_rewritten}) and
(\ref{eq:betadFdn5_rewritten})--(\ref{eq:betadFdnepsilon_rewritten}) for~$n(\bx)$
as the parameters vary. Our approach is based on~\cite{Doedel1991} for the
pseudo-arclength continuation method, and we use the approach advocated
by~\cite{Kelley2003} to solve the large linear systems at each Newton step.
The main parameters are the chemical potential~$\mu$, the parameters in the
linear operators~$\cL$ and $\cLgrad$, and the domain size. In this discussion,
we focus on $\mu$ as the parameter that is varied.

\subsection{Pseudo-arclength continuation}

The main idea behind pseudo-arclength continuation is to suppose that we are 
looking to calculate a branch of solutions $n(\bx)$ depending on the 
parameter~$\mu$. The branch may have folds (as in 
Fig.~\ref{fig:DDFTPFCsummary}), and the 
method should be able to go around these. The method defines a 
parameter (the arclength~$s$) that increases or decreases monotonically along 
the branch (including its folds) such that both $n(\bx)$ and $\mu$ can be
regarded as single-valued functions of~$s$. Then the equation to be solved is
 \begin{equation}
 \cG(n(s),\mu(s)) = 0,
 \label{eqApp:equationtosolve}
 \end{equation}
where $\cG$ represents the equation we are solving for~$n$. Instead of thinking
of $n$ as being a function of position~$\bx$, we represent $n$ as a series of
values~$n_i$ on $N$~grid points $\bx_i$ ($i=1,\dots,N$), so $n$ is now a vector
in~$\mathbb{R}^N$, and $\cG$ is a function from
$\mathbb{R}^{N+1}\rightarrow\mathbb{R}^{N}$.
Equation~(\ref{eqApp:equationtosolve}) represents $N$ equations for $N+1$
unknowns, and so it is supplemented by an orthogonality condition, that the
next point on a branch should lie in a plane orthogonal to a line connecting
the two previous points. It is this that allows the branch following technique
to go around folds. If we have two points on the branch $(n(s),\mu(s))$ at
$s_0$ and $s_1$, then we take the derivatives of $n$ and $\mu$ with respect to
the arclength to be approximately
 \begin{equation}
 \frac{\rd{n}}{\rd{s}} = S \frac{n(s_1)-n(s_0)}{s_1-s_0},
 \quad
 \frac{\rd\mu}{\rd{s}} = S \frac{\mu(s_1)-\mu(s_0)}{s_1-s_0},
 \label{eqApp:pseudoarclengthstep}
 \end{equation}
with the scaling factor~$S$ chosen so as to 
satisfy
 \begin{equation}
 \frac{1}{N}\sum_{i=1}^N \left(\frac{\rd{n_i}}{\rd{s}}\right)^2 + 
 \left(\frac{\rd\mu}{\rd{s}}\right)^2 = 1.
 \label{eqApp:arclengthnormalization}
 \end{equation}
The $\frac{1}{N}$ prefactor means that the parameterization of the branch
by the arclength is essentially independent of the number of grid points. 

The method then proceeds in a predictor--corrector fashion. The predictor step,
with a target stepsize~$\Delta{s}$ provides $(n_2,\mu_2)$:
 \begin{equation}
 n_2 = n(s_1) + \Delta{s}\frac{\rd{n}}{\rd{s}},
 \quad
 \mu_2 = \mu(s_1) + \Delta{s}\frac{\rd\mu}{\rd{s}}.
 \label{eqApp:pseudoarclengthpredictor}
 \end{equation}
Then, $(n_2,\mu_2)$ is used as an initial iterate for a Newton solver for
equation~(\ref{eqApp:equationtosolve}), supplemented by the condition that the
Newton iterates lie in a plane orthogonal to the line given
in~(\ref{eqApp:pseudoarclengthpredictor}), parameterised by~$\Delta{s}$. This means 
that we are solving $\cH(n,\mu)=0$, where $\cH$ is 
$\mathbb{R}^{N+1}\rightarrow\mathbb{R}^{N+1}$, with the first~$N$ equations 
in~$\cH$ being the same as~$\cG$, and
the last equation being
 \begin{equation}
 (n - n_2) \cdot \frac{\rd{n}}{\rd{s}} + (\mu - \mu_2) \frac{\rd\mu}{\rd{s}} = 0.
 \label{eqApp:pseudoarcorthogonality}
 \end{equation}
To be precise, we take
 \begin{equation}
 \cH(n,\mu)=\begin{pmatrix}
 \frac{1}{\sqrt{N}} P\cdot \cG(n,\mu)\\
 \textrm{Eq.~(\ref{eqApp:pseudoarcorthogonality})}
 \end{pmatrix},
 \label{eqApp:fullsystem}
 \end{equation}
where $P$~is a linear preconditioner for $\cG$ (see below). The $\sqrt{N}$ scaling
means that the norm $\lVert\cH(n,\mu)\rVert$ 
(the square root of the sum of the squares of its components)
is independent of the number
of grid points~$N$, and it also means that the equations in $\cG$ and the
orthogonality condition~(\ref{eqApp:pseudoarcorthogonality}) are given a
similar weighting by the Newton method.

Solving $\cH(n,\mu)=0$ results in a new point on the branch of solutions, 
$(n(s_2),\mu(s_2))$, where $s_2$ is given by
 \begin{equation}
 s_2 = s_1 + \frac{1}{N}(n(s_2)-n(s_1))\cdot\frac{\rd{n}}{\rd{s}}
           +            (\mu(s_2)-\mu(s_1))\frac{\rd\mu}{\rd{s}},
 \label{eqApp:newpseudoarc}
 \end{equation}
with $\frac{\rd{n}}{\rd{s}}$ and $\frac{\rd\mu}{\rd{s}}$ given 
by~(\ref{eqApp:pseudoarclengthstep}). This last equation comes from replacing 
$n_2$ by $n(s_2)$ and $\mu_2$ by $\mu(s_2)$ in~(\ref{eqApp:pseudoarclengthpredictor})
and finding a $\Delta{s}=s_2-s_1$ from $1/N$ times the first equation plus the 
second equation. This is not quite the actual change to the arclength that was
achieved in the step, and the approximation is the reason that the method is
called the pseudo-arclength method.

\subsection{Newton's method}

For Newton's method, we define $X=(n,\mu)$ and solve $\cH(X)=0$. We start with
$X_0$ given by the predictor step above in~(\ref{eqApp:pseudoarclengthpredictor}),
and follow~\cite{Kelley2003}, at each step solving the linear equation
 \begin{equation}
 \frac{\partial\cH}{\partial{X}} \cdot \deltaX_n = - \cH(X_n),
 \label{eqApp:linearNewtonEqns}
 \end{equation}
where $\frac{\partial\cH}{\partial{X}}$ is the $(N+1)\times(N+1)$ matrix of 
derivatives of $\cH$ with respect to~$X$,
and then improving our estimate of the root by using~$\deltaX$. 

The Newton method proceeds until convergence, defined by
 \begin{equation}
 \lVert\cH(X_n)\rVert < N_\text{abs} + N_\text{rel}\lVert\cH(X_0)\rVert,
 \end{equation}
where $N_\text{abs}$ and $N_\text{rel}$ are the Newton
absolute and relative convergence tolerances respectively, typically $10^{-10}$
and~$10^{-8}$. We also monitored the maximum of $|\cH(n(\bx),\mu)|$
across the domain, and this was typically no larger than ten
times~$N_\text{abs}$, so the equations are well 
satisfied at each point in space as well as in norm.

The linear equations in~(\ref{eqApp:linearNewtonEqns}) are solved to 
find~$\deltaX_n$ using
\textsc{Matlab}'s biconjugate gradient stabilized~(l) (\texttt{bicgstabl}) method. This
allows the matrix--vector multiplications to be evaluated using a function
(rather than by explicitly computing a large matrix). The method is iterative, 
and proceeds until 
 \begin{equation}
 \left\lVert\frac{\partial\cH}{\partial{X}} \cdot \deltaX_n + \cH(X_n)\right\rVert <
 L_\text{rel}\lVert\cH(X_n)\rVert,
 \end{equation}
where the relative tolerance $L_\text{rel}$ of the linear solver
is chosen so as to balance the number of Newton steps against the number of
\texttt{bicgstabl} iterations. Based on~\cite{Kelley2003}, we choose
 \begin{equation}
 L_\text{rel} = 0.1\sqrt{\lVert\cH(X_n)\rVert} 
            + \frac{N_\text{abs}}{\lVert\cH(X_n)\rVert},
 \end{equation} 
subject to the constraint that $L_\text{rel}$ should be no larger
than~$0.1$. The effect of this is that in the initial first or second of the
Newton iterations, when $\lVert\cH(X_n)\rVert$ is at its largest, the linear
solver is not asked to work too hard to solve~(\ref{eqApp:linearNewtonEqns}),
since any reasonably good approximate solution is likely to improve the
estimate of the root, and an absolutely perfect solution isn't going to do much
better. In the middle stages of the Newton iterations, when
$\lVert\cH(X_n)\rVert$ is about $10^{-6}$, the tolerance
$L_\text{rel}$ is about $2\times10^{-4}$, which is not good enough
for quadratic convergence of Newton's method, but is good enough to provide two
or three orders of magnitude improvement to the quality of the solution at a
considerably lower cost. In the final stages of the Newton iterations,
$L_\text{rel}=0.1$, good enough for polishing the solution to the
tolerance $N_\text{abs}$ while not attempting to solve the linear
problem down to round-off error.

Also based on~\cite{Kelley2003}, we implement the Armijo rule, which ensures 
that each Newton step results in an improvement to the solution. The idea is 
that the solution of the linear equation~(\ref{eqApp:linearNewtonEqns}) should 
give the correct direction for improving the solution of $\cH(X)=0$, but taking a 
full step may not actually result in an improvement, so instead we set
 \begin{equation}
 X_{n+1} = X_n + 2^{-j} \deltaX_n,
 \label{eqApp:armijo}
 \end{equation}
where $j=0,1,2,\dots$ is chosen to be the smallest such that 
 \begin{equation}
 \lVert\cH(X_{n+1})\rVert <
 \lVert\cH(X_{n})\rVert.
 \label{eqApp:armijocondition}
 \end{equation}
In most cases, the first ($j=0$) Armijo step 
satisfies~(\ref{eqApp:armijocondition}), equivalent to the normal Newton 
method, but when the density is close to zero in the DDFT calculations, and 
small changes in density lead to large changes in its logarithm, the Armijo 
rule is helpful.

We do not use a preconditioner in the GEM-4 calculations (so $P$
in~(\ref{eqApp:fullsystem}) is the identity), but in the gradient expansion calculations,
a preconditioner is helpful. In Fourier space, $\cLgrad$ can easily be inverted, so the
preconditioner is $\cLgradinv$ when the absolute value of the eigenvalue $\sigma(k)$ is
greater than~$1$, otherwise the preconditioner is the identity. This has the effect of
reducing the number of iterations needed to solve~(\ref{eqApp:linearNewtonEqns}) by a
factor of 10 or even~100.

A sample \textsc{Matlab} code to solve Eq.~(\ref{eq:betadFdn3_rewritten}) 
for DDFT-3 by Newton's method (without the continuation aspect) is given in the
supplementary material.

\subsection{Additional considerations}

We start the computation of each branch close to $n=0$ and $\mu=0$ using an 
approximate solution derived from weakly nonlinear theory. For example, for 
hexagons in DDFT-3, we take
 \begin{equation}
 n(\bx) = \frac{\mu}{\sigma(0)}\left(-1 +
              e^{i\bk_1\cdot\bx} + e^{i\bk_2\cdot\bx} + e^{i\bk_3\cdot\bx} + 
              \text{c.c.}\right),
 \label{eqApp:weaklynonlinearhexagons}
 \end{equation}
where the initial value of~$\mu$ is small, $\sigma(0)$ comes from 
Table~\ref{tabApp:GEM4parameters}, $\bk_1=(1,0)$, 
$\bk_2=(-\frac{1}{2},\frac{\sqrt{3}}{2})$,
$\bk_3=(-\frac{1}{2},-\frac{\sqrt{3}}{2})$, and
$\text{c.c.}$ stands for complex conjugate.

The equations are posed on periodic domains and we use $N_x\times{N_y}$ grid
points, depending on the number of wavelengths in the domain and the nature of
the solution. The GEM-4 hexagonal calculations require
$8\times\frac{8}{\sqrt{3}}$ wavelength domains, with resolutions starting at
$80\times48$ Fourier modes close to onset. At larger amplitude, $512\times320$
Fourier modes (or even more) are needed, especially if the density maxima are
sharply peaked (in~DDFT-3) or if the density minima are very close to zero
(in~DDFT-5). In order to accommodate the changing needs for resolution along a
branch, we monitor whether the solutions are well resolved and implemented
automatic regridding, so as to maintain enough grid points to resolve the
solution well, regardless of what features emerge as~$\mu$ is varied. Typically we
require that the amplitudes of the highest-wavenumber Fourier modes be no
higher than $10^{-10}$ times the largest Fourier amplitude.

We also implement automatic pseudo-arclength stepsize control: $\Delta{s}$ is 
increased by a factor of~1.1 (up to a maximum of~0.1) if the Newton method converges
quickly (in fewer than 5~iterations), or is decreased by a factor of~2 (down to a minimum
of~$10^{-6}$) if it converges slowly (more than 8~iterations) or fails.

Finally, we adjust the domain size continuously along each branch so as to minimise the
specific grand potential~$\Omega/A$. This is done by adding an extra parameter (the
domain stretch factor~$K$), so that the real domain size is $KL_x\times{K}L_y$ instead of
an unstretched $L_x\times{L_y}$. The number of grid points is not altered. Then the real
GEM-4 potential is proportional to $\psi(|\bx|)=\exp\left(-K^4|\bx|^4/R^4\right)$, where
$\bx$ is the unstretched coordinate on the unaltered grid.  Quantities like the mean
value of~$n$ are the sum of the values of~$n$ at each grid point divided by the number of
grid points, and so are not affected by the stretch factor. The only parts of $\Omega/A$
that are affected are the convolutions (for~$\cL$ and the GEM-4 potential) and the
spatial derivatives (for~$\cLgrad$). In the case of the GEM-4 potential, when considering 
the specific 
grand potential $\Omega_3/A$ arising from~(\ref{eq:DDFTF3}) for example, 
and $\cL$ given by~(\ref{eq:defnL_GEM4}), 
the integrals in
$\Omega_3/A$ are proportional to $K^2$, with additional $K$~dependence coming from the
GEM-4 potential itself. Therefore,
 \begin{equation}
 \frac{\rd(\Omega_3/A)}{\rd{K}} \propto
    \int n(\bx) \left(2 K \psi \otimes n + K^2 \frac{\partial\psi}{\partial K} \otimes n\right)\,\rdbx,
 \end{equation}
where $\otimes$ represents the convolution integral evaluated on the unstretched grid, and the
$\bx$ integral is also on the unstretched grid. In the case of~$\cLgrad$, Laplacians on
the real grid are a factor of $K^{-2}$ times Laplacians on the unstretched grid, so
$\rd(\Omega/A)/\rd{K}$ is evaluated accordingly. In both cases, an extra equation
($\rd(\Omega/A)/\rd{K}=0$) is added to~$\cH$ in~(\ref{eqApp:fullsystem}), and this is
solved alongside all the other equations. The Jacobian also needs to be evaluated.
In practice this made little difference in the cases considered here, and
the domain stretch factor largely stayed between $0.98$ and~$1.02$.

%\bibliographystyle{apsrev}
%\bibliography{pfc_ddft}

\begin{thebibliography}{68}%
\makeatletter
\providecommand \@ifxundefined [1]{%
 \@ifx{#1\undefined}
}%
\providecommand \@ifnum [1]{%
 \ifnum #1\expandafter \@firstoftwo
 \else \expandafter \@secondoftwo
 \fi
}%
\providecommand \@ifx [1]{%
 \ifx #1\expandafter \@firstoftwo
 \else \expandafter \@secondoftwo
 \fi
}%
\providecommand \natexlab [1]{#1}%
\providecommand \enquote  [1]{``#1''}%
\providecommand \bibnamefont  [1]{#1}%
\providecommand \bibfnamefont [1]{#1}%
\providecommand \citenamefont [1]{#1}%
\providecommand \href@noop [0]{\@secondoftwo}%
\providecommand \href [0]{\begingroup \@sanitize@url \@href}%
\providecommand \@href[1]{\@@startlink{#1}\@@href}%
\providecommand \@@href[1]{\endgroup#1\@@endlink}%
\providecommand \@sanitize@url [0]{\catcode `\\12\catcode `\$12\catcode
  `\&12\catcode `\#12\catcode `\^12\catcode `\_12\catcode `\%12\relax}%
\providecommand \@@startlink[1]{}%
\providecommand \@@endlink[0]{}%
\providecommand \url  [0]{\begingroup\@sanitize@url \@url }%
\providecommand \@url [1]{\endgroup\@href {#1}{\urlprefix }}%
\providecommand \urlprefix  [0]{URL }%
\providecommand \Eprint [0]{\href }%
\providecommand \doibase [0]{http://dx.doi.org/}%
\providecommand \selectlanguage [0]{\@gobble}%
\providecommand \bibinfo  [0]{\@secondoftwo}%
\providecommand \bibfield  [0]{\@secondoftwo}%
\providecommand \translation [1]{[#1]}%
\providecommand \BibitemOpen [0]{}%
\providecommand \bibitemStop [0]{}%
\providecommand \bibitemNoStop [0]{.\EOS\space}%
\providecommand \EOS [0]{\spacefactor3000\relax}%
\providecommand \BibitemShut  [1]{\csname bibitem#1\endcsname}%
\let\auto@bib@innerbib\@empty
%</preamble>
\bibitem [{\citenamefont {Elder}\ \emph {et~al.}(2002)\citenamefont {Elder},
  \citenamefont {Katakowski}, \citenamefont {Haataja},\ and\ \citenamefont
  {Grant}}]{Elder2002}%
  \BibitemOpen
  \bibfield  {author} {\bibinfo {author} {\bibfnamefont {K.~R.}\ \bibnamefont
  {Elder}}, \bibinfo {author} {\bibfnamefont {M.}~\bibnamefont {Katakowski}},
  \bibinfo {author} {\bibfnamefont {M.}~\bibnamefont {Haataja}}, \ and\
  \bibinfo {author} {\bibfnamefont {M.}~\bibnamefont {Grant}},\ }\bibfield
  {title} {\enquote {\bibinfo {title} {Modeling elasticity in crystal
  growth},}\ }\href@noop {} {\bibfield  {journal} {\bibinfo  {journal} {Phys.
  Rev. Lett.}\ }\textbf {\bibinfo {volume} {88}},\ \bibinfo {pages} {245701}
  (\bibinfo {year} {2002})}\BibitemShut {NoStop}%
\bibitem [{\citenamefont {Elder}\ and\ \citenamefont
  {Grant}(2004)}]{Elder2004}%
  \BibitemOpen
  \bibfield  {author} {\bibinfo {author} {\bibfnamefont {K.~R.}\ \bibnamefont
  {Elder}}\ and\ \bibinfo {author} {\bibfnamefont {M.}~\bibnamefont {Grant}},\
  }\bibfield  {title} {\enquote {\bibinfo {title} {Modeling elastic and plastic
  deformations in nonequilibrium processing using phase field crystals},}\
  }\href@noop {} {\bibfield  {journal} {\bibinfo  {journal} {Phys. Rev. E}\
  }\textbf {\bibinfo {volume} {70}},\ \bibinfo {pages} {051605} (\bibinfo
  {year} {2004})}\BibitemShut {NoStop}%
\bibitem [{\citenamefont {Backofen}\ and\ \citenamefont
  {Voigt}(2010)}]{Backofen2010}%
  \BibitemOpen
  \bibfield  {author} {\bibinfo {author} {\bibfnamefont {R.}~\bibnamefont
  {Backofen}}\ and\ \bibinfo {author} {\bibfnamefont {A.}~\bibnamefont
  {Voigt}},\ }\bibfield  {title} {\enquote {\bibinfo {title} {A
  phase-field-crystal approach to critical nuclei},}\ }\href@noop {} {\bibfield
   {journal} {\bibinfo  {journal} {J. Phys.: Condens. Matter}\ }\textbf
  {\bibinfo {volume} {22}},\ \bibinfo {pages} {364104} (\bibinfo {year}
  {2010})}\BibitemShut {NoStop}%
\bibitem [{\citenamefont {Toth}\ \emph {et~al.}({2010})\citenamefont {Toth},
  \citenamefont {Tegze}, \citenamefont {Pusztai}, \citenamefont {Toth},\ and\
  \citenamefont {Granasy}}]{Toth2010}%
  \BibitemOpen
  \bibfield  {author} {\bibinfo {author} {\bibfnamefont {Gyula~I.}\
  \bibnamefont {Toth}}, \bibinfo {author} {\bibfnamefont {Gyoergy}\
  \bibnamefont {Tegze}}, \bibinfo {author} {\bibfnamefont {Tamas}\ \bibnamefont
  {Pusztai}}, \bibinfo {author} {\bibfnamefont {Gergely}\ \bibnamefont {Toth}},
  \ and\ \bibinfo {author} {\bibfnamefont {Laszlo}\ \bibnamefont {Granasy}},\
  }\bibfield  {title} {\enquote {\bibinfo {title} {Polymorphism, crystal
  nucleation and growth in the phase-field crystal model in 2{D} and 3{D}},}\
  }\href@noop {} {\bibfield  {journal} {\bibinfo  {journal} {J. Phys.: Condens.
  Matter}\ }\textbf {\bibinfo {volume} {{22}}},\ \bibinfo {pages} {{364101}}
  (\bibinfo {year} {{2010}})}\BibitemShut {NoStop}%
\bibitem [{\citenamefont {Archer}\ \emph {et~al.}({2012})\citenamefont
  {Archer}, \citenamefont {Robbins}, \citenamefont {Thiele},\ and\
  \citenamefont {Knobloch}}]{Archer2012}%
  \BibitemOpen
  \bibfield  {author} {\bibinfo {author} {\bibfnamefont {A.~J.}\ \bibnamefont
  {Archer}}, \bibinfo {author} {\bibfnamefont {M.~J.}\ \bibnamefont {Robbins}},
  \bibinfo {author} {\bibfnamefont {U.}~\bibnamefont {Thiele}}, \ and\ \bibinfo
  {author} {\bibfnamefont {E.}~\bibnamefont {Knobloch}},\ }\bibfield  {title}
  {\enquote {\bibinfo {title} {Solidification fronts in supercooled liquids:
  {H}ow rapid fronts can lead to disordered glassy solids},}\ }\href {\doibase
  {10.1103/PhysRevE.86.031603}} {\bibfield  {journal} {\bibinfo  {journal}
  {Phys. Rev. E}\ }\textbf {\bibinfo {volume} {{86}}},\ \bibinfo {pages}
  {031603} (\bibinfo {year} {{2012}})}\BibitemShut {NoStop}%
\bibitem [{\citenamefont {Berry}\ \emph {et~al.}(2008)\citenamefont {Berry},
  \citenamefont {Elder},\ and\ \citenamefont {Grant}}]{Berry2008a}%
  \BibitemOpen
  \bibfield  {author} {\bibinfo {author} {\bibfnamefont {J.}~\bibnamefont
  {Berry}}, \bibinfo {author} {\bibfnamefont {K.~R.}\ \bibnamefont {Elder}}, \
  and\ \bibinfo {author} {\bibfnamefont {M.}~\bibnamefont {Grant}},\ }\bibfield
   {title} {\enquote {\bibinfo {title} {Simulation of an atomistic dynamic
  field theory for monatomic liquids: Freezing and glass formation},}\
  }\href@noop {} {\bibfield  {journal} {\bibinfo  {journal} {Phys. Rev. E}\
  }\textbf {\bibinfo {volume} {77}},\ \bibinfo {pages} {061506} (\bibinfo
  {year} {2008})}\BibitemShut {NoStop}%
\bibitem [{\citenamefont {Emmerich}\ \emph {et~al.}(2012)\citenamefont
  {Emmerich}, \citenamefont {L{\"o}wen}, \citenamefont {Wittkowski},
  \citenamefont {Gruhn}, \citenamefont {T{\'o}th}, \citenamefont {Tegze},\ and\
  \citenamefont {Gr{\'a}n{\'a}sy}}]{Emmerich2012}%
  \BibitemOpen
  \bibfield  {author} {\bibinfo {author} {\bibfnamefont {Heike}\ \bibnamefont
  {Emmerich}}, \bibinfo {author} {\bibfnamefont {Hartmut}\ \bibnamefont
  {L{\"o}wen}}, \bibinfo {author} {\bibfnamefont {Raphael}\ \bibnamefont
  {Wittkowski}}, \bibinfo {author} {\bibfnamefont {Thomas}\ \bibnamefont
  {Gruhn}}, \bibinfo {author} {\bibfnamefont {Gyula~I.}\ \bibnamefont
  {T{\'o}th}}, \bibinfo {author} {\bibfnamefont {Gy{\"o}rgy}\ \bibnamefont
  {Tegze}}, \ and\ \bibinfo {author} {\bibfnamefont {L{\'a}szl{\'o}}\
  \bibnamefont {Gr{\'a}n{\'a}sy}},\ }\bibfield  {title} {\enquote {\bibinfo
  {title} {Phase-field-crystal models for condensed matter dynamics on atomic
  length and diffusive time scales: an overview},}\ }\href {\doibase
  10.1080/00018732.2012.737555} {\bibfield  {journal} {\bibinfo  {journal}
  {Adv. Phys.}\ }\textbf {\bibinfo {volume} {61}},\ \bibinfo {pages} {665--743}
  (\bibinfo {year} {2012})}\BibitemShut {NoStop}%
\bibitem [{\citenamefont {Boettinger}\ \emph {et~al.}(2002)\citenamefont
  {Boettinger}, \citenamefont {Warren}, \citenamefont {Beckermann},\ and\
  \citenamefont {Karma}}]{Boettinger2002}%
  \BibitemOpen
  \bibfield  {author} {\bibinfo {author} {\bibfnamefont {William~J.}\
  \bibnamefont {Boettinger}}, \bibinfo {author} {\bibfnamefont {James~A.}\
  \bibnamefont {Warren}}, \bibinfo {author} {\bibfnamefont {Christoph}\
  \bibnamefont {Beckermann}}, \ and\ \bibinfo {author} {\bibfnamefont {Alain}\
  \bibnamefont {Karma}},\ }\bibfield  {title} {\enquote {\bibinfo {title}
  {Phase-field simulation of solidification},}\ }\href@noop {} {\bibfield
  {journal} {\bibinfo  {journal} {Annu. Rev. Mater. Res.}\ }\textbf {\bibinfo
  {volume} {32}},\ \bibinfo {pages} {163--194} (\bibinfo {year}
  {2002})}\BibitemShut {NoStop}%
\bibitem [{\citenamefont {Jaatinen}\ \emph {et~al.}(2009)\citenamefont
  {Jaatinen}, \citenamefont {Achim}, \citenamefont {Elder},\ and\ \citenamefont
  {Ala-Nissila}}]{Jaatinen2009}%
  \BibitemOpen
  \bibfield  {author} {\bibinfo {author} {\bibfnamefont {A.}~\bibnamefont
  {Jaatinen}}, \bibinfo {author} {\bibfnamefont {C.~V.}\ \bibnamefont {Achim}},
  \bibinfo {author} {\bibfnamefont {K.~R.}\ \bibnamefont {Elder}}, \ and\
  \bibinfo {author} {\bibfnamefont {T.}~\bibnamefont {Ala-Nissila}},\
  }\bibfield  {title} {\enquote {\bibinfo {title} {Thermodynamics of bcc metals
  in phase-field-crystal models},}\ }\href@noop {} {\bibfield  {journal}
  {\bibinfo  {journal} {Phys. Rev. E}\ }\textbf {\bibinfo {volume} {80}},\
  \bibinfo {pages} {031602} (\bibinfo {year} {2009})}\BibitemShut {NoStop}%
\bibitem [{\citenamefont {Pisutha-Arnond}\ \emph {et~al.}({2013})\citenamefont
  {Pisutha-Arnond}, \citenamefont {Chan}, \citenamefont {Iyer}, \citenamefont
  {Gavini},\ and\ \citenamefont {Thornton}}]{Pisutha-Arnond2013b}%
  \BibitemOpen
  \bibfield  {author} {\bibinfo {author} {\bibfnamefont {N.}~\bibnamefont
  {Pisutha-Arnond}}, \bibinfo {author} {\bibfnamefont {V.~W.~L.}\ \bibnamefont
  {Chan}}, \bibinfo {author} {\bibfnamefont {M.}~\bibnamefont {Iyer}}, \bibinfo
  {author} {\bibfnamefont {V.}~\bibnamefont {Gavini}}, \ and\ \bibinfo {author}
  {\bibfnamefont {K.}~\bibnamefont {Thornton}},\ }\bibfield  {title} {\enquote
  {\bibinfo {title} {{Classical density functional theory and the phase-field
  crystal method using a rational function to describe the two-body direct
  correlation function}},}\ }\href {\doibase {10.1103/PhysRevE.87.013313}}
  {\bibfield  {journal} {\bibinfo  {journal} {{Phys. Rev. E}}\ }\textbf
  {\bibinfo {volume} {{87}}},\ \bibinfo {pages} {{013313}} (\bibinfo {year}
  {{2013}})}\BibitemShut {NoStop}%
\bibitem [{\citenamefont {Wu}\ \emph {et~al.}(2010)\citenamefont {Wu},
  \citenamefont {Adland},\ and\ \citenamefont {Karma}}]{Wu2010}%
  \BibitemOpen
  \bibfield  {author} {\bibinfo {author} {\bibfnamefont {Kuo-An}\ \bibnamefont
  {Wu}}, \bibinfo {author} {\bibfnamefont {Ari}\ \bibnamefont {Adland}}, \ and\
  \bibinfo {author} {\bibfnamefont {Alain}\ \bibnamefont {Karma}},\ }\bibfield
  {title} {\enquote {\bibinfo {title} {Phase-field-crystal model for fcc
  ordering},}\ }\href {\doibase 10.1103/PhysRevE.81.061601} {\bibfield
  {journal} {\bibinfo  {journal} {Phys. Rev. E}\ }\textbf {\bibinfo {volume}
  {81}},\ \bibinfo {pages} {061601} (\bibinfo {year} {2010})}\BibitemShut
  {NoStop}%
\bibitem [{\citenamefont {Barkan}\ \emph {et~al.}(2011)\citenamefont {Barkan},
  \citenamefont {Diamant},\ and\ \citenamefont {Lifshitz}}]{Barkan2011}%
  \BibitemOpen
  \bibfield  {author} {\bibinfo {author} {\bibfnamefont {Kobi}\ \bibnamefont
  {Barkan}}, \bibinfo {author} {\bibfnamefont {Haim}\ \bibnamefont {Diamant}},
  \ and\ \bibinfo {author} {\bibfnamefont {Ron}\ \bibnamefont {Lifshitz}},\
  }\bibfield  {title} {\enquote {\bibinfo {title} {Stability of quasicrystals
  composed of soft isotropic particles},}\ }\href@noop {} {\bibfield  {journal}
  {\bibinfo  {journal} {Phys. Rev. B}\ }\textbf {\bibinfo {volume} {83}},\
  \bibinfo {pages} {172201} (\bibinfo {year} {2011})}\BibitemShut {NoStop}%
\bibitem [{\citenamefont {Achim}\ \emph {et~al.}(2014)\citenamefont {Achim},
  \citenamefont {Schmiedeberg},\ and\ \citenamefont {L\"owen}}]{Achim2014}%
  \BibitemOpen
  \bibfield  {author} {\bibinfo {author} {\bibfnamefont {C.~V.}\ \bibnamefont
  {Achim}}, \bibinfo {author} {\bibfnamefont {M.}~\bibnamefont {Schmiedeberg}},
  \ and\ \bibinfo {author} {\bibfnamefont {H.}~\bibnamefont {L\"owen}},\
  }\bibfield  {title} {\enquote {\bibinfo {title} {Growth modes of
  quasicrystals},}\ }\href {\doibase 10.1103/PhysRevLett.112.255501} {\bibfield
   {journal} {\bibinfo  {journal} {Phys. Rev. Lett.}\ }\textbf {\bibinfo
  {volume} {112}},\ \bibinfo {pages} {255501} (\bibinfo {year}
  {2014})}\BibitemShut {NoStop}%
\bibitem [{\citenamefont {Subramanian}\ \emph {et~al.}(2016)\citenamefont
  {Subramanian}, \citenamefont {Archer}, \citenamefont {Knobloch},\ and\
  \citenamefont {Rucklidge}}]{Subramanian2016}%
  \BibitemOpen
  \bibfield  {author} {\bibinfo {author} {\bibfnamefont {P.}~\bibnamefont
  {Subramanian}}, \bibinfo {author} {\bibfnamefont {A.~J.}\ \bibnamefont
  {Archer}}, \bibinfo {author} {\bibfnamefont {E.}~\bibnamefont {Knobloch}}, \
  and\ \bibinfo {author} {\bibfnamefont {A.~M.}\ \bibnamefont {Rucklidge}},\
  }\bibfield  {title} {\enquote {\bibinfo {title} {Three-dimensional
  icosahedral phase field quasicrystal},}\ }\href {\doibase
  10.1103/PhysRevLett.117.075501} {\bibfield  {journal} {\bibinfo  {journal}
  {Phys. Rev. Lett.}\ }\textbf {\bibinfo {volume} {117}},\ \bibinfo {pages}
  {075501} (\bibinfo {year} {2016})}\BibitemShut {NoStop}%
\bibitem [{\citenamefont {Jiang}\ \emph {et~al.}(2017)\citenamefont {Jiang},
  \citenamefont {Zhang},\ and\ \citenamefont {Shi}}]{Jiang2017}%
  \BibitemOpen
  \bibfield  {author} {\bibinfo {author} {\bibfnamefont {Kai}\ \bibnamefont
  {Jiang}}, \bibinfo {author} {\bibfnamefont {Pingwen}\ \bibnamefont {Zhang}},
  \ and\ \bibinfo {author} {\bibfnamefont {An-Chang}\ \bibnamefont {Shi}},\
  }\bibfield  {title} {\enquote {\bibinfo {title} {Stability of icosahedral
  quasicrystals in a simple model with two-length scales},}\ }\href@noop {}
  {\bibfield  {journal} {\bibinfo  {journal} {J. Phys.: Condens. Matter}\
  }\textbf {\bibinfo {volume} {29}},\ \bibinfo {pages} {124003} (\bibinfo
  {year} {2017})}\BibitemShut {NoStop}%
\bibitem [{\citenamefont {Savitz}\ \emph {et~al.}(2018)\citenamefont {Savitz},
  \citenamefont {Babadi},\ and\ \citenamefont {Lifshitz}}]{Savitz2018}%
  \BibitemOpen
  \bibfield  {author} {\bibinfo {author} {\bibfnamefont {Samuel}\ \bibnamefont
  {Savitz}}, \bibinfo {author} {\bibfnamefont {Mehrtash}\ \bibnamefont
  {Babadi}}, \ and\ \bibinfo {author} {\bibfnamefont {Ron}\ \bibnamefont
  {Lifshitz}},\ }\bibfield  {title} {\enquote {\bibinfo {title}
  {{Multiple-scale structures: from Faraday waves to soft-matter
  quasicrystals}},}\ }\href {\doibase 10.1107/S2052252518001161} {\bibfield
  {journal} {\bibinfo  {journal} {IUCrJ}\ }\textbf {\bibinfo {volume} {5}},\
  \bibinfo {pages} {247--268} (\bibinfo {year} {2018})}\BibitemShut {NoStop}%
\bibitem [{\citenamefont {Swift}\ and\ \citenamefont
  {Hohenberg}(1977)}]{Swift1977}%
  \BibitemOpen
  \bibfield  {author} {\bibinfo {author} {\bibfnamefont {J.}~\bibnamefont
  {Swift}}\ and\ \bibinfo {author} {\bibfnamefont {P.~C.}\ \bibnamefont
  {Hohenberg}},\ }\bibfield  {title} {\enquote {\bibinfo {title} {Hydrodynamic
  fluctuations at the convective instability},}\ }\href@noop {} {\bibfield
  {journal} {\bibinfo  {journal} {Phys. Rev. A}\ }\textbf {\bibinfo {volume}
  {15}},\ \bibinfo {pages} {319--328} (\bibinfo {year} {1977})}\BibitemShut
  {NoStop}%
\bibitem [{\citenamefont {Thiele}\ \emph {et~al.}(2013)\citenamefont {Thiele},
  \citenamefont {Archer}, \citenamefont {Robbins}, \citenamefont {Gomez},\ and\
  \citenamefont {Knobloch}}]{Thiele2013}%
  \BibitemOpen
  \bibfield  {author} {\bibinfo {author} {\bibfnamefont {U.}~\bibnamefont
  {Thiele}}, \bibinfo {author} {\bibfnamefont {A.~J.}\ \bibnamefont {Archer}},
  \bibinfo {author} {\bibfnamefont {M.~J.}\ \bibnamefont {Robbins}}, \bibinfo
  {author} {\bibfnamefont {H.}~\bibnamefont {Gomez}}, \ and\ \bibinfo {author}
  {\bibfnamefont {E.}~\bibnamefont {Knobloch}},\ }\bibfield  {title} {\enquote
  {\bibinfo {title} {Localized states in the conserved {S}wift--{H}ohenberg
  equation with cubic nonlinearity},}\ }\href {\doibase
  10.1103/PhysRevE.87.042915} {\bibfield  {journal} {\bibinfo  {journal} {Phys.
  Rev. E}\ }\textbf {\bibinfo {volume} {87}},\ \bibinfo {pages} {042915}
  (\bibinfo {year} {2013})}\BibitemShut {NoStop}%
\bibitem [{\citenamefont {Sagui}\ and\ \citenamefont
  {Desai}(1994)}]{Sagui1994}%
  \BibitemOpen
  \bibfield  {author} {\bibinfo {author} {\bibfnamefont {Celeste}\ \bibnamefont
  {Sagui}}\ and\ \bibinfo {author} {\bibfnamefont {Rashmi~C.}\ \bibnamefont
  {Desai}},\ }\bibfield  {title} {\enquote {\bibinfo {title} {Kinetics of phase
  separation in two-dimensional systems with competing interactions},}\ }\href
  {\doibase 10.1103/PhysRevE.49.2225} {\bibfield  {journal} {\bibinfo
  {journal} {Phys. Rev. E}\ }\textbf {\bibinfo {volume} {49}},\ \bibinfo
  {pages} {2225--2244} (\bibinfo {year} {1994})}\BibitemShut {NoStop}%
\bibitem [{\citenamefont {Knobloch}(2015)}]{Knobloch2015}%
  \BibitemOpen
  \bibfield  {author} {\bibinfo {author} {\bibfnamefont {E.}~\bibnamefont
  {Knobloch}},\ }\bibfield  {title} {\enquote {\bibinfo {title} {Spatial
  localization in dissipative systems},}\ }\href {\doibase
  10.1146/annurev-conmatphys-031214-014514} {\bibfield  {journal} {\bibinfo
  {journal} {Annu. Rev. Condens. Matter Phys.}\ }\textbf {\bibinfo {volume}
  {6}},\ \bibinfo {pages} {325--359} (\bibinfo {year} {2015})}\BibitemShut
  {NoStop}%
\bibitem [{\citenamefont {Matthews}\ and\ \citenamefont
  {Cox}(2000)}]{Matthews2000}%
  \BibitemOpen
  \bibfield  {author} {\bibinfo {author} {\bibfnamefont {P.~C.}\ \bibnamefont
  {Matthews}}\ and\ \bibinfo {author} {\bibfnamefont {S.~M.}\ \bibnamefont
  {Cox}},\ }\bibfield  {title} {\enquote {\bibinfo {title} {Pattern formation
  with a conservation law},}\ }\href@noop {} {\bibfield  {journal} {\bibinfo
  {journal} {Nonlinearity}\ }\textbf {\bibinfo {volume} {13}},\ \bibinfo
  {pages} {1293--1320} (\bibinfo {year} {2000})}\BibitemShut {NoStop}%
\bibitem [{\citenamefont {Elder}\ \emph {et~al.}(2007)\citenamefont {Elder},
  \citenamefont {Provatas}, \citenamefont {Berry}, \citenamefont {Stefanovic},\
  and\ \citenamefont {Grant}}]{Elder2007}%
  \BibitemOpen
  \bibfield  {author} {\bibinfo {author} {\bibfnamefont {K.~R.}\ \bibnamefont
  {Elder}}, \bibinfo {author} {\bibfnamefont {N.}~\bibnamefont {Provatas}},
  \bibinfo {author} {\bibfnamefont {J.}~\bibnamefont {Berry}}, \bibinfo
  {author} {\bibfnamefont {P.}~\bibnamefont {Stefanovic}}, \ and\ \bibinfo
  {author} {\bibfnamefont {M.}~\bibnamefont {Grant}},\ }\bibfield  {title}
  {\enquote {\bibinfo {title} {Phase-field crystal modeling and classical
  density functional theory of freezing},}\ }\href@noop {} {\bibfield
  {journal} {\bibinfo  {journal} {Phys. Rev. B}\ }\textbf {\bibinfo {volume}
  {75}},\ \bibinfo {pages} {064107} (\bibinfo {year} {2007})}\BibitemShut
  {NoStop}%
\bibitem [{\citenamefont {van Teeffelen}\ \emph {et~al.}(2009)\citenamefont
  {van Teeffelen}, \citenamefont {Backofen}, \citenamefont {Voigt},\ and\
  \citenamefont {Lowen}}]{Teeffelen2009}%
  \BibitemOpen
  \bibfield  {author} {\bibinfo {author} {\bibfnamefont {S.}~\bibnamefont {van
  Teeffelen}}, \bibinfo {author} {\bibfnamefont {R.}~\bibnamefont {Backofen}},
  \bibinfo {author} {\bibfnamefont {A.}~\bibnamefont {Voigt}}, \ and\ \bibinfo
  {author} {\bibfnamefont {H.}~\bibnamefont {Lowen}},\ }\bibfield  {title}
  {\enquote {\bibinfo {title} {Derivation of the phase-field-crystal model for
  colloidal solidification},}\ }\href@noop {} {\bibfield  {journal} {\bibinfo
  {journal} {Phys. Rev. E}\ }\textbf {\bibinfo {volume} {79}},\ \bibinfo
  {pages} {051404} (\bibinfo {year} {2009})}\BibitemShut {NoStop}%
\bibitem [{\citenamefont {Huang}\ \emph {et~al.}(2010)\citenamefont {Huang},
  \citenamefont {Elder},\ and\ \citenamefont {Provatas}}]{Huang2010a}%
  \BibitemOpen
  \bibfield  {author} {\bibinfo {author} {\bibfnamefont {Z.~F.}\ \bibnamefont
  {Huang}}, \bibinfo {author} {\bibfnamefont {K.~R.}\ \bibnamefont {Elder}}, \
  and\ \bibinfo {author} {\bibfnamefont {N.}~\bibnamefont {Provatas}},\
  }\bibfield  {title} {\enquote {\bibinfo {title} {Phase-field-crystal dynamics
  for binary systems: Derivation from dynamical density functional theory,
  amplitude equation formalism, and applications to alloy heterostructures},}\
  }\href@noop {} {\bibfield  {journal} {\bibinfo  {journal} {Phys. Rev. E}\
  }\textbf {\bibinfo {volume} {82}},\ \bibinfo {pages} {021605} (\bibinfo
  {year} {2010})}\BibitemShut {NoStop}%
\bibitem [{\citenamefont {Evans}(1979)}]{Evans1979a}%
  \BibitemOpen
  \bibfield  {author} {\bibinfo {author} {\bibfnamefont {R.}~\bibnamefont
  {Evans}},\ }\bibfield  {title} {\enquote {\bibinfo {title} {The nature of the
  liquid-vapour interface and other topics in the statistical mechanics of
  non-uniform, classical fluids},}\ }\href {\doibase 10.1080/00018737900101365}
  {\bibfield  {journal} {\bibinfo  {journal} {Adv. Phys.}\ }\textbf {\bibinfo
  {volume} {28}},\ \bibinfo {pages} {143--200} (\bibinfo {year}
  {1979})}\BibitemShut {NoStop}%
\bibitem [{\citenamefont {Evans}(1992)}]{Evans1992}%
  \BibitemOpen
  \bibfield  {author} {\bibinfo {author} {\bibfnamefont {R.}~\bibnamefont
  {Evans}},\ }\bibfield  {title} {\enquote {\bibinfo {title} {Density
  functionals in the theory of non-uniform fluids},}\ }in\ \href@noop {} {\emph
  {\bibinfo {booktitle} {Fundamentals of Inhomogeneous Fluids}}},\ \bibinfo
  {editor} {edited by\ \bibinfo {editor} {\bibfnamefont {D.}~\bibnamefont
  {Henderson}}}\ (\bibinfo  {publisher} {Marcel Dekker},\ \bibinfo {address}
  {New York},\ \bibinfo {year} {1992})\ Chap.~\bibinfo {chapter} {3}, pp.\
  \bibinfo {pages} {85--175}\BibitemShut {NoStop}%
\bibitem [{\citenamefont {Hansen}\ and\ \citenamefont
  {McDonald}(1992)}]{Hansen2013}%
  \BibitemOpen
  \bibfield  {author} {\bibinfo {author} {\bibfnamefont {J.-P.}\ \bibnamefont
  {Hansen}}\ and\ \bibinfo {author} {\bibfnamefont {I.~R.}\ \bibnamefont
  {McDonald}},\ }\href@noop {} {\emph {\bibinfo {title} {Theory of Simple
  Liquids with Applications to Soft Matter: Fourth Edition}}}\ (\bibinfo
  {publisher} {Elsevier},\ \bibinfo {address} {Oxford},\ \bibinfo {year}
  {1992})\BibitemShut {NoStop}%
\bibitem [{\citenamefont {Marconi}\ and\ \citenamefont
  {Tarazona}(1999)}]{Marconi1999}%
  \BibitemOpen
  \bibfield  {author} {\bibinfo {author} {\bibfnamefont {U.~Marini~Bettolo}\
  \bibnamefont {Marconi}}\ and\ \bibinfo {author} {\bibfnamefont
  {P.}~\bibnamefont {Tarazona}},\ }\bibfield  {title} {\enquote {\bibinfo
  {title} {Dynamic density functional theory of fluids},}\ }\href@noop {}
  {\bibfield  {journal} {\bibinfo  {journal} {J. Chem. Phys.}\ }\textbf
  {\bibinfo {volume} {110}},\ \bibinfo {pages} {8032} (\bibinfo {year}
  {1999})}\BibitemShut {NoStop}%
\bibitem [{\citenamefont {Marconi}\ and\ \citenamefont
  {Tarazona}(2000)}]{Marconi2000}%
  \BibitemOpen
  \bibfield  {author} {\bibinfo {author} {\bibfnamefont {U.~Marini~Bettolo}\
  \bibnamefont {Marconi}}\ and\ \bibinfo {author} {\bibfnamefont
  {P.}~\bibnamefont {Tarazona}},\ }\bibfield  {title} {\enquote {\bibinfo
  {title} {Dynamic density functional theory of fluids},}\ }\href@noop {}
  {\bibfield  {journal} {\bibinfo  {journal} {J. Phys.: Condens. Matter}\
  }\textbf {\bibinfo {volume} {12}},\ \bibinfo {pages} {A413} (\bibinfo {year}
  {2000})}\BibitemShut {NoStop}%
\bibitem [{\citenamefont {Archer}\ and\ \citenamefont
  {Rauscher}(2004)}]{Archer2004}%
  \BibitemOpen
  \bibfield  {author} {\bibinfo {author} {\bibfnamefont {Andrew~J.}\
  \bibnamefont {Archer}}\ and\ \bibinfo {author} {\bibfnamefont {Markus}\
  \bibnamefont {Rauscher}},\ }\bibfield  {title} {\enquote {\bibinfo {title}
  {Dynamical density functional theory for interacting brownian particles:
  stochastic or deterministic?}}\ }\href@noop {} {\bibfield  {journal}
  {\bibinfo  {journal} {J. Phys. A: Math. Gen.}\ }\textbf {\bibinfo {volume}
  {37}},\ \bibinfo {pages} {9325--9333} (\bibinfo {year} {2004})}\BibitemShut
  {NoStop}%
\bibitem [{\citenamefont {Archer}\ and\ \citenamefont
  {Evans}(2004)}]{Archer2004a}%
  \BibitemOpen
  \bibfield  {author} {\bibinfo {author} {\bibfnamefont {Andrew~J.}\
  \bibnamefont {Archer}}\ and\ \bibinfo {author} {\bibfnamefont {Robert}\
  \bibnamefont {Evans}},\ }\bibfield  {title} {\enquote {\bibinfo {title}
  {Dynamical density functional theory and its application to spinodal
  decomposition},}\ }\href {\doibase 10.1063/1.1778374} {\bibfield  {journal}
  {\bibinfo  {journal} {J. Chem. Phys.}\ }\textbf {\bibinfo {volume} {121}},\
  \bibinfo {pages} {4246--4254} (\bibinfo {year} {2004})}\BibitemShut {NoStop}%
\bibitem [{\citenamefont {Archer}(2006)}]{Archer2006}%
  \BibitemOpen
  \bibfield  {author} {\bibinfo {author} {\bibfnamefont {Andrew~J.}\
  \bibnamefont {Archer}},\ }\bibfield  {title} {\enquote {\bibinfo {title}
  {Dynamical density functional theory for dense atomic liquids},}\ }\href@noop
  {} {\bibfield  {journal} {\bibinfo  {journal} {J. Phys.: Condens. Matter}\
  }\textbf {\bibinfo {volume} {18}},\ \bibinfo {pages} {5617} (\bibinfo {year}
  {2006})}\BibitemShut {NoStop}%
\bibitem [{\citenamefont {Archer}(2009)}]{Archer2009}%
  \BibitemOpen
  \bibfield  {author} {\bibinfo {author} {\bibfnamefont {Andrew~J.}\
  \bibnamefont {Archer}},\ }\bibfield  {title} {\enquote {\bibinfo {title}
  {Dynamical density functional theory for molecular and colloidal fluids: A
  microscopic approach to fluid mechanics},}\ }\href@noop {} {\bibfield
  {journal} {\bibinfo  {journal} {J. Chem. Phys.}\ }\textbf {\bibinfo {volume}
  {130}},\ \bibinfo {pages} {014509} (\bibinfo {year} {2009})}\BibitemShut
  {NoStop}%
\bibitem [{\citenamefont {Goddard}\ \emph {et~al.}(2012)\citenamefont
  {Goddard}, \citenamefont {Pavliotis},\ and\ \citenamefont
  {Kalliadasis}}]{Goddard2012}%
  \BibitemOpen
  \bibfield  {author} {\bibinfo {author} {\bibfnamefont {Benjamin~D.}\
  \bibnamefont {Goddard}}, \bibinfo {author} {\bibfnamefont {Grigorios~A.}\
  \bibnamefont {Pavliotis}}, \ and\ \bibinfo {author} {\bibfnamefont {Serafim}\
  \bibnamefont {Kalliadasis}},\ }\bibfield  {title} {\enquote {\bibinfo {title}
  {The overdamped limit of dynamic density functional theory: Rigorous
  results},}\ }\href@noop {} {\bibfield  {journal} {\bibinfo  {journal}
  {Multiscale Modeling \& Simulation}\ }\textbf {\bibinfo {volume} {10}},\
  \bibinfo {pages} {633--663} (\bibinfo {year} {2012})}\BibitemShut {NoStop}%
\bibitem [{\citenamefont {Goddard}\ \emph {et~al.}(2013)\citenamefont
  {Goddard}, \citenamefont {Nold}, \citenamefont {Savva}, \citenamefont
  {Yatsyshin},\ and\ \citenamefont {Kalliadasis}}]{Goddard2013}%
  \BibitemOpen
  \bibfield  {author} {\bibinfo {author} {\bibfnamefont {B.~D.}~\bibnamefont
  {Goddard}}, \bibinfo {author} {\bibfnamefont {A.}~\bibnamefont {Nold}},
  \bibinfo {author} {\bibfnamefont {N.}~\bibnamefont {Savva}}, \bibinfo {author}
  {\bibfnamefont {P.}~\bibnamefont {Yatsyshin}}, \ and\ \bibinfo {author}
  {\bibfnamefont {S.}~\bibnamefont {Kalliadasis}},\ }\bibfield  {title}
  {\enquote {\bibinfo {title} {Unification of dynamic density functional theory
  for colloidal fluids to include inertia and hydrodynamic interactions:
  derivation and numerical experiments},}\ }\href@noop {} {\bibfield  {journal}
  {\bibinfo  {journal} {J. Phys.: Condens. Matter}\ }\textbf {\bibinfo {volume}
  {25}},\ \bibinfo {pages} {035101} (\bibinfo {year} {2013})}\BibitemShut
  {NoStop}%
\bibitem [{\citenamefont {Dur{\'a}n-Olivencia}\ \emph
  {et~al.}(2017)\citenamefont {Dur{\'a}n-Olivencia}, \citenamefont {Yatsyshin},
  \citenamefont {Goddard},\ and\ \citenamefont
  {Kalliadasis}}]{DuranOlivencia2017}%
  \BibitemOpen
  \bibfield  {author} {\bibinfo {author} {\bibfnamefont {Miguel~A.}\
  \bibnamefont {Dur{\'a}n-Olivencia}}, \bibinfo {author} {\bibfnamefont
  {Peter}\ \bibnamefont {Yatsyshin}}, \bibinfo {author} {\bibfnamefont
  {Benjamin~D.}\ \bibnamefont {Goddard}}, \ and\ \bibinfo {author}
  {\bibfnamefont {Serafim}\ \bibnamefont {Kalliadasis}},\ }\bibfield  {title}
  {\enquote {\bibinfo {title} {General framework for fluctuating dynamic
  density functional theory},}\ }\href@noop {} {\bibfield  {journal} {\bibinfo
  {journal} {New J. Phys.}\ }\textbf {\bibinfo {volume} {19}},\ \bibinfo
  {pages} {123022} (\bibinfo {year} {2017})}\BibitemShut {NoStop}%
\bibitem [{\citenamefont {Schmidt}(2018)}]{Schmidt2018}%
  \BibitemOpen
  \bibfield  {author} {\bibinfo {author} {\bibfnamefont {Matthias}\
  \bibnamefont {Schmidt}},\ }\bibfield  {title} {\enquote {\bibinfo {title}
  {Power functional theory for newtonian many-body dynamics},}\ }\href@noop {}
  {\bibfield  {journal} {\bibinfo  {journal} {J. Chem. Phys.}\ }\textbf
  {\bibinfo {volume} {148}},\ \bibinfo {pages} {044502} (\bibinfo {year}
  {2018})}\BibitemShut {NoStop}%
\bibitem [{\citenamefont {Imperio}\ and\ \citenamefont
  {Reatto}(2004)}]{Imperio2004}%
  \BibitemOpen
  \bibfield  {author} {\bibinfo {author} {\bibfnamefont {A.}~\bibnamefont
  {Imperio}}\ and\ \bibinfo {author} {\bibfnamefont {L.}~\bibnamefont
  {Reatto}},\ }\bibfield  {title} {\enquote {\bibinfo {title} {A bidimensional
  fluid system with competing interactions: spontaneous and induced pattern
  formation},}\ }\href@noop {} {\bibfield  {journal} {\bibinfo  {journal} {J.
  Phys.: Condens. Matter}\ }\textbf {\bibinfo {volume} {16}},\ \bibinfo {pages}
  {S3769} (\bibinfo {year} {2004})}\BibitemShut {NoStop}%
\bibitem [{\citenamefont {Imperio}\ and\ \citenamefont
  {Reatto}(2006)}]{Imperio2006}%
  \BibitemOpen
  \bibfield  {author} {\bibinfo {author} {\bibfnamefont {A.}~\bibnamefont
  {Imperio}}\ and\ \bibinfo {author} {\bibfnamefont {L.}~\bibnamefont
  {Reatto}},\ }\bibfield  {title} {\enquote {\bibinfo {title} {Microphase
  separation in two-dimensional systems with competing interactions},}\
  }\href@noop {} {\bibfield  {journal} {\bibinfo  {journal} {J. Chem. Phys.}\
  }\textbf {\bibinfo {volume} {124}},\ \bibinfo {pages} {164712} (\bibinfo
  {year} {2006})}\BibitemShut {NoStop}%
\bibitem [{\citenamefont {Archer}\ and\ \citenamefont
  {Wilding}(2007)}]{Archer2007}%
  \BibitemOpen
  \bibfield  {author} {\bibinfo {author} {\bibfnamefont {Andrew~J.}\
  \bibnamefont {Archer}}\ and\ \bibinfo {author} {\bibfnamefont {Nigel~B.}\
  \bibnamefont {Wilding}},\ }\bibfield  {title} {\enquote {\bibinfo {title}
  {Phase behavior of a fluid with competing attractive and repulsive
  interactions},}\ }\href {\doibase 10.1103/PhysRevE.76.031501} {\bibfield
  {journal} {\bibinfo  {journal} {Phys. Rev. E}\ }\textbf {\bibinfo {volume}
  {76}},\ \bibinfo {pages} {031501} (\bibinfo {year} {2007})}\BibitemShut
  {NoStop}%
\bibitem [{\citenamefont {Mladek}\ \emph {et~al.}(2006)\citenamefont {Mladek},
  \citenamefont {Gottwald}, \citenamefont {Kahl}, \citenamefont {Neumann},\
  and\ \citenamefont {Likos}}]{Mladek2006}%
  \BibitemOpen
  \bibfield  {author} {\bibinfo {author} {\bibfnamefont {Bianca~M.}\
  \bibnamefont {Mladek}}, \bibinfo {author} {\bibfnamefont {Dieter}\
  \bibnamefont {Gottwald}}, \bibinfo {author} {\bibfnamefont {Gerhard}\
  \bibnamefont {Kahl}}, \bibinfo {author} {\bibfnamefont {Martin}\ \bibnamefont
  {Neumann}}, \ and\ \bibinfo {author} {\bibfnamefont {Christos~N.}\
  \bibnamefont {Likos}},\ }\bibfield  {title} {\enquote {\bibinfo {title}
  {Formation of polymorphic cluster phases for a class of models of purely
  repulsive soft spheres},}\ }\href {\doibase 10.1103/PhysRevLett.96.045701}
  {\bibfield  {journal} {\bibinfo  {journal} {Phys. Rev. Lett.}\ }\textbf
  {\bibinfo {volume} {96}},\ \bibinfo {pages} {045701} (\bibinfo {year}
  {2006})}\BibitemShut {NoStop}%
\bibitem [{\citenamefont {Prestipino}\ and\ \citenamefont
  {Saija}(2014)}]{Prestipino2014}%
  \BibitemOpen
  \bibfield  {author} {\bibinfo {author} {\bibfnamefont {Santi}\ \bibnamefont
  {Prestipino}}\ and\ \bibinfo {author} {\bibfnamefont {Franz}\ \bibnamefont
  {Saija}},\ }\bibfield  {title} {\enquote {\bibinfo {title} {Hexatic phase and
  cluster crystals of two-dimensional {GEM4} spheres},}\ }\href@noop {}
  {\bibfield  {journal} {\bibinfo  {journal} {J. Chem. Phys.}\ }\textbf
  {\bibinfo {volume} {141}},\ \bibinfo {pages} {184502} (\bibinfo {year}
  {2014})}\BibitemShut {NoStop}%
\bibitem [{\citenamefont {Likos}(2001)}]{Likos2001}%
  \BibitemOpen
  \bibfield  {author} {\bibinfo {author} {\bibfnamefont {C.~N.}\ \bibnamefont
  {Likos}},\ }\bibfield  {title} {\enquote {\bibinfo {title} {Effective
  interactions in soft condensed matter physics},}\ }\href@noop {} {\bibfield
  {journal} {\bibinfo  {journal} {Phys. Rep.}\ }\textbf {\bibinfo {volume}
  {348}},\ \bibinfo {pages} {267--439} (\bibinfo {year} {2001})}\BibitemShut
  {NoStop}%
\bibitem [{\citenamefont {Mladek}\ \emph {et~al.}(2007)\citenamefont {Mladek},
  \citenamefont {Gottwald}, \citenamefont {Kahl}, \citenamefont {Neumann},\
  and\ \citenamefont {Likos}}]{Mladek2007}%
  \BibitemOpen
  \bibfield  {author} {\bibinfo {author} {\bibfnamefont {Bianca~M.}\
  \bibnamefont {Mladek}}, \bibinfo {author} {\bibfnamefont {Dieter}\
  \bibnamefont {Gottwald}}, \bibinfo {author} {\bibfnamefont {Gerhard}\
  \bibnamefont {Kahl}}, \bibinfo {author} {\bibfnamefont {Martin}\ \bibnamefont
  {Neumann}}, \ and\ \bibinfo {author} {\bibfnamefont {Christos~N.}\
  \bibnamefont {Likos}},\ }\bibfield  {title} {\enquote {\bibinfo {title}
  {Clustering in the absence of attractions: Density functional theory and
  computer simulations},}\ }\href@noop {} {\bibfield  {journal} {\bibinfo
  {journal} {J Phys. Chem. B}\ }\textbf {\bibinfo {volume} {111}},\ \bibinfo
  {pages} {12799--12808} (\bibinfo {year} {2007})}\BibitemShut {NoStop}%
\bibitem [{\citenamefont {Archer}\ \emph {et~al.}(2014)\citenamefont {Archer},
  \citenamefont {Walters}, \citenamefont {Thiele},\ and\ \citenamefont
  {Knobloch}}]{Archer2014}%
  \BibitemOpen
  \bibfield  {author} {\bibinfo {author} {\bibfnamefont {Andrew~J.}\
  \bibnamefont {Archer}}, \bibinfo {author} {\bibfnamefont {Morgan~C.}\
  \bibnamefont {Walters}}, \bibinfo {author} {\bibfnamefont {Uwe}\ \bibnamefont
  {Thiele}}, \ and\ \bibinfo {author} {\bibfnamefont {Edgar}\ \bibnamefont
  {Knobloch}},\ }\bibfield  {title} {\enquote {\bibinfo {title} {Solidification
  in soft-core fluids: Disordered solids from fast solidification fronts},}\
  }\href@noop {} {\bibfield  {journal} {\bibinfo  {journal} {Phys. Rev. E}\
  }\textbf {\bibinfo {volume} {90}},\ \bibinfo {pages} {042404} (\bibinfo
  {year} {2014})}\BibitemShut {NoStop}%
\bibitem [{\citenamefont {Wang}\ \emph {et~al.}({2018})\citenamefont {Wang},
  \citenamefont {Huang},\ and\ \citenamefont {Liu}}]{Wang2018b}%
  \BibitemOpen
  \bibfield  {author} {\bibinfo {author} {\bibfnamefont {Zi-Le}\ \bibnamefont
  {Wang}}, \bibinfo {author} {\bibfnamefont {Zhi-Feng}\ \bibnamefont {Huang}}, \
  and\ \bibinfo {author} {\bibfnamefont {Zhirong}\ \bibnamefont {Liu}},\
  }\bibfield  {title} {\enquote {\bibinfo {title} {{Elastic constants of 
  stressed and unstressed materials in the phase-field crystal model}},}\ }\href
  {\doibase {10.1103/PhysRevB.97.144112}} {\bibfield  {journal} {\bibinfo
  {journal} {Phys. Rev. B}\ }\textbf {\bibinfo {volume} {{97}}},\ \bibinfo
  {pages} {{144112}} (\bibinfo {year} {{2018}})}\BibitemShut {NoStop}%
\bibitem [{\citenamefont {Somerville}\ \emph {et~al.}(2018)\citenamefont
  {Somerville}, \citenamefont {Stokes}, \citenamefont {Adawi}, \citenamefont
  {Horozov}, \citenamefont {Archer},\ and\ \citenamefont
  {Buzza}}]{Somerville2018}%
  \BibitemOpen
  \bibfield  {author} {\bibinfo {author} {\bibfnamefont {W.~R.~C.}\
  \bibnamefont {Somerville}}, \bibinfo {author} {\bibfnamefont {J.~L.}\
  \bibnamefont {Stokes}}, \bibinfo {author} {\bibfnamefont {A.~M.}\
  \bibnamefont {Adawi}}, \bibinfo {author} {\bibfnamefont {T.~S.}\ \bibnamefont
  {Horozov}}, \bibinfo {author} {\bibfnamefont {Andrew~J.}\ \bibnamefont
  {Archer}}, \ and\ \bibinfo {author} {\bibfnamefont {D.~M.~A.}\ \bibnamefont
  {Buzza}},\ }\bibfield  {title} {\enquote {\bibinfo {title} {Density
  functional theory for the crystallization of two-dimensional dipolar
  colloidal alloys},}\ }\href@noop {} {\bibfield  {journal} {\bibinfo
  {journal} {J. Phys.: Condens. Matter}\ }\textbf {\bibinfo {volume} {30}},\
  \bibinfo {pages} {405102} (\bibinfo {year} {2018})}\BibitemShut {NoStop}%
\bibitem [{\citenamefont {Trudu}\ \emph {et~al.}(2006)\citenamefont {Trudu},
  \citenamefont {Donadio},\ and\ \citenamefont {Parrinello}}]{Trudu2006}%
  \BibitemOpen
  \bibfield  {author} {\bibinfo {author} {\bibfnamefont {Federica}\
  \bibnamefont {Trudu}}, \bibinfo {author} {\bibfnamefont {Davide}\
  \bibnamefont {Donadio}}, \ and\ \bibinfo {author} {\bibfnamefont {Michele}\
  \bibnamefont {Parrinello}},\ }\bibfield  {title} {\enquote {\bibinfo {title}
  {Freezing of a {L}ennard--{J}ones fluid: From nucleation to spinodal
  regime},}\ }\href {\doibase 10.1103/PhysRevLett.97.105701} {\bibfield
  {journal} {\bibinfo  {journal} {Phys. Rev. Lett.}\ }\textbf {\bibinfo
  {volume} {97}},\ \bibinfo {pages} {105701} (\bibinfo {year}
  {2006})}\BibitemShut {NoStop}%
\bibitem [{\citenamefont {Ramakrishnan}\ and\ \citenamefont
  {Yussouff}(1979)}]{Ramakrishnan1979}%
  \BibitemOpen
  \bibfield  {author} {\bibinfo {author} {\bibfnamefont {T.~V.}\ \bibnamefont
  {Ramakrishnan}}\ and\ \bibinfo {author} {\bibfnamefont {M.}~\bibnamefont
  {Yussouff}},\ }\bibfield  {title} {\enquote {\bibinfo {title}
  {First-principles order-parameter theory of freezing},}\ }\href {\doibase
  10.1103/PhysRevB.19.2775} {\bibfield  {journal} {\bibinfo  {journal} {Phys.
  Rev. B}\ }\textbf {\bibinfo {volume} {19}},\ \bibinfo {pages} {2775--2794}
  (\bibinfo {year} {1979})}\BibitemShut {NoStop}%
\bibitem [{\citenamefont {Robbins}\ \emph {et~al.}({2012})\citenamefont
  {Robbins}, \citenamefont {Archer}, \citenamefont {Thiele},\ and\
  \citenamefont {Knobloch}}]{Robbins2012a}%
  \BibitemOpen
  \bibfield  {author} {\bibinfo {author} {\bibfnamefont {M.~J.}\ \bibnamefont
  {Robbins}}, \bibinfo {author} {\bibfnamefont {A.~J.}\ \bibnamefont {Archer}},
  \bibinfo {author} {\bibfnamefont {U.}~\bibnamefont {Thiele}}, \ and\ \bibinfo
  {author} {\bibfnamefont {E.}~\bibnamefont {Knobloch}},\ }\bibfield  {title}
  {\enquote {\bibinfo {title} {{Modeling the structure of liquids and crystals
  using one- and two-component modified phase-field crystal models}},}\ }\href
  {\doibase {10.1103/PhysRevE.85.061408}} {\bibfield  {journal} {\bibinfo
  {journal} {{Phys. Rev. E}}\ }\textbf {\bibinfo {volume} {{85}}},\ \bibinfo
  {pages} {{061408}} (\bibinfo {year} {{2012}})}\BibitemShut {NoStop}%
\bibitem [{\citenamefont {Alster}\ \emph {et~al.}(2017)\citenamefont {Alster},
  \citenamefont {Elder}, \citenamefont {Hoyt},\ and\ \citenamefont
  {Voorhees}}]{Alster2017a}%
  \BibitemOpen
  \bibfield  {author} {\bibinfo {author} {\bibfnamefont {Eli}\ \bibnamefont
  {Alster}}, \bibinfo {author} {\bibfnamefont {K.~R.}\ \bibnamefont {Elder}},
  \bibinfo {author} {\bibfnamefont {Jeffrey~J.}\ \bibnamefont {Hoyt}}, \ and\
  \bibinfo {author} {\bibfnamefont {Peter~W.}\ \bibnamefont {Voorhees}},\
  }\bibfield  {title} {\enquote {\bibinfo {title} {Phase-field-crystal model
  for ordered crystals},}\ }\href {\doibase 10.1103/PhysRevE.95.022105}
  {\bibfield  {journal} {\bibinfo  {journal} {Phys. Rev. E}\ }\textbf {\bibinfo
  {volume} {95}},\ \bibinfo {pages} {022105} (\bibinfo {year}
  {2017})}\BibitemShut {NoStop}%
\bibitem [{\citenamefont {Elder}\ \emph {et~al.}(2018)\citenamefont {Elder},
  \citenamefont {Seymour}, \citenamefont {Lee}, \citenamefont {Hilke},\ and\
  \citenamefont {Provatas}}]{Elder2017a}%
  \BibitemOpen
  \bibfield  {author} {\bibinfo {author} {\bibfnamefont {K.~L.~M.}\
  \bibnamefont {Elder}}, \bibinfo {author} {\bibfnamefont {M.}~\bibnamefont
  {Seymour}}, \bibinfo {author} {\bibfnamefont {M.}~\bibnamefont {Lee}},
  \bibinfo {author} {\bibfnamefont {M.}~\bibnamefont {Hilke}}, \ and\ \bibinfo
  {author} {\bibfnamefont {N.}~\bibnamefont {Provatas}},\ }\bibfield  {title}
  {\enquote {\bibinfo {title} {Two-component structural phase-field crystal
  models for graphene symmetries},}\ }\href {\doibase 10.1098/rsta.2017.0211}
  {\bibfield  {journal} {\bibinfo  {journal} {Phil. Trans. Roy. Soc. Lond. A}\
  }\textbf {\bibinfo {volume} {376}} (\bibinfo {year} {2018})}\BibitemShut {NoStop}%
\bibitem [{\citenamefont {Wang}\ \emph {et~al.}({2018})\citenamefont {Wang},
  \citenamefont {Liu},\ and\ \citenamefont {Huang}}]{Wang2018c}%
  \BibitemOpen
  \bibfield  {author} {\bibinfo {author} {\bibfnamefont {Zi-Le}\ \bibnamefont
  {Wang}}, \bibinfo {author} {\bibfnamefont {Zhirong}\ \bibnamefont {Liu}}, \
  and\ \bibinfo {author} {\bibfnamefont {Zhi-Feng}\ \bibnamefont {Huang}},\
  }\bibfield  {title} {\enquote {\bibinfo {title} {{Angle-adjustable density
  field formulation for the modeling of crystalline microstructure}},}\ }\href
  {\doibase {10.1103/PhysRevB.97.180102}} {\bibfield  {journal} {\bibinfo
  {journal} {Phys. Rev. B}\ }\textbf {\bibinfo {volume} {{97}}},\ \bibinfo
  {pages} {{180102(R)}} (\bibinfo {year} {{2018}})}\BibitemShut {NoStop}%
\bibitem [{\citenamefont {Wu}\ \emph {et~al.}({2010})\citenamefont {Wu},
  \citenamefont {Plapp},\ and\ \citenamefont {Voorhees}}]{Wu2010a}%
  \BibitemOpen
  \bibfield  {author} {\bibinfo {author} {\bibfnamefont {Kuo-An}\ \bibnamefont
  {Wu}}, \bibinfo {author} {\bibfnamefont {Mathis}\ \bibnamefont {Plapp}}, \
  and\ \bibinfo {author} {\bibfnamefont {Peter~W.}\ \bibnamefont {Voorhees}},\
  }\bibfield  {title} {\enquote {\bibinfo {title} {{Controlling crystal
  symmetries in phase-field crystal models}},}\ }\href {\doibase
  {10.1088/0953-8984/22/36/364102}} {\bibfield  {journal} {\bibinfo  {journal}
  {{J. Phys.: Condens. Matter}}\ }\textbf {\bibinfo {volume} {{22}}},\ \bibinfo
  {pages} {{364102}} (\bibinfo {year} {{2010}})}\BibitemShut {NoStop}%
\bibitem [{\citenamefont {Jaatinen}\ and\ \citenamefont
  {Ala-Nissila}({2010})}]{Jaatinen2010a}%
  \BibitemOpen
  \bibfield  {author} {\bibinfo {author} {\bibfnamefont {A.}~\bibnamefont
  {Jaatinen}}\ and\ \bibinfo {author} {\bibfnamefont {T.}~\bibnamefont
  {Ala-Nissila}},\ }\bibfield  {title} {\enquote {\bibinfo {title}
  {{Eighth-order phase-field-crystal model for two-dimensional
  crystallization}},}\ }\href {\doibase {10.1103/PhysRevE.82.061602}}
  {\bibfield  {journal} {\bibinfo  {journal} {{Phys. Rev. E}}\ }\textbf
  {\bibinfo {volume} {{82}}},\ \bibinfo {pages} {{061602}} (\bibinfo {year}
  {{2010}})}\BibitemShut {NoStop}%
\bibitem [{\citenamefont {Evans}(2010)}]{Evans2009}%
  \BibitemOpen
  \bibfield  {author} {\bibinfo {author} {\bibfnamefont {R.}~\bibnamefont
  {Evans}},\ }\bibfield  {title} {\enquote {\bibinfo {title} {Density
  functional theory for inhomogeneous fluids i: Simple fluids in
  equilibrium},}\ }in\ \href@noop {} {\emph {\bibinfo {booktitle} {3rd Warsaw
  School of Statistical Physics}}},\ \bibinfo {editor} {edited by\ \bibinfo
  {editor} {\bibfnamefont {B.}~\bibnamefont {Cichocki}}, \bibinfo {editor}
  {\bibfnamefont {M.}~\bibnamefont {Napi\'orkowski}}, \ and\ \bibinfo {editor}
  {\bibfnamefont {J.}~\bibnamefont {Piasechi}}}\ (\bibinfo  {publisher} {Warsaw
  University Press},\ \bibinfo {address} {Warsaw},\ \bibinfo {year} {2010})\
  Chap.~\bibinfo {chapter} {2}, pp.\ \bibinfo {pages} {43--85}\BibitemShut
  {NoStop}%
\bibitem [{\citenamefont {L{\"o}wen}(2010)}]{Lowen2009}%
  \BibitemOpen
  \bibfield  {author} {\bibinfo {author} {\bibfnamefont {Hartmut}\ \bibnamefont
  {L{\"o}wen}},\ }\bibfield  {title} {\enquote {\bibinfo {title} {Density
  functional theory for inhomogeneous fluids ii: Statics, dynamics and
  applications},}\ }in\ \href@noop {} {\emph {\bibinfo {booktitle} {3rd Warsaw
  School of Statistical Physics}}},\ \bibinfo {editor} {edited by\ \bibinfo
  {editor} {\bibfnamefont {B.}~\bibnamefont {Cichocki}}, \bibinfo {editor}
  {\bibfnamefont {M.}~\bibnamefont {Napi\'orkowski}}, \ and\ \bibinfo {editor}
  {\bibfnamefont {J.}~\bibnamefont {Piasechi}}}\ (\bibinfo  {publisher} {Warsaw
  University Press},\ \bibinfo {address} {Warsaw},\ \bibinfo {year} {2010})\
  Chap.~\bibinfo {chapter} {3}, pp.\ \bibinfo {pages} {87--12}\BibitemShut
  {NoStop}%
\bibitem [{\citenamefont {Likos}(2006)}]{Likos2006}%
  \BibitemOpen
  \bibfield  {author} {\bibinfo {author} {\bibfnamefont {C.~N.}\ \bibnamefont
  {Likos}},\ }\bibfield  {title} {\enquote {\bibinfo {title} {Soft matter with
  soft particles},}\ }\href@noop {} {\bibfield  {journal} {\bibinfo  {journal}
  {Soft Matter}\ }\textbf {\bibinfo {volume} {2}},\ \bibinfo {pages} {478--498}
  (\bibinfo {year} {2006})}\BibitemShut {NoStop}%
\bibitem [{\citenamefont {Lenz}\ \emph {et~al.}(2012)\citenamefont {Lenz},
  \citenamefont {Blaak}, \citenamefont {Likos},\ and\ \citenamefont
  {Mladek}}]{Lenz2012}%
  \BibitemOpen
  \bibfield  {author} {\bibinfo {author} {\bibfnamefont {Dominic~A.}\
  \bibnamefont {Lenz}}, \bibinfo {author} {\bibfnamefont {Ronald}\ \bibnamefont
  {Blaak}}, \bibinfo {author} {\bibfnamefont {Christos~N.}\ \bibnamefont
  {Likos}}, \ and\ \bibinfo {author} {\bibfnamefont {Bianca~M.}\ \bibnamefont
  {Mladek}},\ }\bibfield  {title} {\enquote {\bibinfo {title} {Microscopically
  resolved simulations prove the existence of soft cluster crystals},}\ }\href
  {\doibase 10.1103/PhysRevLett.109.228301} {\bibfield  {journal} {\bibinfo
  {journal} {Phys. Rev. Lett.}\ }\textbf {\bibinfo {volume} {109}},\ \bibinfo
  {pages} {228301} (\bibinfo {year} {2012})}\BibitemShut {NoStop}%
\bibitem [{\citenamefont {Wu}\ and\ \citenamefont {Karma}({2007})}]{Wu2007}%
  \BibitemOpen
  \bibfield  {author} {\bibinfo {author} {\bibfnamefont {Kuo-An}\ \bibnamefont
  {Wu}}\ and\ \bibinfo {author} {\bibfnamefont {Alain}\ \bibnamefont {Karma}},\
  }\bibfield  {title} {\enquote {\bibinfo {title} {Phase-field crystal modeling
  of equilibrium bcc-liquid interfaces},}\ }\href@noop {} {\bibfield  {journal}
  {\bibinfo  {journal} {Phys. Rev. B}\ }\textbf {\bibinfo {volume} {{76}}},\
  \bibinfo {pages} {{184107}} (\bibinfo {year} {{2007}})}\BibitemShut {NoStop}%
\bibitem [{\citenamefont {Bodenschatz}\ \emph {et~al.}(2000)\citenamefont
  {Bodenschatz}, \citenamefont {Pesch},\ and\ \citenamefont
  {Ahlers}}]{Bodenschatz2000}%
  \BibitemOpen
  \bibfield  {author} {\bibinfo {author} {\bibfnamefont {E.}~\bibnamefont
  {Bodenschatz}}, \bibinfo {author} {\bibfnamefont {W.}~\bibnamefont {Pesch}},
  \ and\ \bibinfo {author} {\bibfnamefont {G.}~\bibnamefont {Ahlers}},\
  }\bibfield  {title} {\enquote {\bibinfo {title} {Recent developments in
  {R}ayleigh--{B}{\'e}nard convection},}\ }\href@noop {} {\bibfield  {journal}
  {\bibinfo  {journal} {Annu. Rev. Fluid Mech.}\ }\textbf {\bibinfo {volume}
  {32}},\ \bibinfo {pages} {709--778} (\bibinfo {year} {2000})}\BibitemShut
  {NoStop}%
\bibitem [{\citenamefont {Archer}\ and\ \citenamefont
  {Malijevsk{\`y}}(2016)}]{Archer2016}%
  \BibitemOpen
  \bibfield  {author} {\bibinfo {author} {\bibfnamefont {Andrew~J.}\
  \bibnamefont {Archer}}\ and\ \bibinfo {author} {\bibfnamefont {Alexandr}\
  \bibnamefont {Malijevsk{\`y}}},\ }\bibfield  {title} {\enquote {\bibinfo
  {title} {Crystallization of soft matter under confinement at interfaces and
  in wedges},}\ }\href@noop {} {\bibfield  {journal} {\bibinfo  {journal} {J.
  Phys.: Condens. Matter}\ }\textbf {\bibinfo {volume} {28}},\ \bibinfo {pages}
  {244017} (\bibinfo {year} {2016})}\BibitemShut {NoStop}%
\bibitem [{\citenamefont {Fan}\ \emph {et~al.}({2017})\citenamefont {Fan},
  \citenamefont {Hirvonen}, \citenamefont {Pereira}, \citenamefont {Ervasti},
  \citenamefont {Elder}, \citenamefont {Donadio}, \citenamefont {Harju},\ and\
  \citenamefont {Ala-Nissila}}]{Fan2017}%
  \BibitemOpen
  \bibfield  {author} {\bibinfo {author} {\bibfnamefont {Zheyong}\ \bibnamefont
  {Fan}}, \bibinfo {author} {\bibfnamefont {Petri}\ \bibnamefont {Hirvonen}},
  \bibinfo {author} {\bibfnamefont {Luiz Felipe~C.}\ \bibnamefont {Pereira}},
  \bibinfo {author} {\bibfnamefont {Mikko~M.}\ \bibnamefont {Ervasti}},
  \bibinfo {author} {\bibfnamefont {Ken~R.}\ \bibnamefont {Elder}}, \bibinfo
  {author} {\bibfnamefont {Davide}\ \bibnamefont {Donadio}}, \bibinfo {author}
  {\bibfnamefont {Ari}\ \bibnamefont {Harju}}, \ and\ \bibinfo {author}
  {\bibfnamefont {Tapio}\ \bibnamefont {Ala-Nissila}},\ }\bibfield  {title}
  {\enquote {\bibinfo {title} {{Bimodal grain-size scaling of thermal transport
  in polycrystalline graphene from large-scale molecular dynamics
  dimulations}},}\ }\href {\doibase {10.1021/acs.nanolett.7b01742}} {\bibfield
  {journal} {\bibinfo  {journal} {{Nano Lett.}}\ }\textbf {\bibinfo {volume}
  {{17}}},\ \bibinfo {pages} {{5919--5924}} (\bibinfo {year}
  {{2017}})}\BibitemShut {NoStop}%
\bibitem [{\citenamefont {Carrillo}\ \emph {et~al.}(2018)\citenamefont
  {Carrillo}, \citenamefont {Gvalani}, \citenamefont {Pavliotis},\ and\
  \citenamefont {Schlichting}}]{Carrillo2018}%
  \BibitemOpen
  \bibfield  {author} {\bibinfo {author} {\bibfnamefont {J.~A.}\ \bibnamefont
  {Carrillo}}, \bibinfo {author} {\bibfnamefont {R.~S.}\ \bibnamefont {Gvalani}},
  \bibinfo {author} {\bibfnamefont {G.~A.}\ \bibnamefont {Pavliotis}}, \ and\
  \bibinfo {author} {\bibfnamefont {A.}~\bibnamefont {Schlichting}},\ }\bibfield
   {title} {\enquote {\bibinfo {title} {Long-time behaviour and phase
  transitions for the {M}c{K}ean--{V}lasov equation on the torus},}\
  }\href@noop {} {\bibfield  {journal} {\bibinfo  {journal} {arXiv preprint
  arXiv:1806.01719}\ } (\bibinfo {year} {2018})}\BibitemShut {NoStop}%
\bibitem [{\citenamefont {Gomes}\ \emph {et~al.}(2019)\citenamefont {Gomes},
  \citenamefont {Kalliadasis}, \citenamefont {Pavliotis},\ and\ \citenamefont
  {Yatsyshin}}]{Gomes2019}%
  \BibitemOpen
  \bibfield  {author} {\bibinfo {author} {\bibfnamefont {Susana~N.}\
  \bibnamefont {Gomes}}, \bibinfo {author} {\bibfnamefont {Serafim}\
  \bibnamefont {Kalliadasis}}, \bibinfo {author} {\bibfnamefont {Grigorios~A.}\
  \bibnamefont {Pavliotis}}, \ and\ \bibinfo {author} {\bibfnamefont {Petr}\
  \bibnamefont {Yatsyshin}},\ }\bibfield  {title} {\enquote {\bibinfo {title}
  {Dynamics of the {D}esai--{Z}wanzig model in multiwell and random energy
  landscapes},}\ }\href@noop {} {\bibfield  {journal} {\bibinfo  {journal}
  {Phys. Rev. E}\ }\textbf {\bibinfo {volume} {99}},\ \bibinfo {pages} {032109}
  (\bibinfo {year} {2019})}\BibitemShut {NoStop}%
\bibitem [{\citenamefont {Bernis}\ \emph {et~al.}(1992)\citenamefont {Bernis},
  \citenamefont {Peletier},\ and\ \citenamefont {Williams}}]{Bernis1992}%
  \BibitemOpen
  \bibfield  {author} {\bibinfo {author} {\bibfnamefont {Francisco}\
  \bibnamefont {Bernis}}, \bibinfo {author} {\bibfnamefont {Lambertus~A.}\
  \bibnamefont {Peletier}}, \ and\ \bibinfo {author} {\bibfnamefont {S.~M.}\
  \bibnamefont {Williams}},\ }\bibfield  {title} {\enquote {\bibinfo {title}
  {Source type solutions of a fourth order nonlinear degenerate parabolic
  equation},}\ }\href@noop {} {\bibfield  {journal} {\bibinfo  {journal}
  {Nonlinear Analysis}\ }\textbf {\bibinfo {volume} {18}},\ \bibinfo {pages}
  {217--234} (\bibinfo {year} {1992})}\BibitemShut {NoStop}%
\bibitem [{\citenamefont {Doedel}\ \emph {et~al.}(1991)\citenamefont {Doedel},
  \citenamefont {Keller},\ and\ \citenamefont {Kernevez}}]{Doedel1991}%
  \BibitemOpen
  \bibfield  {author} {\bibinfo {author} {\bibfnamefont {E.}~\bibnamefont
  {Doedel}}, \bibinfo {author} {\bibfnamefont {H.~B.}\ \bibnamefont {Keller}},
  \ and\ \bibinfo {author} {\bibfnamefont {J.~P.}\ \bibnamefont {Kernevez}},\
  }\bibfield  {title} {\enquote {\bibinfo {title} {Numerical analysis and
  control of bifurcation problems ({I}): Bifurcation in finite dimensions},}\
  }\href {\doibase 10.1142/S0218127491000397} {\bibfield  {journal} {\bibinfo
  {journal} {Int. J. Bifurcat. Chaos}\ }\textbf {\bibinfo {volume} {1}},\
  \bibinfo {pages} {493--520} (\bibinfo {year} {1991})}\BibitemShut {NoStop}%
\bibitem [{\citenamefont {Kelley}(2003)}]{Kelley2003}%
  \BibitemOpen
  \bibfield  {author} {\bibinfo {author} {\bibfnamefont {C.~T.}\ \bibnamefont
  {Kelley}},\ }\href@noop {} {\emph {\bibinfo {title} {Solving Nonlinear
  Equations with {N}ewton's Method}}}\ (\bibinfo  {publisher} {SIAM},\ \bibinfo
  {address} {Philadelphia},\ \bibinfo {year} {2003})\BibitemShut {NoStop}%
\end{thebibliography}

%\begin{thebibliography}{99}

%\end{thebibliography}


\end{document}